\documentclass{pasa}%

\usepackage{graphicx}
\usepackage{hyperref}
\usepackage[table]{xcolor}
\usepackage{geometry}
\usepackage{tabularx}
\usepackage{subcaption}
\usepackage{mdframed}
\usepackage{diagbox}
\usepackage{enumitem}
\usepackage{comment}
\usepackage{tikz}
\usepackage[T1]{fontenc}
\usepackage[utf8]{inputenc}
\usepackage[french,spanish,german,english]{babel}
\def\checkmark{\tikz\fill[scale=0.4](0,.35) -- (.25,0) -- (1,.7) -- (.25,.15) -- cycle;}
\setlist{nolistsep}

\jid{PASA}
\doi{10.1017/pas.\the\year.xxx}
\jyear{\the\year}

\usepackage{aas_macros}
\usepackage{hyperref} 
\hypersetup{colorlinks,citecolor=blue,linkcolor=blue,urlcolor=blue}

\definecolor{blue}{HTML}{008ED7}
\definecolor{mygray}{gray}{0.75}
\definecolor{lightBlue}{HTML}{e5f7ff}

\title[Deep Learning for galaxies and cosmology]{The Dawes Review 10: The impact of deep learning for the analysis of galaxy surveys}

\author[M. Huertas-Company and F. Lanusse]
{M. Huertas-Company$^{1,2,3}$ and F. Lanusse$^2$\\
\affil{$^1$Instituto de Astrof\'isica de Canarias, c/ V\'ia L\'actea sn, 38025, La Laguna, Spain}
\affil{$^2$Universidad de La Laguna. Avda. Astrof\'isico Fco. Sanchez, La Laguna, Tenerife,Spain}
\affil{$^3$LERMA, Observatoire de Paris, CNRS, PSL, Universit\'e Paris-Cit\'e, France}
\affil{$^4$AIM, CEA, CNRS, Universit\'e Paris-Saclay, Universit\'e Paris-Cit\'e, Sorbonne Paris Cit\'e, F-91191 Gif-sur-Yvette, France}}

\begin{document}

\begin{frontmatter}
\maketitle

\begin{abstract}
The amount and complexity of data delivered by modern galaxy surveys has been steadily increasing over the past years. New facilities will soon provide imaging and spectra of hundreds of millions of galaxies.  Extracting coherent scientific information from these large and multi-modal data sets remains an open issue for the community and data driven approaches such as deep learning have rapidly emerged as a potentially powerful solution to some long lasting challenges. This enthusiasm is reflected in an unprecedented exponential growth of publications using neural networks, which have gone from a handful of works in 2015 to an average of one paper per week in 2021 in the area of galaxy surveys. Half a decade after the first published work in astronomy mentioning deep learning, and shortly before new big-data sets such as Euclid and LSST start becoming available, we believe it is timely to review what has been the real impact of this new technology in the field and its potential to solve key challenges raised by the size and complexity of the new datasets. The purpose of this review is thus two-fold. We first aim at summarizing, in a common document, the main applications of deep learning for galaxy surveys that have emerged so far. We then extract the major achievements and lessons learned and highlight key open questions and limitations, which in our opinion, will require particular attention in the coming years. Overall, state-of-the art deep learning methods are rapidly adopted by the astronomical community, reflecting a democratization of these methods. This review shows that the majority of works using deep learning up to date are oriented to computer vision tasks (e.g. classification, segmentation). This is also the domain of application where deep learning has brought the most important breakthroughs so far. However, we also report that the applications are becoming more diverse and deep learning is used for estimating galaxy properties, identifying outliers or constraining the cosmological model. Most of these works remain at the exploratory level though which could partially explain the limited impact in terms of citations. Some common challenges will most likely need to be addressed before moving to the next phase of massive deployment of deep learning in the processing of future surveys; e.g. uncertainty quantification, interpretability, data labeling and domain shift issues from training with simulations, which constitutes a common practice in astronomy. 

\end{abstract}

\end{frontmatter}

\section{Introduction}
\label{sec:intro}

Most fields in astronomy are rapidly changing. Unprecedentedly large observational data exists or will soon become available. Modern spectro-photometric surveys such as the Legacy Survey of Space and Time (LSST;~\citealp{Ivezic2019}) or Euclid~\citep{Laureijs2011} will provide high quality spectra and images for hundreds of millions of galaxies. Integral field spectroscopic surveys at low and high redshift are reaching statistically relevant sizes (e.g. MaNGA - \citealp{Bundy2015}) enabling to resolve the internal structure of galaxies beyond integrated properties. In addition, new facilities like the James Webb Space Telescope (JWST) are opening the window to a completely new redshift and stellar mass regime both in imaging and spectroscopy and we will be able to witness the emergence of the first galaxies in the universe. X-ray and radio facilities (e.g. SKA, Athena) will  probe cold and hot gas in galaxies with improved resolution. On the theory side, computing power has evolved to the extent that we can now generate realistic simulations of galaxies in a cosmological context spanning most of the Universe's history (e.g. TNG - \citealp{Pillepich2018}) which properly reproduce a large number of observable properties. In this context of growing complexity and rapid increase of data volumes, it has become a new challenge for the community to combine and accurately extract scientifically relevant information from these datasets. 

\begin{figure}
\centering
    \includegraphics[width=\linewidth]{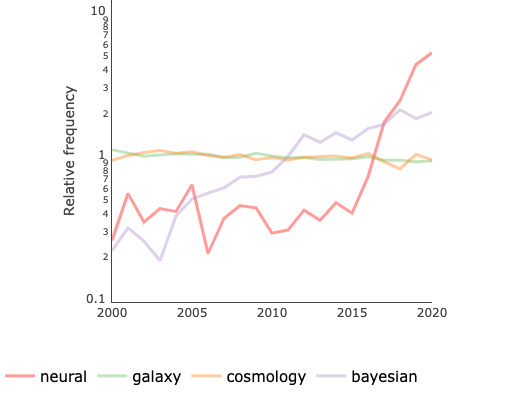}
    \caption{Relative  change of the number of papers on arXiv:astro-ph with different keywords in the abstract as a function of time. The number of works mentioning neural networks in the abstract has experienced an unprecedented growth in the last $\sim6$ years, significantly steeper than other topic in astrophysics. Source: ArXivSorter}
    \label{fig:dl_papers}
\end{figure}

Although Machine Learning applications to astronomy exist since at least thirty years ago (see~\autoref{sec:history}), the past years have witnessed an unprecedented increase of deep learning methods translated on an exponential increase of publications (\autoref{fig:dl_papers}). This \emph{revival} is fueled by significant breakthroughs in the field of Machine Learning since the  popularization of Convolutional Neural Networks (CNNs) a decade ago~\citep{Krizhevsky2012}.
The first published work mentioning deep learning in astronomy is from 2015 in which CNNs were applied for the classification of galaxy morphology. Since then, the number of works using deep learning in astrophysics has been growing exponentially, being the fastest growth of other topics in the field (\autoref{fig:dl_papers}). The generalization of deep learning represents to some extent a change of paradigm in the way we approach data analysis. By using gradient-based optimization techniques to extract meaningful features directly from the data, we move from an approach based on specific algorithms and features to a fully data driven one. It has potentially profound implications for astronomy and science in general. 

After half a decade of this new wave of applications of deep learning to astrophysics, we thus think it is timely to look back at the impact that this new technology has had in our field. Given the large amount of existing publications, we will only review works focusing on the analysis of galaxy surveys and we will restrict to recent works after the so called deep learning boom. We believe however that most of the lessons learned can be extrapolated to other areas of astrophysics. 

 It is not the goal of this review to provide technical details about how deep learning techniques work, but to describe its applications to cosmology and galaxy formation in a unique reference document.  Over the past years, deep learning has been used for a large variety of tasks such as classification, object detection, but also to derive physical properties of galaxies such as photometric redshifts, to identify anomalous objects or to accelerate simulations and constrain cosmology among many others. The review is thus structured following major areas of application. We provide a quick description and some keywords about the technical solutions adopted for each science case but we emphasize it is not the goal of this work to focus on technical aspects. The reader is encouraged to read the publications referring to each method - which are provided in a best effort basis - to obtain a complete and formal description of the deep learning methods. For completeness and easy access, we provide in the \autoref{app:acron} a list of different methods mentioned in the review with the corresponding references. A list of current and future galaxy surveys which are referred in this work is also provided in \autoref{app:acron}.
 
 Following this idea, we have divided the applications of deep learning into four major categories: 
 \begin{itemize}
     \item Deep learning for general computer vision tasks. These are applications that we consider closest to standard computer vision applications for natural images for which deep learning has been shown to generally outperform other traditional approaches.  It typically includes classification and segmentation tasks.
     \item Deep learning to derive physical properties of galaxies (both posteriors and point estimates). These are applications in which deep learning is used to estimate galaxy properties such as photometric redshifts or stellar populations properties. Neural Networks are typically used to replace existing algorithms with a faster and more efficient solution, hence more suited for large data volumes. In addition, we also review applications in which deep learning is employed to derive properties of galaxies which are not directly accessible with known observables, i.e. to find new relations between observable quantities and physical properties of galaxies from simulations. 
     \item Deep learning for assisted discovery. Neural networks are used here for data exploration and visualization of complex datasets in lower dimension. We include also in this category, efforts to automatically identify potentially interesting new objects, i.e. anomalies or outliers.
     \item Deep learning for cosmology.  Cosmological simulations including baryonic physics are computationally expensive. Deep learning can be used as a fast emulator of the galaxy-halo connection by populating dark matter halos. In addition, a second major application is cosmological inference. Cosmological models are traditionally constrained using summary statistics (e.g. 2 point statistics). Deep learning has been used to bypass these summary statistics and constrain models using all available data.  
 \end{itemize}
 
 This is of course a subjective division of the applications of deep learning to cosmology and galaxy formation. There necessarily exist overlaps between the different categories. The review is organized such that, for each family of applications, we  review the state of the art and key publications, highlight where the limitations are and what could be the most promising research lines for the future (\autoref{sec:low_level} to~\autoref{sec:acc}).  In the final sections (\autoref{sec:final_thoughts}), we assess the impact of deep learning for galaxy surveys and extract some global lessons learned. We have tried to provide a fair and complete description of the different works. However, as previously stated, the field has exploded in the last years and it has become more and more difficult to keep track of all new publications. This is partly one of the motivations for this review. It is also implies that we might easily miss some relevant works. We apologize in advance.
 
 \section{A very brief historical overview - or what we are not covering in this review}
 \label{sec:history}
 
Before we start discussing the most relevant results and applications, we would like to clarify that this review focuses on recent applications of deep learning, essentially after the first applications of CNNs to astronomy. As described in the introduction section, we consider as \emph{deep learning} all recent developments around neural networks which have arisen in the last decade approximately, since the first practical application of convolutional neural networks for image classification. Deep learning generally designates gradient-based optimization techniques of modular architectures of varying complexity; it is therefore a sub field of the more general machine learning discipline. There is a long history of machine learning applications in astronomy which started since well before the more recent deep learning boom. Different types of machine learning algorithms including early Artificial Neural Networks (ANNs), Decision Trees (DTs), Random Forests (RFs) or kernel algorithms such as Support Vector Machines (SVMs) have been applied to different areas of astrophysics since the second half of the past century. For example ANNs, decision trees and Self Organizing Maps (SOMs) have been extensively applied to the classification of stars and galaxies (e.g.~\citealp{Odewahn1992,Weir1995,Miller1996,Bazell1998,Andreon2000,Qin2003,Ball2006}); an ANN being the primary way of identifying point sources in the commonly used SExtractor software~\cite{Bertin1996} for segmentation of astronomical images. The problem of galaxy morphology classification has also been subject to a significant amount of machine learning related works led in particular by the group of O. Lahav and collaborators using ANNs and DTs (e.g.~\citealp{StorrieLombardi1992,Lahav1995,Lahav1996,Odewahn1996,Naim1997,Madgwick2003,Cohen2003}).~\cite{ball2004} is likely the first work to use ANNs to classify galaxies in the SDSS. In the first decade of the present century SVMs became more popular and were also used to provide catalogs of galaxy morphology (e.g.~\citealp{HuertasCompany2008,HuertasCompany2011}). Decision Trees have also been applied to other classification tasks such as AGN/galaxy separation (e.g.~\citealp{White2000,Gao2008}). Beyond classification, machine learning, and especially ANNs have been extensively applied to the problem of estimating photometric redshifts (e.g.~\citealp{DAbrusco2007,Li2007,Banerji2008}).  This review will not describe these works though. We refer the reader to~\cite{Ball2010} and~\cite{Baron2019} for a complete and extensive review of \textit{pre-deep learning} machine learning techniques applied to astronomy. This obviously does not mean that other machine learning approaches are less interesting for astrophysics. There have been recently very relevant applications of RFs for example for anomaly detection (see the works by~\citealp{Baron2017}) and to assess the main causes of star formation quenching in galaxies (e.g.~\citealp{Bluck2022}) among many others. However, we have made the choice not to include a detailed description of these works in this review. We will focus essentially on how deep learning has changed the landscape in the past half decade. 

What is different with deep learning and why a dedicated review? In many aspects, deep learning represents a change in the way we approach data analysis. Because, we now have access to large datasets and the computing resources are powerful enough - especially thanks to Graphic Processing Units - we can move from an algorithmic-centered approach relying on manually engineered features to a fully data-driven unsupervised feature learning approach. This implies that instead of developing advanced domain specific algorithms for each task, we rely on a generic optimization algorithm to extract the most meaningful features in an end-to-end training loop. This is a new approach to data in astrophysics and in science in general. This change of paradigm has enabled in fact tremendous progress in the computer vision community, especially for image classification, but also for many other tasks such as translation, speech recognition or image segmentation over the past ten years. The purpose of this review is therefore to assess what has been the impact so far of this new approach for data processing in the fields of galaxy formation and cosmology. 

\section{Deep learning for computer vision tasks in astronomy}
\label{sec:low_level}
We begin by reviewing applications close to standard computer vision problems, for which deep learning approaches have been demonstrated to be very efficient. We focus on classification and source detection. 

\subsection{Classification}
\label{sec:class}

Source classification is a basic first order processing step in most deep surveys, for which deep learning has had a noticeable impact in the past years. The rapid penetration of deep learning can be naturally explained because it is arguably the most straightforward out-of-the box application. Deep learning started in fact to attract the attention of the computer vision community, when convolutional neural networks first won the ImageNet contest of image classification~\citep{Krizhevsky2012}. 

 One of the first tasks scientists do when confronted with a complex problem is to identify objects that look morphologically similar. In extra galactic astronomy, object classification can be of different flavours. For imaging, the most common applications which we review in the remaining of this section, are galaxy morphology classification, star-galaxy separation, and strong lenses detection. We understand this is not an exhaustive list of all image classification applications, however the techniques and approaches used are representative. From the spectroscopic point of view, there have been some works attempting to classify galaxy spectra. However, this remains less common than images. Finally, the classification of transients is something which has been extensively explored over the last years, especially in view of the LSST survey from the Rubin Observatory.

\subsubsection{Optical/NIR galaxy morphology}
\label{sec:opt_mor}
Galaxy morphological classification is a paradigmatic example of a science case where deep learning has rapidly become the state-of-the art. This task was first done by E. Hubble who classified galaxies in the well known Hubble sequence. The classification scheme, which is now more than a 100 years old, establishes that galaxies in the Universe today come essentially in two flavors. On one side, there are disk galaxies, like our own Milky Way; on the other side elliptical like galaxies. We now know that, besides morphology, the broad classes present different physical properties. Understanding the origin of the morphological diversity remains an open issue in the field of galaxy formation.


Therefore, galaxy morphological classification is still performed in almost all extra galactic imaging surveys. The traditional way to estimate galaxy morphology has been through visual inspection. However, this approach became prohibitively time consuming in the last decade with the advent of large extra galactic imaging surveys such as the Sloan Digital Sky Survey (SDSS). Approaches to overcome this limitation came in two fronts essentially. Citizen science approaches, of which the Galaxy Zoo project is the more popular example~\citep{Lintott2008}, were developed to classify large samples of galaxies.  Automation through Machine Learning has been always on the table as early as 1990's~(e.g. \citealp{Spiekermann1992}) and continued in the 2000's (e.g.~\citealp{HuertasCompany2008}).~\cite{HuertasCompany2011} was the first to provide the community with a Machine Learning based classification of SDSS galaxies. However the accuracy reached by these early approaches based on manually engineered features remained moderate -  especially when dealing with detailed morphological features such as bars or spiral arms -  hampering their penetration in the community. The main limitation is that the features typically used by these methods present only weak correlations with detailed morphological structures and are also very dependent on noise and spatial resolution.

\begin{figure*}
\centering
    \includegraphics[width=0.30\linewidth]{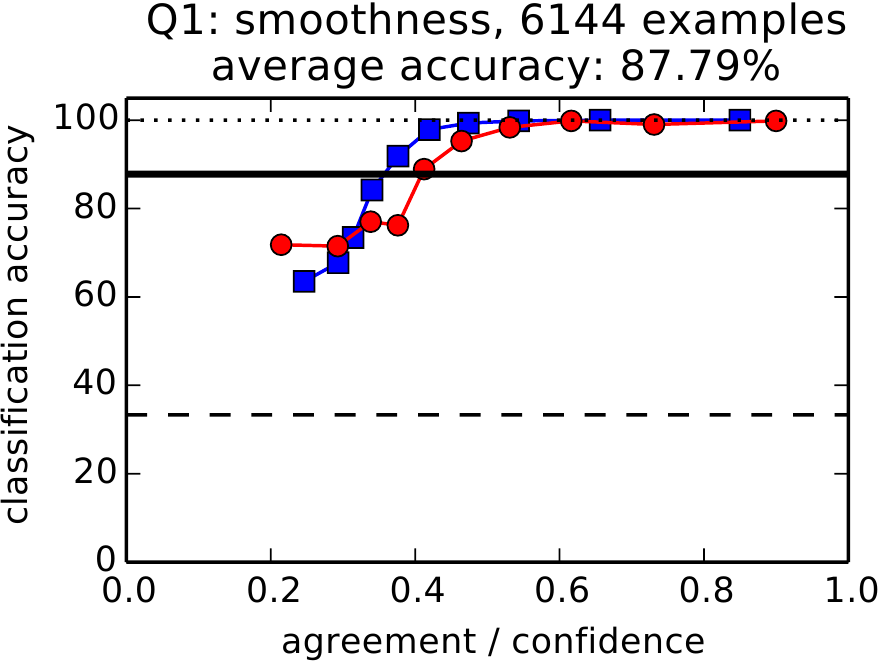}
    \includegraphics[width=0.30\linewidth]{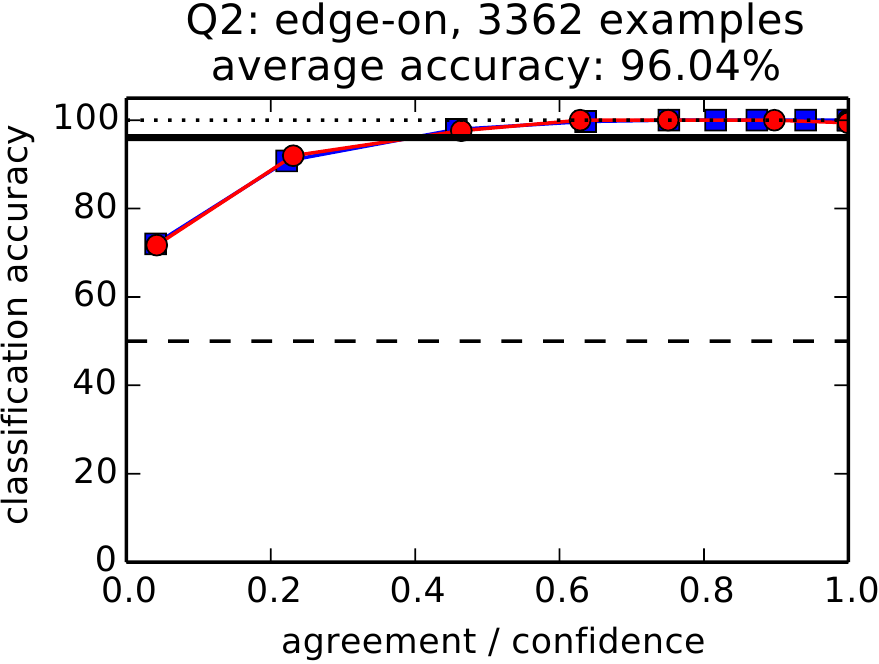}
    \includegraphics[width=0.30\linewidth]{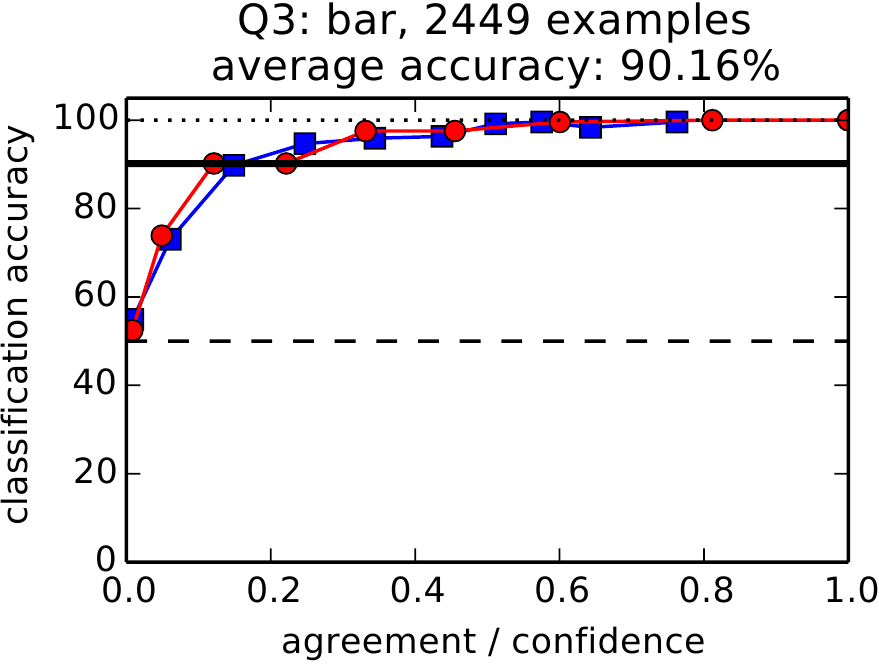}
    \caption{  Example of level of agreement (red circles) and model confidence (blue squares) versus classification accuracy. Each panel shows a different question in the Galaxy Zoo classification tree (smoothness, edge-on, bar). The authors quote an unprecedented accuracy of $>90\%$. This is the first work that uses CNNs in astrophysics. The figure is adapted from~\cite{dieleman2015}}
    \label{fig:diel1}
\end{figure*}

In this context, it is not a surprise that the first works using Convolutional Neural Networks in astronomy focus on galaxy morphology~\citep{dieleman2015,huertascompany2015}. The first one used labeled images from the Galaxy Zoo2 sample and trained a supervised CNN to estimate the morphological properties of SDSS galaxies going from global morphology to more detailed properties such as the number of spiral arms. The work by~\cite{dieleman2015} was the winner of a public challenge on the Kaggle platform~\footnote{\url{https://www.kaggle.com/}}. It achieved unprecedented classification accuracy of $>90\%$ in most of the tasks (\autoref{fig:diel1}).

\begin{figure}
\centering
    \includegraphics[width=\linewidth]{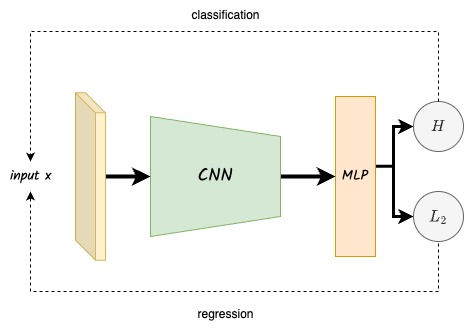}
    \caption{Schematic view of a simple \emph{Vanilla} type Convolutional Neural Network, the most common approach for binary, multi class classification and regression in extragalactic imaging. The input, which is typically an image is fed to a series of convolutional layers. The resulting embedding is used an input of a Multi Layer Perceptron which outputs a float or array of floats. If the problem is a classification, the standard loss function is the crossentropy ($H$), while if it is a regression the quadratic loss ($L_2$) is usually employed. }
    \label{fig:vanilla}
\end{figure}

This is a major improvement compared to previous approaches and marks the beginning of the penetration of deep learning techniques in astrophysics. 

   

Soon after,~\cite{huertascompany2015} applied a similar architecture to high redshift galaxies observed with the Hubble Space Telescope (HST), demonstrating again a similar improvement as compared to other approaches, including feature based Machine Learning. Using CNNs to classify galaxy images represents in some sense a change of paradigm in the way we approach the classification problem, analogous to what happened with natural images. Instead of manually trying to identify specific features that correlate with the classes to distinguish, the features are learnt simultaneously with the classification process. This of course entails loosing some interpretability, since the features learned by the network are no longer directly associated with physical properties. Interpretability is a major issue in deep learning applications to natural sciences that we will address in \autoref{sec:challenges}.  

\paragraph{CNNs as state-of-the art approach}
In the past years, the number of works using deep learning to classify galaxies based on their morphology has exploded, and CNNs have been used to classify the morphologies of galaxies in a variety of optical/Near Infrared surveys (we address the radio domain in \autoref{sec:radio}). It demonstrates that deep learning has fast become the state-of-the-art approach to estimate galaxy morphology in big datasets. Figure~\ref{fig:vanilla} illustrates the most common approach used for morphological classification. Images are fed into a series of convolutional layers which extract some summary statistics. The extracted statistics are then fed into a Multi Layer Perceptron which maps them into a class. The employed loss function is usually a cross-entropy loss. We notice that in the original approach by~\cite{dieleman2015}, a series of siamese networks were introduced to add rotational invariance. This approach has not been used in other works. ~\cite{dominguezsanchez2018} revisited the proof-of-concept work by~\cite{dieleman2015} by carefully cleaning the training sets and released the first deep learning catalog of galaxy morphologies in SDSS.~\cite{ghosh2020} explored the classification of distant galaxies from the CANDELS survey based on their bulge-to-total ratios.~\cite{Goddard2020} applied a similar strategy for Pan-STARRS. ~\cite{VegaFerrero2021} and~\cite{Cheng2021a} used CNNs to classify galaxies in the Dark Energy Survey.~\cite{Bom2021} followed a similar strategy for the S-PLUS survey and~\cite{walmsley2022} classified galaxies in the DECALS survey. In addition to observations, CNNs have also been extended to classify images from cosmological simulations in order to assess the realism of galaxy morphologies~\citep{HuertasCompany2019, Varma2022}. In such applications, the CNN is trained on labeled observations and then applied to mock images. In addition to global morphology, ~\cite{Tadaki2020} used also a similar supervised CNN setting to classify spiral galaxies based on their resolved properties (i.e. type of spiral arms).

\subsubsection{Radio galaxy morphology}
\label{sec:radio}
In addition to the Optical and Near Infrared domains, the radio astronomy community has also been very active in developing and testing deep learning techniques for the classification of radio galaxies. These efforts are motivated by the forthcoming arrival of new radio facilities such as SKA\footnote{\url{https://www.skatelescope.org/the-ska-project/}} that will change the landscape by detecting hundreds of thousands of new radio galaxies. Similarly to what happened in the optical, the availability of large datasets with labels has enabled the community to extensively test deep learning for classification. The Radio Galaxy Zoo project~\citep{Banfield2015} used indeed citizen science to determine the host galaxy of the radio emission and the radio morphology of $\sim170,000$ galaxies. This therefore constitutes an excellent database for applying neural networks. It highlights the importance of the preparatory work done by the community for accelerating the adoption of deep learning techniques.

The first work exploring CNNs for classification of radio galaxies is~\cite{Aniyan2017}. They use a simple sequential CNN (\autoref{fig:vanilla}) and conclude that an accuracy up to $\sim95\%$ can be achieved 
in classifying galaxies in three main classes - Fanaroff-Riley I (FRI), Fanaroff-Riley II (FRII) and bent-tailed galaxies. A number of works have followed. ~\cite{Alhassan2018} also reported similar accuracy when using CNNs to classify compact and extended radio sources observed in the FIRST radio survey (see also~\citealp{MaslejKresnakova2021} for similar conclusions).~\cite{Lukic2018} explores different network configurations and concludes that a three layer network is typically enough to reach more than $90\%$ accuracy. \cite{Wu2019a} explored Faster Region-based Convolutional Neutral Networks to detect and classify radio sources from the Radio Galaxy Zoo project.

\subsubsection{Strong Lenses}

\label{sec:stronglenses}

Deep Learning based supervised classification has been widely extended to other extragalactic classification tasks. An example of application which has significantly benefited from the advent of deep learning is the detection of strong gravitational lenses (e.g.~\citealp{jacobs2017,petrillo2017,Lanusse2018, davies2019,schaefer2018,metcalf2019, petrillo2019,jacobs2019, li2020, huang2020}). Gravitational lenses produce characteristic distortions of the light of background sources, caused by the presence of a foreground massive galaxy or cluster in the same line of sight. The analysis of strong lenses provides information about the total matter distribution of the foreground system. Strong lenses provide therefore a unique probe of the dark matter distribution in galaxies. The first step consists in identifying the lenses on large samples of galaxy images. 

\begin{figure}
\centering
    \includegraphics[width=\linewidth]{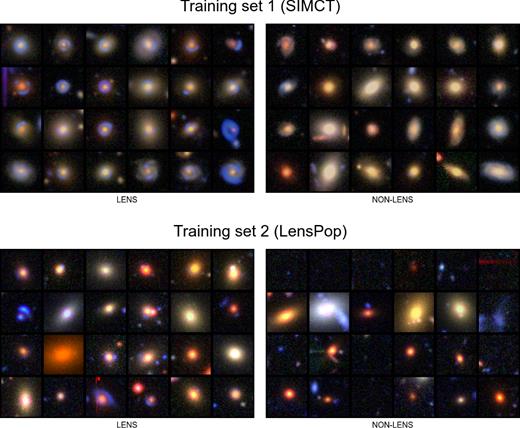}
    \caption{Example of two different simulated samples of strong lenses used for training a CNN. These simulations were used to detect strong lenses in the CFHTLS survey. Figure adapted from~\cite{jacobs2017}}
    \label{fig:lens1}
\end{figure}

Similarly to what has been described for galaxy morphology, the usual method to identify lenses is through Convolutional Neural Networks as done for galaxy morphology (\autoref{fig:vanilla}). However, there are some specific issues related to strong lensing detection since the problem is severely unbalanced. The number densities of strong lenses are indeed several orders of magnitude smaller than the ones of regular galaxies. This poses two main problems. First, it is impossible to build a large enough training sample of observed lenses. Second, in order to be scientifically useful, the classifier needs to reach extremely high purity values because even a small contamination from negative examples provokes that the sample of lenses is dominated by false positives. The community has proposed two sorts of solutions to these problems which appear in most of the works. To cope with the lack of training examples, the CNNs are usually trained on simulations. The physics of strong lenses is sufficiently well known so that lenses can be simulated with some degree of realism (see~\autoref{fig:lens1}). This practice is rather common in astrophysical applications as we will describe in the forthcoming sections and especially in section~\autoref{sec:final_thoughts}. It does not come free of biases though. The work by ~\cite{Lanusse2018} highlights  the importance of using realistically complex simulations for training in order to limit the potential biases.

The problem with false positives is in general more difficult to solve. The usual solution consists in visually inspecting the strong lenses candidates to remove the false positives. In that context, deep learning helps to reduce by several orders of magnitude the amount of required visual inspections but does not completely get rid of them.

Despite these issues, deep learning has rapidly become the state-of-the art technique to find strong lenses in large surveys. 
It has been successfully applied to multiple imaging surveys following very similar strategies as just described. The first work using CNNs for identifying lenses focused on the CFHTLS survey~\cite{jacobs2017}.~\cite{petrillo2017} and~\cite{petrillo2019} applied a similar strategy for KIDS galaxies and~\cite{jacobs2019} extended the approach to DES.~\cite{huang2020} focused on the DECALS datasets. Some other works focus more on simulations only in view of preparing future surveys.~\cite{Lanusse2018} demonstrated the performance of CNNs to detect lenses in LSST images and~\cite{davies2019} focused on Euclid.

 Overall, the conclusions are shared among all different works. Deep learning approaches are shown to improve more traditional techniques and therefore will likely be used on future surveys. In support for this, the work by ~\citealp{metcalf2019} shows the results of a strong lensing detection challenge in the framework of the Euclid survey where the five best algorithms were based on CNNs.    

\subsubsection{Open issues}
\label{sec:openissuesmorph}
In summary, deep learning techniques have rapidly replaced traditional approaches for classification of astronomical images to the point that it will most likely be the adopted approach to classify galaxies in forthcoming surveys across the electromagnetic spectrum. The main advantages are speed and accuracy. Convolutional Neural Networks in particular have demonstrated to be more accurate for these classification tasks than other feature based automated methods as seen in other disciplines. Classification is also one of the less risky applications in the sense that it is - in most cases - very close to the application of deep learning in the computer vision community. Some open issues remain still.

\begin{figure}
\centering
    \includegraphics[width=\linewidth]{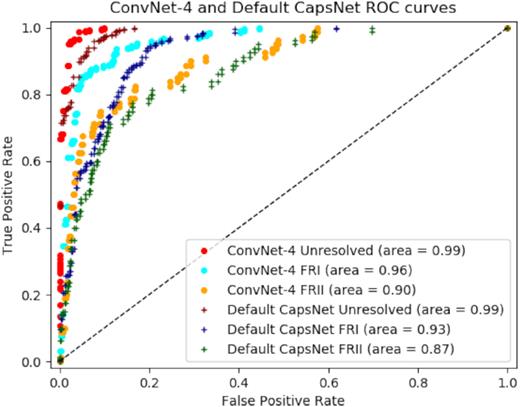}
    \caption{Comparison of Capsule Networks and CNNs applied to classify morphologies of radio galaxies. The ROC curves show the different performances. Capsule Networks do not offer a significant gain in this context. Figure adapted from~\cite{Lukic2019}}
    \label{fig:radio}
\end{figure}

\paragraph{Beyond vanilla CNNs}
Even though standard vanilla CNNs provide in general accurate results, several works have explored more complex configurations commonly used in computer vision such  as ResNets (e.g.~\citealp{Zhu2019, Kalvankar2020}).Overall, the results are promising but do not significantly improve over approaches based on simpler CNNs. Some works have systematically compared different neural network architectures on the same dataset (e.g.~\citealp{Fielding2021,Cavanagh2021}). The general conclusion is that, even if there are some differences in accuracy and in efficiency, all architectures fall in the same ballpark. A possible explanation for this is that astronomical images present in general less diversity than natural ones and hence relatively simple CNNs suffice to extract the relevant information. The works by~\cite{Katebi2019} and \cite{Lukic2019} are among the few works in astronomy exploring the use of Capsule Networks~\citep{Sabour2017}. Capsule Networks are proposed as an alternative to CNNs that incorporates spatial information 
about the features present in the images. Very briefly, it uses a sort of \textit{inverse rendering} to encode the presence of a given object in an image. It therefore encodes information, not only about the presence or not of a given object - which is what a CNN would do - but also  
information about where the object is and how it is oriented. We do not provide a detailed representation of the architectures given that this type of approach has only been marginally used. The conclusion of the work by~\cite{Lukic2019} is that, in the case of radio galaxy classification, Capsule Networks perform less well than more standard CNNs, reaching only an accuracy of $\sim75\%$ (\autoref{fig:radio}). One possible explanation for that is that Capsule Networks were initially thought to identify scenes which do not look realistic. For example a CNN would learn to recognize faces based on features like eyes or noses. However, they do not take into account the position of these features in the image. Capsule Network do, but this is not a common issue for astronomical imaging. This might be one of the reasons why Capsule Networks have not been very used in astronomy so far.~\cite{Becker2021} did the exercise of systematically testing the performance of CNNs for radio galaxy classification using multiple performance metrics such as inference time, model complexity, computational complexity, and mean per class accuracy. They report three main types of architectures that perform best but they are all variations of sequential CNNs. In a recent work, ~\cite{Tang2021} explores the use of multi-branch CNNs to simultaneously learn from multiple survey inputs (NVSS and FIRST). Interestingly they confirm that including multi-domain information allows to reduce the number of miss classifications by $\sim40\%$.  

\begin{figure}
\centering
    \includegraphics[width=\linewidth]{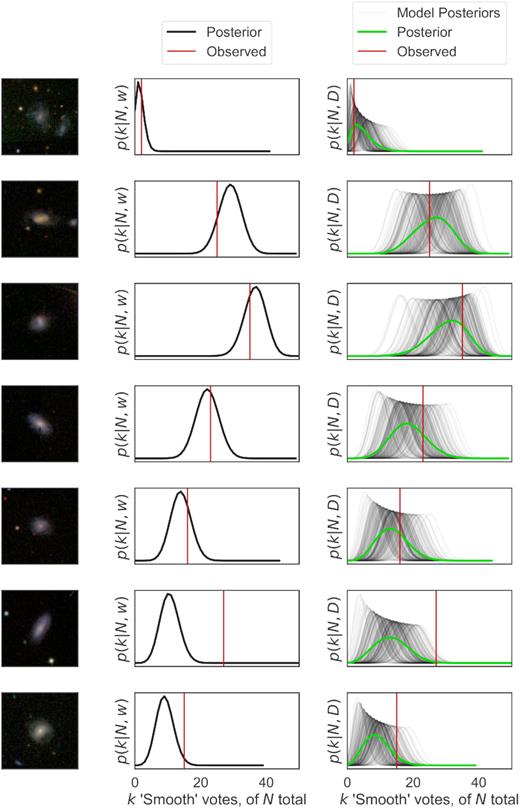}
    \caption{Example of posterior distributions of morphologies estimated from the votes of different classifiers. The leftmost column shows an image of a galaxy. The middle column shows the posterior predicted by a single network (black), while the right column shows the posterior averaged over 30 Montecarlo dropout approximated networks. The red vertical lines indicate the ground truth value, which generally shows a good overlap with the posterior distribution. Figure adapted from~\cite{walmsley2020}}
    \label{fig:bayes_walmsley}
\end{figure}

\begin{figure}
\centering
    \includegraphics[width=\linewidth]{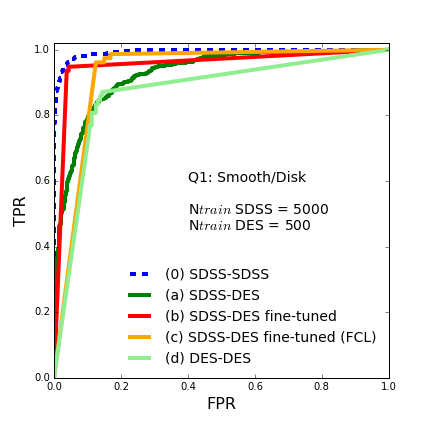}
    \includegraphics[width=\linewidth]{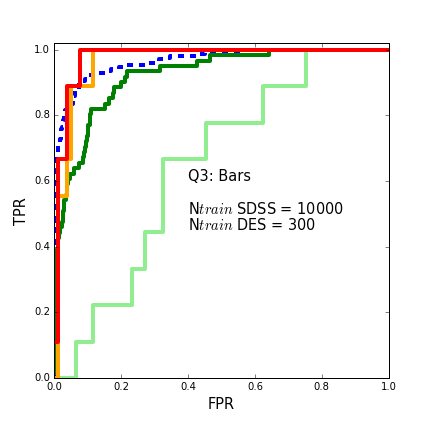}
    \caption{Example of Transfer Learning adapted to galaxy morphological classification exploring how to use a CNN trained on SDSS to the DES survey. Each panel shows a different morphological feature (smooth/disk, bars from top to bottom). The different lines show the ROC curves using different training models and test sets. The work concludes that when using a small subset of DES galaxies to refine the weights of a CNN trained on SDSS, the accuracy of the classification becomes optimal (red solid lines compared to blue dashed lines). Figure adapted from~\cite{dominguezsanchez2019}}
    \label{fig:ds2019}
\end{figure}

\paragraph{Labeled data}
A common bottleneck for deep learning based classifcation is the availability of labeled samples to train the supervised algorithms. In astronomy this is particularly delicate since the data properties (e.g. noise, resolution) change from one dataset to another so in theory the labeling process should be repeated for every new dataset.  A number of works have addressed this issue with different approaches. The work by~\cite{walmsley2020} explores Active Learning as a way to reduce the amount of required examples for training. Active learning allows one to select the most informative examples for the model which are the ones showed to human classifiers. The work by~\cite{walmsley2020} is also the first to explore Bayesian deep learning as a way to both estimate uncertainties an also identify the most informative examples (\autoref{fig:bayes_walmsley}).  ~\cite{dominguezsanchez2019} and~\cite{ghosh2020} explored transfer learning, which consists on refining the weights of a neural network trained on a similar labeled dataset to reduce the need of large training samples (\autoref{fig:ds2019}). They showed that the amount of labeled examples can be reduced by a factor of a few with this approach. The issue of the size of the training set that is needed is also investigated by~\cite{Samudre2022}. The authors explore whether reliable morphological classifications can be obtained with a small sample of $2000$ labeled images. They namely test transfer learning but also few-shot learning techniques based on twin networks. The conclusion is that even with small datasets, reliable classifications can be obtained using CNNs, with an adapted training strategy. The recent work by~\cite{Walmsley2021} explores another version of transfer learning. They show that the features learned by the CNNs for a given task can be recycled to estimate other morphological properties. ~\cite{VegaFerrero2021} used instead a simulated training set built from an observational sample from SDSS to classify more distant galaxies from the Dark Energy Survey. 

Despite some remaining issues, some of which are common to most of deep learning application - see \autoref{sec:final_thoughts} for a more detailed discussion - it is probably safe to argue that the community has accepted that deep learning will be employed to classify galaxies from future surveys such as LSST and Euclid.

\subsubsection{Transient Astronomy}
\label{sec:trans_class}
The field of transient astronomy is about to change dramatically with the arrival of synoptic sky surveys such as the LSST survey by the Rubin Observatory, which will observe large areas of the sky with an unprecedented frequency to find variable and transient astronomical sources. The number of detections per night is expected to easily exceed several thousands. The community has seen in machine learning techniques and particularly in deep learning a promising way of classifying the detected objects and filtering the most potentially (unknown) interesting candidates (see the report by~\citealp{Ishida2019}). We will address the discovery of new types of transients in \autoref{sec:discovery}. We focus here one the supervised classification of variable sources.

One key science topic for cosmology is the detection and characterization of SuperNovae light curves. There are different types of Supernovae and not all are useful for the same purposes. For example, SNIa are used for cosmology. Rapidly identifying the type of object saves - among other things - telescope time. Although this is ideally done with spectroscopy, it is unfeasible to perform a spectroscopic follow up of all the sources that will be detected. Therefore the community started as early as 2010 to prepare for this data deluge with the creation of simulated datasets such as the Supernovae Photometric Classification Challenge (SPCC) or the Photometric LSST Astronomical Time-Series Classification Challenge (PLAsTiCC)\footnote{\url{https://plasticc.org/}}. Because of these early efforts, there exists a consequent literature using \emph{pre deep learning} machine learning methods (e.g. SVMs, RFs and ANNs) to address the problem of SN light curve classification (e.g.~\citealp{Lochner2016,Villar2019,Hosseinzadeh2020,VargasdosSantos2020}). 

\begin{figure}
\centering
    \includegraphics[width=\linewidth]{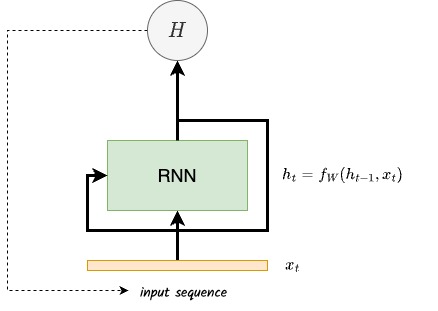}

    \caption{Schematic view of a Recursive Neural Network (RNN) which has been used for classification of light curves in several works. The photometric sequence is fed to a recursive block and trained with a crossentropy loss. The RNN blocks keep a memory ($h_t$) of the previous time step which make them suitable for handling time sequences.}
    \label{fig:RNNs}
\end{figure}

\begin{figure}
\centering

     \includegraphics[width=\linewidth]{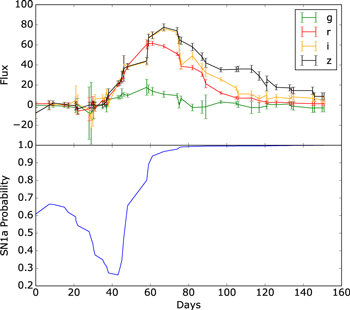}
    \caption{RNNs used for SN photometric light curve classification. The figure shows an example of light curve in five photometric bands (top panel) and the probability of classifications as a function of the time step (bottom panel). After roughly 50 days, the supernova type is classified with high confidence. Figure adapted from~\cite{Charnock2017}}
    \label{fig:RNNs_SN}
\end{figure}

The first work to use deep learning for SN light curve classification is by~\cite{Charnock2017}. They use to that purpose Recurrent Neural Networks (RNNs), which are a type of Neural Network architecture designed to handle sequences of variable length (see~\autoref{fig:RNNs} for a simple illustration). They are called recurrent because they keep a memory of the previous information in the sequence and use it to make the predictions. They are typically used for language modelling. The authors report an average accuracy above $90\%$ for the classification of light curves in three types - supernovae types I, II and III - on the simulated sample from the Supernovae Photometric Classification Challenge (\autoref{fig:RNNs_SN}). Despite the small training set of a hundred data points, RNNs achieved state-of-the-art results compared with a combination of template fits and boosted decision trees~\citep{Lochner2016}.  In addition of not requiring feature engineering, one advantage of RNNs is the ability to classify incomplete light curves.~\cite{Moss2018} also explores RNNs on the same simulated dataset. They propose some improvements such as a stronger data augmentation process to mitigate the effects of small samples and reach $>95\%$ accuracy. A similar conclusion is reached by~\cite{Moeller2020} who also explored RNNs. They report a similar accuracy also on realistic simulations and confirm accuracies above $85\%$ for incomplete light curves. See also~\cite{Burhanudin2021} for similar conclusions. This latter work proposes handling imbalance with a focal loss function.

\begin{figure}
\centering
    \includegraphics[width=\linewidth]{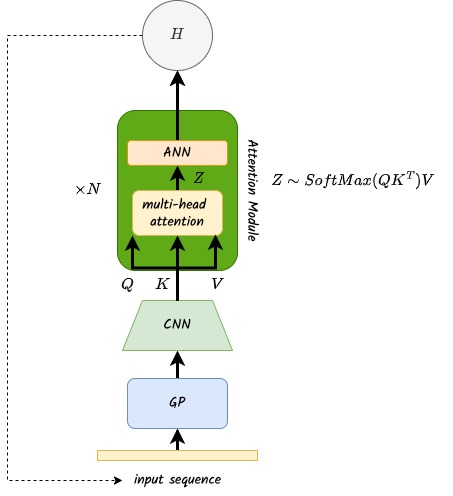}
    \caption{Example of typical transformer architecture for the classification of light curves. A Gaussian Process is first used to homogenize the time sampling and the resulting sequence is fed into a CNN for feature extraction. The attention modules are used to extract the final features which are used to classify.}
    \label{fig:transformer}
\end{figure}

\begin{figure}
\centering
    \includegraphics[width=\linewidth]{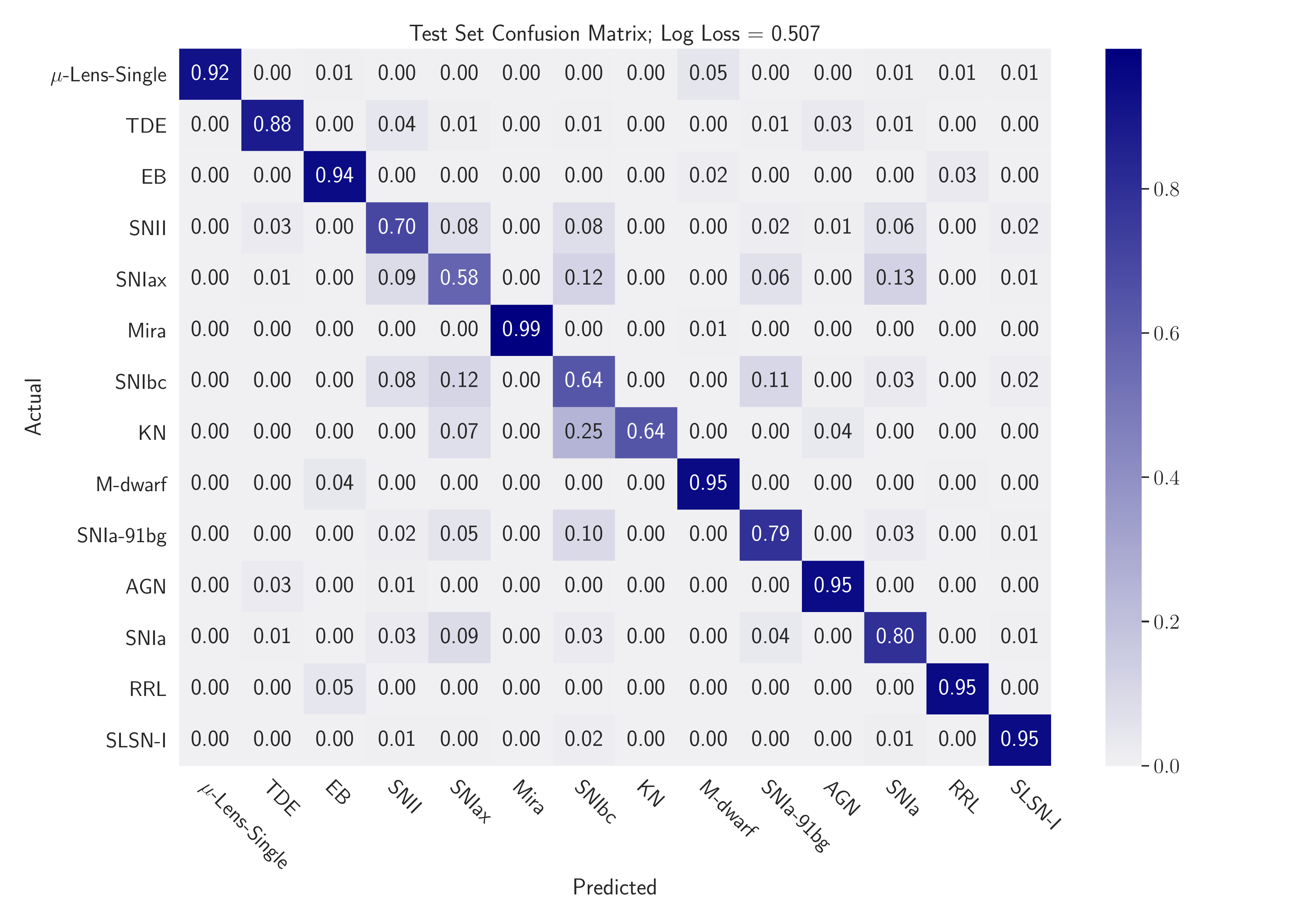}
    \caption{Transformer applied to SN photometric light curve classification. The figure shows a confusion matrix on the PLAsTiCC simulated dataset. Most of the classes are identified with accuracies larger than $90\%$ as seen in the diagonal of the matrix. Figure adapted from~\cite{Allam2021}}
    \label{fig:transf_SN}
\end{figure}

The processing of data in the form of sequences has experienced significant breakthroughs in the machine learning community over the past years. In particular, the so-called attention methods which identify the region of the sequences that contain the most relevant information have been demonstrated to be very powerful for sequence to sequence tasks such as translation and for classification of time series~\citep{Vaswani2017}. This type of attention based architectures are commonly known as Transformers (see~\autoref{fig:transformer}). The application of Transformers to astronomy is still rather limited. However some works have already explored their performance for SN light curve classification and other types of transients.~\cite{Allam2021} use a variation of the original Transformer architecture to classify photometric light curves from the PLAsTiCC simulated dataset. The authors demonstrate they achieve state-of-the art accuracies. They claim the Transformer is able to deal with very unbalanced classes without need of augmentation, 
achieving the lowest logarithmic loss to date (\autoref{fig:transf_SN}). However, as the authors emphasize, the comparison with other methods is not straightforward given that they are evaluated under different conditions. This highlights a general problem for the comparison of different works performing classification. Astronomy lacks in general of standardized datasets on which algorithms can be consistently tested (see~\autoref{sec:final_thoughts} for a general discussion).~\cite{Pimentel2022} is a second work exploring the use of attention mechanisms. They apply their method to real data after a first training on simulations and a fine tuning step. They also conclude that the attention network outperforms classical approaches based on RFs and also RNNs especially for late-classification and early-classification of light curves.

In addition to purely photometric data, the classification of variable sources can also be performed on images. This is traditionally done by subtracting images from different epochs. Several works have explored the use of deep learning to work directly in the image sequences.~\cite{CarrascoDavis2019} used for example Recursive Convolutional Neural Networks (RCNNs) to identify real transients from artifacts.~\cite{Gomez2020} extend the use of RCNNs to a multi-class classification of transients. They also use RCNNs to extract information from both temporal and spatial correlations.  

In summary, deep learning approaches have naturally incorporated to other ML approaches to classify photometric light curves. In particular RNNs and more recently Transformers provide competitive results. However, for this particular task, deep learning does not seem to have dramatically improved previous techniques as it is the case for image classification. The work by~\cite{Hlozek2020} summarizes the results of the PLAsTiCC challenge organized in the Kaggle Platform. It can clearly be seen that both \emph{classical} and deep learning approaches provide competitive results.


\subsubsection{Other classifications}
Similar supervised approaches for other classification tasks have been tested over the past years, reaching also similar conclusions and facing similar challenges.~\cite{Kim2017} used convolutional neural networks for separating stars from galaxies. Star-galaxy separation is a classical task in the analysis of deep surveys. As for other classification tasks, the advantage of CNNs is that they use the pixel level information and do not rely on summary statistics. The general conclusion however is that CNNs offer a marginal gain over more ML feature based approaches for this classification problem.~\cite{Ono2021} used CNNs to distinguish between real Ly$\alpha$ emitters (LAEs) and contaminants using imaging data in six narrow band filters.~\cite{Ackermann2018} made the first tests to classify galaxy mergers trained on Galaxy Zoo and reported significant improvements over state-of-the-art approaches (the case of galaxy mergers is extensively discussed in~\autoref{sec:unobs}). ~\cite{Walmsley2019} and \cite{Tanoglidis2021a} explored the use of CNNs for the classification of low surface brightness (LSB) structures in deep imaging surveys. The systematic exploration of the low surface brightness universe will be enabled by future surveys such as LSST or Euclid. Automatically identifying and classifying LSB structures  is therefore a new challenge. The authors test a CNN model on the Dark Energy Survey data and report a $\sim95\%$ accuracy in separating artifacts from real LSB structures. Only a few works have explored this science case, probably due to the lack of proper labeled datasets.      

\subsection{Segmentation, deblending and pixel-level classification}
\label{sec:segmentation}
Object detection and deblending is another problem on which deep learning techniques have been extensively tested in these past years. The detection of sources to build catalogs with some measured properties is a first standard step in the processing of imaging from deep surveys. In image processing this typically fits into the field of image segmentation, which is precisely the task of identifying the positions and boundaries of different objects in an image. The type of segmentation can be semantic, if the objects belong to different classes, or instance if we aim at detecting objects of the same type. 

\begin{figure}
\centering
    \includegraphics[width=\linewidth]{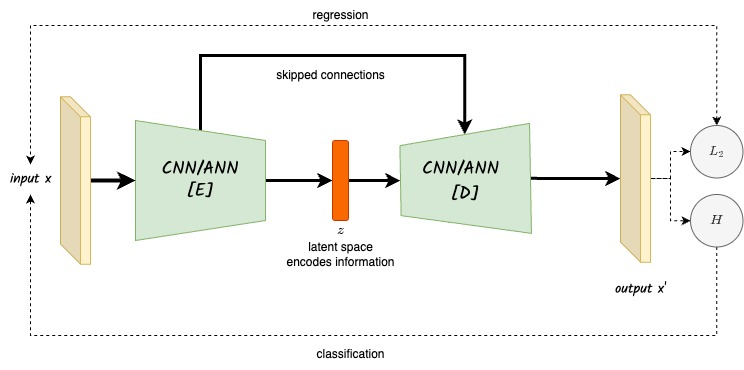}
 
    \caption{Schematic representation of a Unet type of architecture. This is typically used for addressing image to image problems (e.g. segmentation, galaxy properties in 2D). A first CNN (encoder) encodes information into a latent space ($z$), which is then fed to another CNN (decoder) which produces an image. One particularity of the Unet is the presence of skipped connections between the encoding and decoding parts which have been demonstrated to provide higher accuracies. }
    \label{fig:unet}
\end{figure}

Since the popularization of CCDs, object detection in astrophysics is generally done through the software called \textsc{SExtractor}, which implements an advanced multi thresholding technique to detect and separate objects. Although it has been extensively used over the past years, the limitations become more obvious with the advent of deeper surveys in which the confusion between sources becomes very common. It is estimated that $\sim80\%$ of galaxies will be affected by some sort of overlapping or \emph{blending}. Given that blending can severely affect the scientific conclusions, it is important to have reliable methods for detection and deblending (see~\citealp{Melchior2021} for a review on deblending). 

Over the past years, there has been significant progress in the computer vision community on image segmentation by applying deep learning networks. Therefore, similarly to what happens for classification, deep learning segmentation techniques developed for general purpose computer vision applications are available. An out-of-the box implementation is thus expected to provide reasonable results. Astrophysical data has however some key properties which are not found in other types of images. The dynamic range is very large, typically spanning several orders of magnitude from the centers of the objects to the outskirts. Objects do not have clear edges. This is a fundamental difference with respect to natural imaging applications, which makes the segmentation task with neural networks more challenging. 

\begin{figure*}
\centering
    \includegraphics[width=\linewidth]{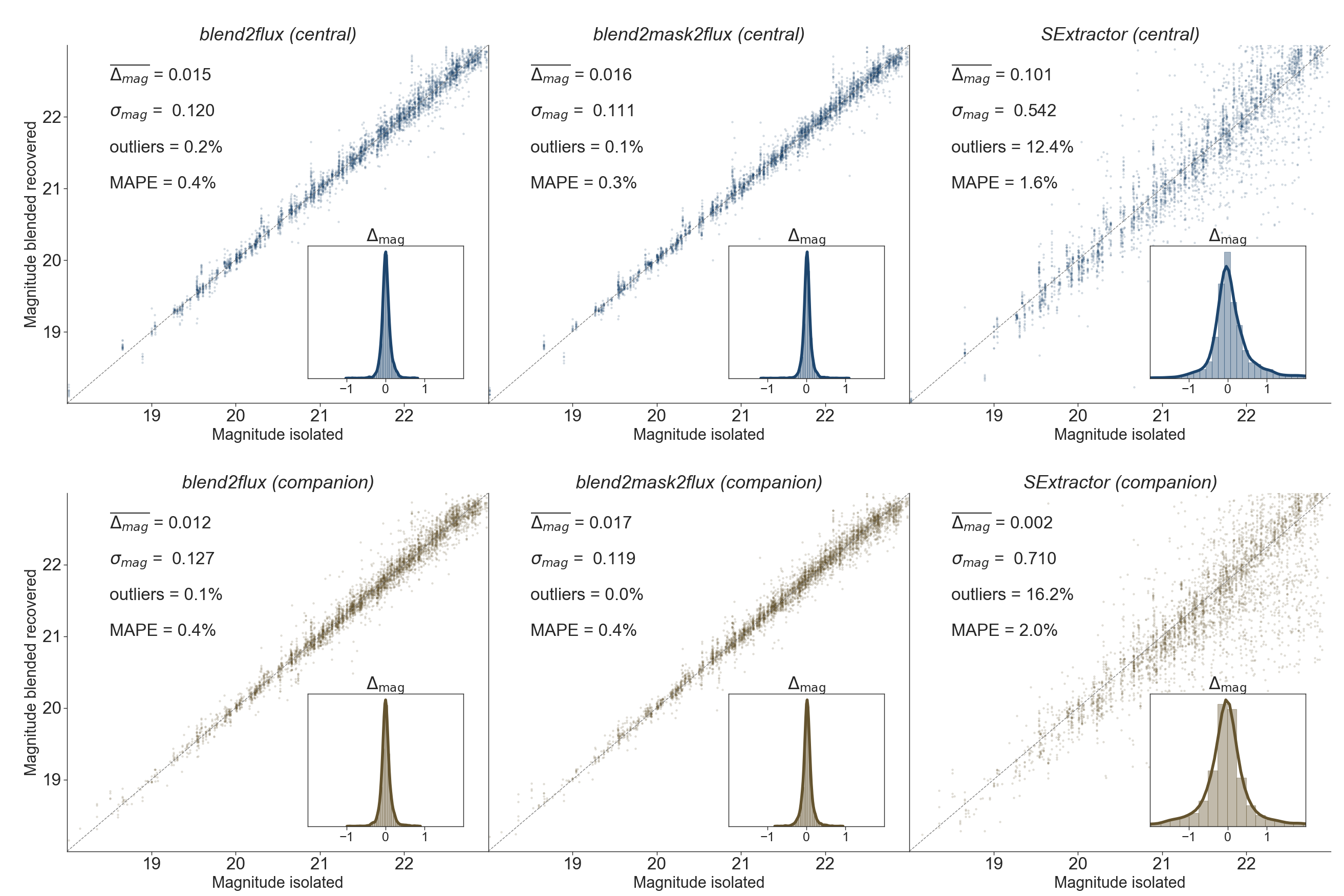}
    \caption{Comparison between the photometry measured with the Unet and with \textsc{SExtractor}. All panels show a comparison between input and recovered photometry. The top row shows the central galaxy in the stamp while the bottom row is the companion one. The two leftmost panels show two different deep learning architectures. The rightmost panel shows the results of the baseline \textsc{SExtractor}. Both dispersion and bias are improved with deep learning. Figure adapted from~\cite{boucaud2020}}
    \label{fig:boucaud_unet}
\end{figure*}

A first very popular approach for object detection is the use of encoder-decoder networks. Unets~\citep{Ronneberger2015}, which incorporate skipped connections between the encoder and decoder branches, and have emerged as one of the state-of-the art segmentation networks (see~\autoref{fig:unet}). Originally designed for medical imaging, they have been commonly applied to astrophysics for detection over the past years.~\cite{boucaud2020} first applied a Unet to detect objects in image stamps and measure the photometry of overlapping sources. In this proof-of-concept work, it is shown that the measured photometry of pairs of galaxies is improved with respect to the standard \textsc{SExtractor} based approach (\autoref{fig:boucaud_unet}). 

\begin{figure}
\centering
    \includegraphics[width=\linewidth]{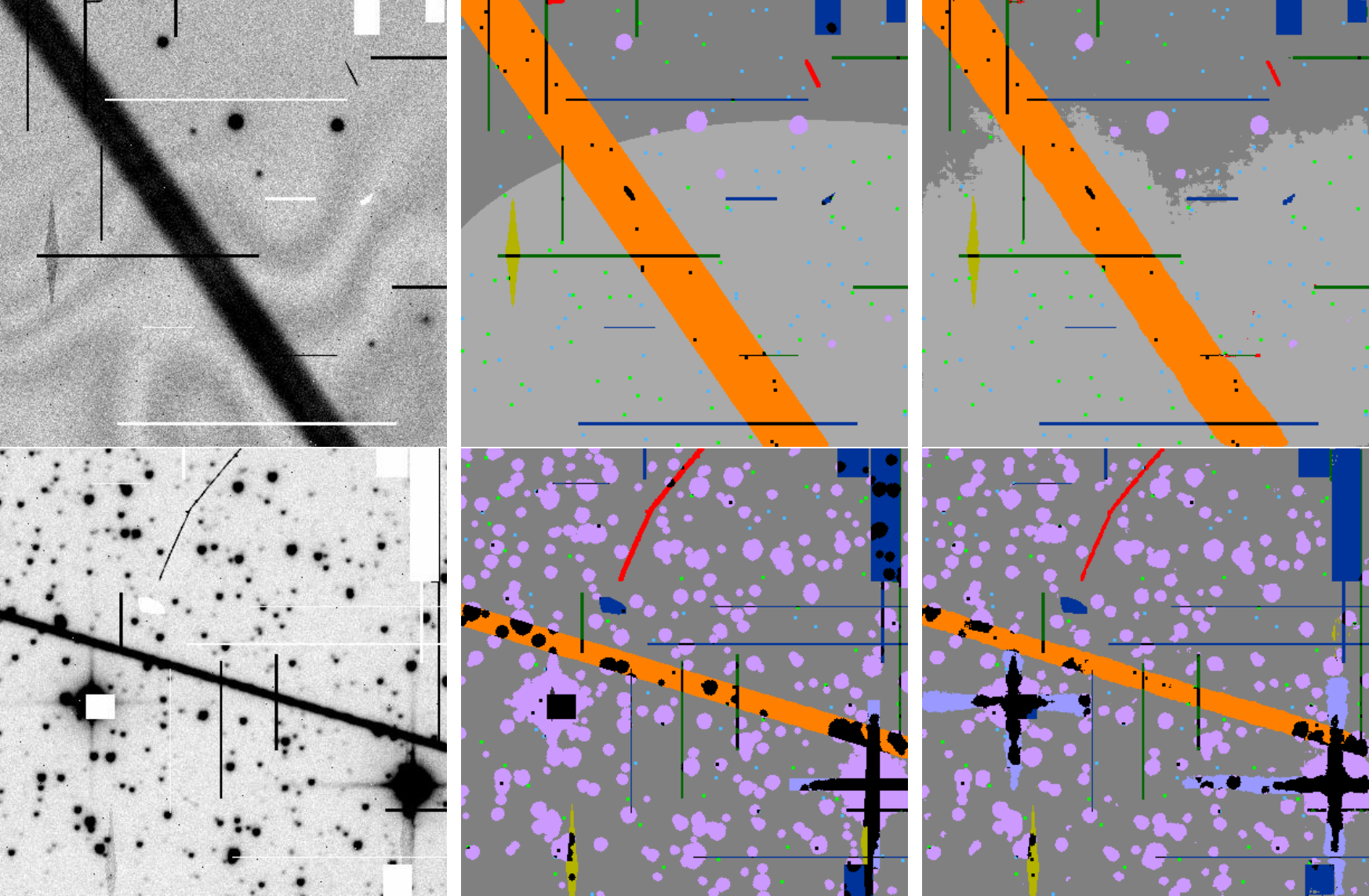}
    \caption{Unet used to segment different types of artifacts on CCD images. The leftmost figures show a field containing different artifacts. The middle panel shows the ground truth and the rightmost panels the predictions obtained by the Unet.  Figure adapted from~\cite{paillassa2020}}
    \label{fig:paillassa}
\end{figure}

However, this was tested in an idealized setting where the stamps only contained two images with one galaxy at the center.~\cite{paillassa2020} explored a similar type of architecture, in a more realistic setting, to identify and classify artifacts in CCD images. In that case, classification and segmentation are performed simultaneously at the pixel level. The proposed approach successfully identifies multiple types of artifacts in an image (\autoref{fig:paillassa}). 

\begin{figure}
\centering
    \includegraphics[width=\linewidth]{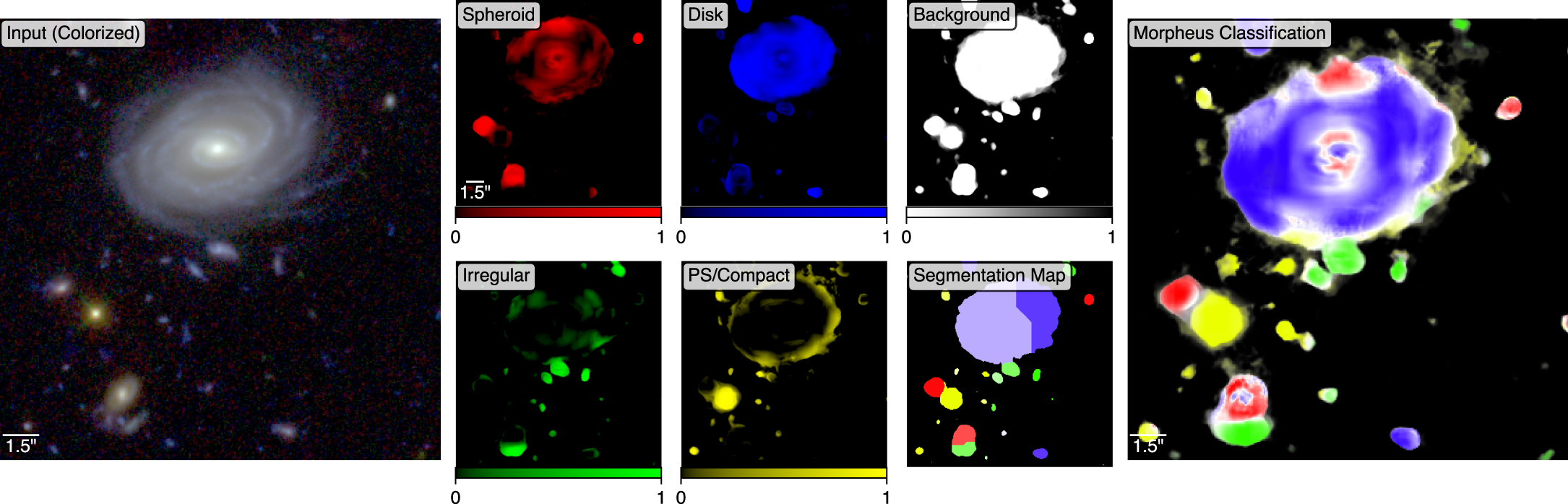}
    
    \caption{ Example of pixel-level classification of a region in the CANDELS survey using a Unet. The leftmost image shows the original field. The six smaller images show different channels with different morphological types and the rightmost image is the combination of all channels with a pixel level morphological classification. Figure adapted from~\cite{hausen2020}}
    \label{fig:morpheus_unet}
\end{figure}

A similar approach is explored by~\cite{hausen2020} combining this time object detection and morphological classification of galaxies at the pixel level. Using a sliding window, the Unet successfully classifies every pixel of the CANDELS survey in five morphological classes (\autoref{fig:morpheus_unet}).~\cite{HuertasCompany2020} used a similar type of architecture to study the resolved properties of galaxies by detecting giant star forming clumps within high-redshift galaxies in the CANDELS survey. ~\cite{burke2019},~\cite{Farias2020} ~\cite{tanoglidis2021} explored an alternative approach based on region based CNN architectures such as Mask R-CNNs to perform similar tasks, i.e. detection, deblending and classification.\\

Overall, these applications all show very promising results and clearly improve on more traditional methods both in terms of speed but also in accuracy, especially when combined with pixel-level classifications. In most cases, the application of out-of-the box architectures is enough to provide accurate results, once the input data are properly rescaled to limit the effect of the dynamic range. However, until now, and with the exception of the work by~\cite{hausen2020}, the majority of the works have focused more on testing and demonstrating the performance of these deep learning based approaches. The application to real data to produce scientifically exploitable data products remains very moderate. These approaches suffer from similar limitations as the classification tasks, i.e. training sets and uncertainty quantification (see~\autoref{sec:final_thoughts}). Finding suitable training sets is more challenging in this case as one need to both label and identify the boundaries of the objects. In astronomy, the definition of object boundaries strongly depends on the noise levels. The adopted solutions so far are either training on simulations or relying on previous detections. None of them seems fully satisfactory. Simulations are usually too simplistic and the generalization to data can induce unexpected behaviors. Using outputs from other software packages tends to propagate the existing biases. Possible solutions include the generation of more realistic training sets with generative models (e.g.~\citealp{Feder2020,Lanusse2021,Collaboration2022}). We will discuss this in more detail in \autoref{sec:data_emulation}. Recent works have also started to explore the implementation of uncertainties in the produced segmentation maps~\citep{bretonniere2021} which seems a promising way to limit the impact of possible catastrophic failures when changing domain from simulations to observations. However, all these works remain at the proof-of-concept stage.    

\begin{figure}
\centering
    \includegraphics[width=\linewidth]{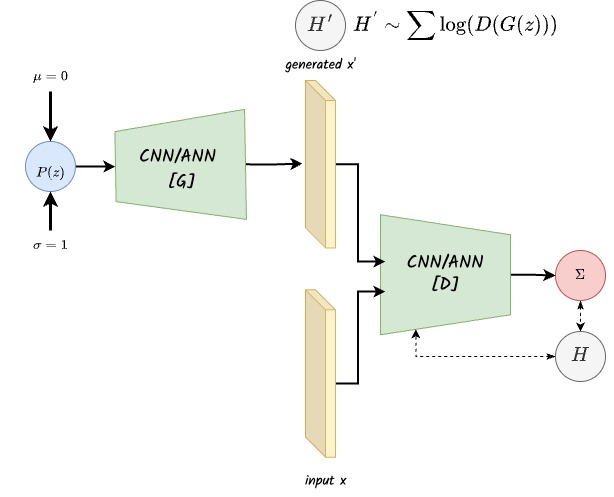}
    \caption{Schematic representation of a standard Generative Adversarial Network (GAN). It has been used as a generative model for deblending, identifying outliers as well as for accelerating cosmological simulations among others. A first CNN (generator) maps a random variable into an image, which is then fed to a second CNN (discriminator) together with real images to distinguish between both datasets. The generator and discriminator networks are trained alternatively.  }
    \label{fig:GAN}
\end{figure}

\begin{figure}
\centering
    \includegraphics[width=\linewidth]{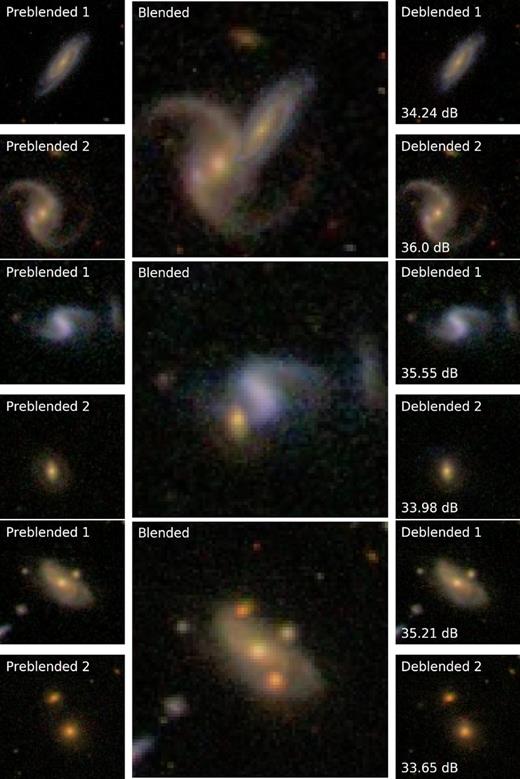}
    \caption{Example of galaxy deblending using GANs. The central images show three overlapping galaxies. On each side of the big images, the original (left) and reconstructed (right) images of the pair of galaxies are shown. Figure adapted from~\cite{Reiman2019}.}
    \label{fig:deb_GAN}
\end{figure}

In addition to these segmentation based approaches, other groups have attempted to go a step forward by reconstructing the surface brightness profiles of overlapping galaxies. This implies moving from a classification to a regression problem, since the output of the networks is the flux at a given pixel. This task typically requires the use of generative models to learn the diversity of galaxy shapes in a data-driven way and then being able to generate likely solutions.~\cite{Reiman2019} first explored the use of Generative Adversarial Networks (GANs) for this purpose (see~\autoref{fig:GAN}). In that case, the network input is a stamp of two overlapping galaxies and the output are two images containing each of the two galaxies separately. An adversarial loss is employed to ensure that the two produced galaxies are realistic and indistinguishable from observed galaxies (see \autoref{fig:deb_GAN}). Following a similar goal,~\cite{Arcelin2021} used Variational Autoencoders (VAEs) to reconstruct the light distribution of blended galaxies applied to simulations of the LSST survey. Recently, ~\cite{Hausen2022} attempted an intermediate solution in between full reconstruction and detection. They proposed a novel approach based on Partial-Attribution Instance Segmentation to estimate the fraction of fluxes in each of the galaxies from the blended system. This is interesting as it provides a solution specifically designed for the astrophysical problem, which remains rare in deep learning applications. 

 Works attempting this task are still at the exploration level as well, even though the results seem very promising (e.g. \autoref{fig:deb_GAN}). ~\cite{Reiman2019} used very simple simulations by just adding two SDSS images; ~\cite{Arcelin2021} employed analytical simulations. As all other deblending efforts, this approach suffers from finding a suitable training set which is close enough to observations without being too simplistic. ~\cite{Arcelin2021} showed that transfer learning from a network trained on simulations is a promising approach. More important, the statistical versus individual accuracy problem becomes more dramatic when generating images instead of masks. Generative Models produce realistic images in a statistical sense, i.e. arising from the same probability density function than observations. However, on an individual basis, artifacts might appear on the  generated images. These are in general very difficult to track down and can therefore induce significant biases. This individual versus statistical accuracy is an inherent problem to machine learning which needs to be taken into account when using ML predictions for scientific analysis.       

In order to limit the black box effect, an interesting approach is proposed by~\cite{lanusse2019}. The authors use an hybrid model that combines a physically motivated model with analytic expressions for known terms, with a data-driven prior for galaxy morphology learned with a generative model. In this approach, the output of the generative model is only used as a prior of the inverse problem, and therefore the impact of unexpected artifacts is reduced. The combination of physically motivated models with data-driven ones appears as an appealing solution which will likely become important in the future.

\subsection{Improving data quality}
\label{sec:data_quality}
Deep learning has also been explored to improve the quality of data, i.e. denoising and deconvolution. Astronomical images are usually noisy and blurred by the effect of the Point Spread Function (PSF), which for ground based data includes both the telescope impulse response and the effect of the atmosphere.  The processes of denoising and deconvolution aim therefore at recovering the information before degradation. This is typically a challenging inverse problem which needs significant regularization. Data-driven approaches have emerged as alternative solutions to more classical deconvolution techniques. ~\cite{Schawinski2017} first explored the use of Generative Adversarial Networks~(\autoref{fig:GAN}) to deconvolve images from SDSS. They show that GANs can recover features even after degradation. This remains however a simple experiment since galaxies were previously degraded.~\cite{Gan2021} built up on a similar idea using GANs to translate between ground and space based observations.~\cite{Jia2021a} proposes an improved solution based on two different GANs which reduces the need of large amount of paired images with and without degradation. ~\cite{Lauritsen2021} extended the idea to the sub millimeter regime by using Variational Autoencoders instead of GANs. Encoder-Decoder networks can also be used for denoising, and some works have explored this for astronomy.~\cite{Vojtekova2021} showed that Unets can effectively increase the exposure time by a factor of two.  

Generally speaking all the attempts show impressive results in solving the long standing problem of deconvolution. They remain however at the proof-of-concept stage and have not been applied for scientific analysis. Similarly to what happens with deblending, the main limitation is robustness. Generative Models such as GANs produce very realistic images but can also introduce artifacts which are statistically meaningful but not necessarily on an image per image basis. These artifacts can introduce uncontrolled biases in the scientific analysis.      

\subsection{Emulating astronomical images}
\label{sec:data_emulation}

In a number of applications, the ability of simulate survey data is very valuable, for instance to test and calibrate measurement pipelines. One of the difficulties faced in simulations of upcoming surveys such as LSST or Euclid is the relatively small set of deep and high-resolution imaging data (essentially limited to HST surveys such as COSMOS) that can be used to provide complex galaxy light profiles as inputs. This is one of the motivations behind the development of Deep Generative Models for galaxy morphology, which can be trained on the existing data and then generate significantly more examples realistic light profiles. 

\begin{figure}
\centering
    \includegraphics[width=\linewidth]{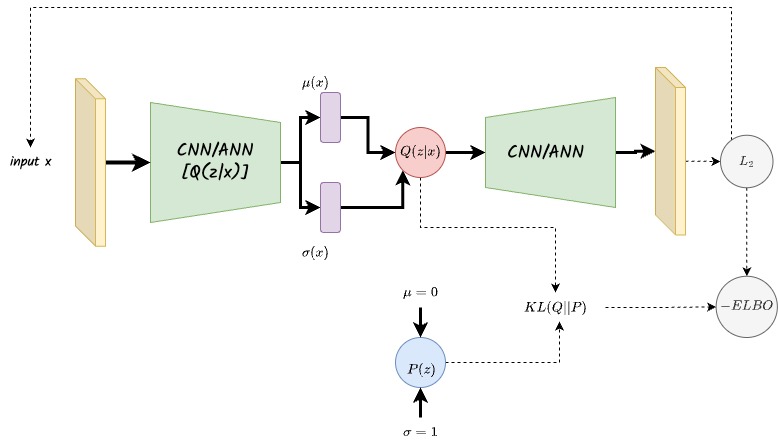}
    
    \caption{Illustration of a Variational Autoencoder (VAE). This generative model has been extensively used in astronomy for multiple emulation tasks. A first CNN maps the input into a distribution, which is then sampled to provide the input to a second CNN which reconstructs the input image. A regularization term is added to ensure that the latent distribution behaves close to a prior distribution. }
    \label{fig:VAE}
\end{figure}

\begin{figure*}
\centering
    \includegraphics[width=\linewidth]{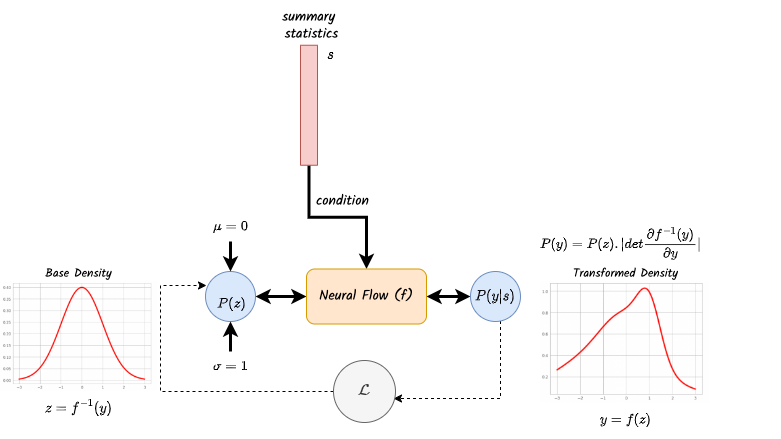}
    
    \caption{Illustration of a Normalizing flow. This density estimator model is used to transform a simple distribution (typically a Normal distribution) into a more complex density using an invertible function $f$ parametrized by a neural network which is optimized by maximizing the log likelihood. The transformation can also be conditioned to an additional set of parameters ($s$ in the figure).   }
    \label{fig:flow_only}
\end{figure*}

In one of the first works following that direction, \cite{ravanbakhsh2017} demonstrated that training relatively simple GANs (\autoref{fig:GAN}) and a Variational Autoencoders (see~\autoref{fig:VAE}) on HST COSMOS postage stamps successfully captured most of the relevant population-level parameters, such as size, magnitude, and ellipticity. In subsequent work, \cite{Lanusse2021} extended that model to explicitly account for the PSF, and proposed a hybrid Normalizing Flow (see~\autoref{fig:flow_only}) and VAE architecture which allowed to achieve diverse and high quality samples, while also making it possible to condition the light-profile generation on galaxy properties. Normalizing Flows are a type of generative models which, as opposed to GANs or VAEs, allow the exact evaluation of the likelihood $p(y)$ and therefore, their weights can be directly learned by maximizing the log  likelihood of the training dataset. The idea is to construct a bijective mapping $f$ such that $y = f(z)$ where z is a variable with a simple base density $p(z)$, typically a Normal distribution. As $f$ is invertible one can evaluate the density $y$ using the change of variable theorem, i.e. simply inverting $f$ and keeping track of the Jacobian of the transformation (see~\autoref{fig:flow_only}). \cite{Collaboration2022} used that generative model to create simulations of the Euclid VIS instrument on a 0.4 deg$^2$ field with complex galaxy morphologies and used those simulations to assess the magnitude limit at which the Euclid surveys (both deep and wide) will be able to resolve the internal morphological structure of galaxies. 

One of the limitations of standard GANs and VAEs for the simulations of such galaxies is that it quickly becomes difficult to generate high quality samples on large stamps. To address these technical difficulties, \cite{fussell2019} developed for instance a StackGAN approach in which a first GAN is trained to generate a low resolution image (e.g. 64x64), which is then up-sampled by a second GAN (e.g. to 128x128). \cite{Smith2019} proposed to use a GAN variant, known as a Spatial-GAN (SGAN) to generate no longer only postage stamps, but entire fields of arbitrary size through a purely convolutional architecture. The authors demonstrated the ability to sample a 87040x87040 pixels image emulating the HST eXtreme Deep Field (XDF). More recently, \cite{Smith2022} explored applying a Denoising Score Matching approach \citep{Song2019} and were able to generate large high resolution postage stamps of size 256x256 of remarkable quality.

\vspace{.3cm}
\begin{mdframed}[backgroundcolor=orange!30] 
{\bf Summary of deep learning for computer vision} \\
\begin{itemize}
    \item Deep learning has rapidly emerged as a solution for the classification of objects in large surveys. Galaxy morphology, strong lenses, and also light curves constitute the main applications. Deep learning based catalogs, especially for galaxy morphology, exist and are being used for scientific analysis.
    \item The most common approach for classification are supervised Convolutional Neural Networks with different degrees of complexity.
    \item Overall, CNNs achieve higher accuracy than previous approaches and are also faster.
    \item The lack of labeled training sets is a common bottleneck. Solutions involving Transfer Learning and/or the use of simulated training sets have been proposed. This implies some additional challenges which we discuss in~\autoref{sec:final_thoughts}
    \item False positives in the case of very unbalanced samples (e.g. for strong lensing) is also a commonly encountered problem. No satisfactory solution has been found to date apart from visual inspection.
    \item Standardized datasets to consistently compare the performances of different classification approaches are  not common in astronomy which limits the possibility of comparing different approaches (see \autoref{sec:final_thoughts}).  
    \item Deep learning has also been explored for object detection and segmentation in images. 
    \item The most popular approach for segmentation are convolutional encoder-decoder networks such as Unets, although more complex architectures have been also employed.
    \item Overall, the results are promising and tend to outperform pre-deep learning approaches for detection.
    \item Most of the works remain still at the proof-of-concept stage. Until now, there are no deep learning based catalogs in major deep imaging surveys. The robustness of such approaches is still a concern, especially for deblending. Physics informed models could be a solution to explore in the future.
\end{itemize}
\end{mdframed}

\section{Deep learning for inferring physical properties of galaxies}
\label{sec:galprop}

\subsection{Neural Networks as fast emulators}
We now move to address the efforts done in the past years to estimate the physical properties of galaxies using deep learning. Compared to the previous section where we described computer vision tasks, these applications are typically more domain specific since they target physical quantities of galaxies from a regression point of view. The general approach followed by the community has been to test neural networks to emulate - replace - more specific tools, developed over the years,  which are usually slow or not fast enough to deal with future big data surveys.

\subsubsection{Photometric Redshifts}

All modern cosmological surveys require a more or less precise estimation of redshifts. When spectroscopy is not available, which is the case for most deep surveys, photometric redshift estimation is the standard way to proceed to measure distances of large numbers of galaxies using a combination of broad and narrow band photometry. Photometric redshift estimation is therefore a non-linear mapping between a set of photometric points and a real number measuring the redshift. The standard way to approach the problem is through the fitting of Spectral Energy Distributions generated from Stellar Population Models (e.g.~\citealp{Benitez2000,Bolzonella2000}). However, since it is - in theory - a well defined problem, it is among the most popular applications of deep learning supervised regression in astrophysics. The first attempts of estimating photometric redshifts with neural networks start well before the deep learning boom, in the early 2000's  \citep{Collister2004,Vanzella2004}. These methods already relied on the idea of learning the mapping between photometry and redshift from data through a Multi-Layer Perceptron trained under a Mean-Squared Error (MSE) loss. The only difference with a modern architecture would be in the depth of the model and the choice of activation function. Perhaps the most successful of these early neural methods for photometric redshifts, ANNz \citep{Collister2004} has continued to evolve over time, with ANNz2 \citep{Sadeh2016} including some quantification of epistemic uncertainties through an ensemble of randomized estimators techniques reminiscent of modern deep ensembles. 

Two significant evolutions of these methods appeared in recent years with the generalization of deep learning: 1. probabilistic modeling of the redshift distribution to estimate posterior probabilities; 2. pixel level models based on CNNs, thus going beyond photometric information and able to use morphology as well.

\paragraph{Probabilistic Modeling of Photometric Redshifts} 

Going beyond a regression task, \citep{Bonnett2015} proposed to use a neural network (still an Multi-Layer Perceptron - MLP), which for a given photometry would output a distribution in the form of a discretized probability density function (PDF). The model would then be trained with a standard cross-entropy loss to predict the redshift bin in which a given galaxy should fall, which in fact mathematically corresponds to estimating the posterior distribution of redshifts given photometry, under a prior given by the selection of the training set. This approach was subsequently reused to train other, more complex, neural networks for photometric redshifts \citep{PasquetItam2018, Pasquet2019}, but can potentially suffer from the discretization needed to represent the PDF. Indeed, the network has no built-in notion that classes with adjacent indices actually correspond to adjacent bins. 

\begin{figure}
	\includegraphics[width=\columnwidth]{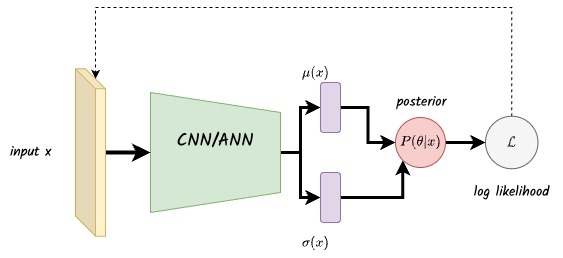}

	\caption{Representation of a Mixture Density Network (MDN). It is a generalization of a standard CNN/ANN in which the output has been replaced by a probability distribution parametrized with some parameters ($\mu$,$\sigma$ in the figure). MDNs provide an estimation of the random (aleatoric) uncertainty. The loss function is the generalized log likelihood.}
	\label{fig:MDN}
\end{figure}

Another approach to model distributions at the output of a neural network is to use a Mixture Density Network \citep[MDN][]{Bishop1994}. MDNs use an MLP to predict the parameters of a mixture of probability densities, and thus provide an analytic and continuous PDF model for a given input (see~\autoref{fig:MDN}). This approach was for instance proposed in \cite{DIsanto2018}, where the neural network outputs are the mean, variance, and weights for a mixture of $n$ one dimensional Gaussians. Overall, the general consensus, is that, when only using photometry as input, neural networks do not provide specially more accurate results than other template based approaches (see e.g.~\citealp{Schmidt2020}). One important challenge is that training sets are generally biased as we will discuss in~\autoref{sec:challenges}. 

\paragraph{Convolutional Photometric Redshift Estimators} 

The second major evolution of neural network-based photometric redshifts has of course been the introduction of CNNs to build a pixel level model capable in principle  of using the entire information content of a multi-band galaxy image to refine a redshift estimate. In the first instance of this approach, \cite{Hoyle2016} used the state-of-the-art model at the time, AlexNet \citep{Krizhevsky2012}, to build a 5-layers deep CNN model taking as inputs a combination of $r$,$g$,$i$,$z$ images of a given galaxy and trained to classify that galaxy into a discrete redshift bin. Interestingly, in this initial study performed over a set of $60,000$ SDSS images, no significant improvements in redshift prediction accuracy were reported when compared to a more traditional photometric feature-based AdaBoost machine learning model. It would take a couple more years for a broader development of CNN based methods, starting with \cite{DIsanto2018} which proposed to combine a simple 3-layers deep convolutional architecture with a mixture density output, but again reported only a mild improvement in terms of accuracy compared to a feature-based approach on an SDSS sample.

\begin{figure}
	\includegraphics[width=\columnwidth]{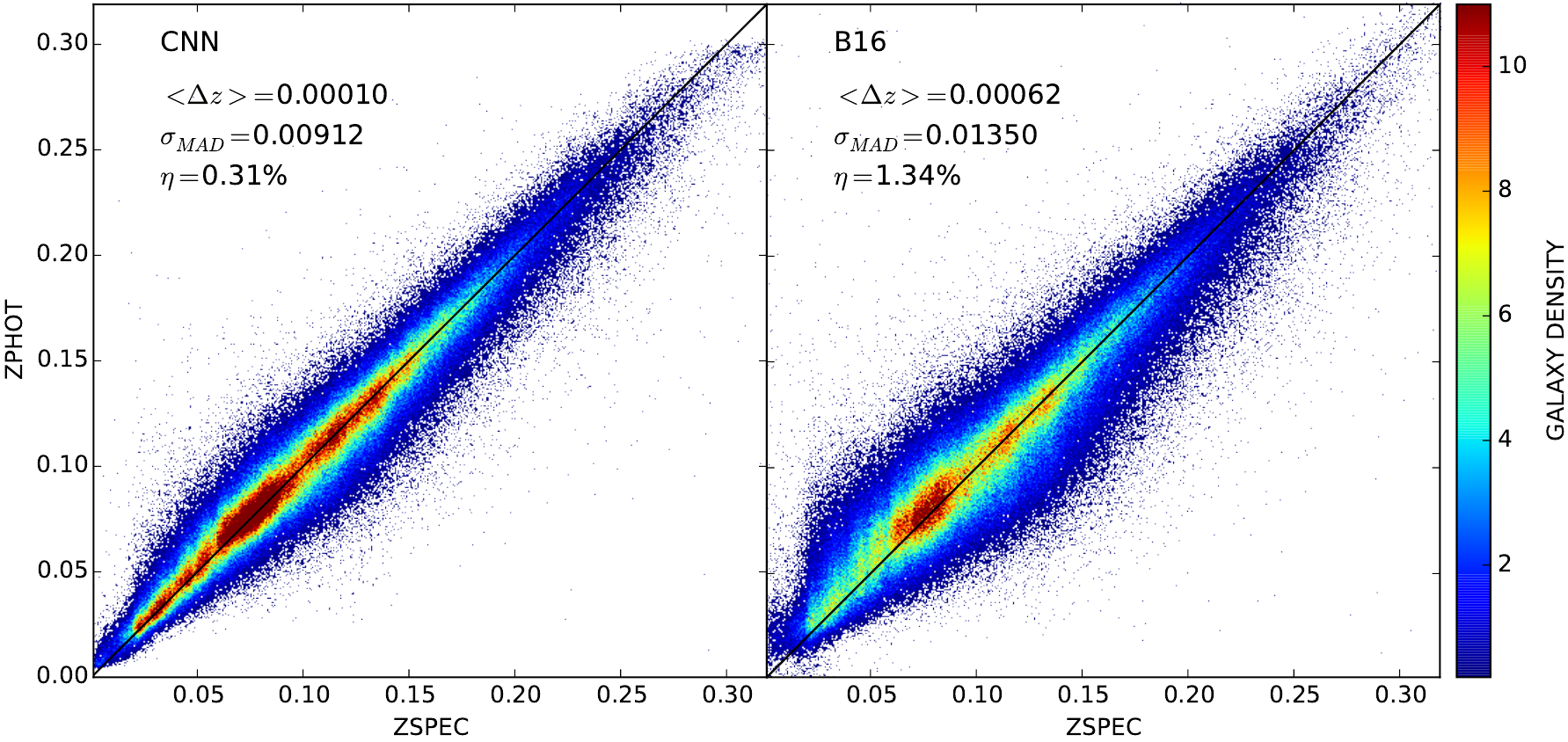}
	\includegraphics[width=\columnwidth]{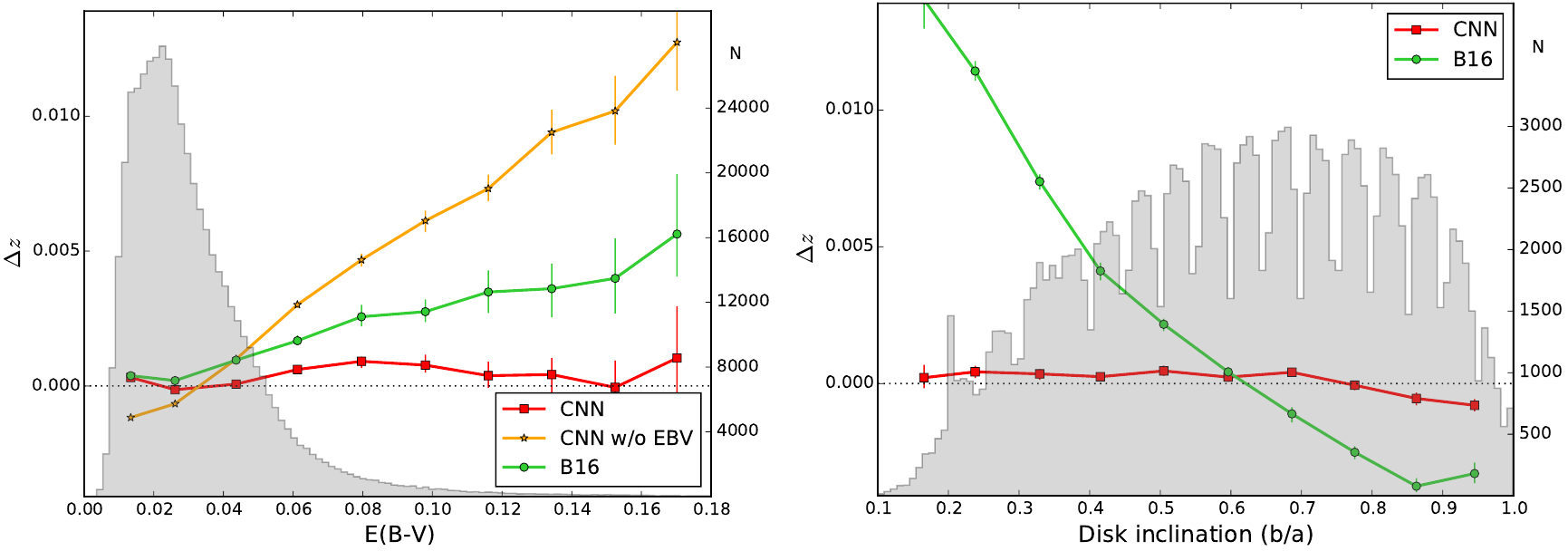}
	\caption{Comparison of photometric redshift performance between a deep CNN and a color-based k-NN (B16) method reported in \citep{Pasquet2019}. The top row shows the predicted redshifts vs spectroscopic redshifts for the CNN (left) and the k-NN method (right). The distribution is noticeably tighter for the CNN with smaller overall bias, scatter, and outlier rate. The bottom row show the photometric redshift bias $\Delta z$ for the two methods, as a function of extinction (left panel) and disk inclination (right panel) of galaxies classified as star-forming or starbust. Having access to the whole image information, the CNN is able to automatically account for the reddening induced by looking at galaxies with high inclination, whereas the k-NN method only using color information is blind to this effect and exhibit a strong bias with inclination.}
	\label{fig:Pasquet2019}
\end{figure}

The benefits of a convolutional approach started to become clear with \cite{Pasquet2019}, which used a much deeper convolutional network, comprised of one input convolution layer and 5 inception blocks, trained under a redshift bin classification loss. These inception blocks \citep{Szegedy2014} essentially replace one convolutional layer by multiple parallel convolutional layers with different kernel sizes, the output of which are concatenated back into a single tensor at the end of the block. This study, based again on galaxies from the SDSS Main Galaxy Sample using $ugriz$ images, illustrated in particular how the CNN is able to automatically make use of pixel-level data to extract information beyond colors, improving redshift estimation. In particular, the bottom row of \autoref{fig:Pasquet2019} shows the comparison between the photometric redshift bias of a standard color-based k-NN photometric redshift estimate and the proposed CNN approach as a function of galaxy ellipticity (proxy for galaxy inclination) for star forming galaxies. As can be seen, the color-based approach exhibits a strong inclination-dependent bias caused by the various amounts of dust attenuation as a function of the viewing angle. The CNN shows however comparatively very little bias, indicating that the model is able to automatically estimate and account for inclination in its prediction by directly drawing that information from the postage stamps. This example illustrates the main advantage of a deep learning approach, it alleviates the need for handcrafted features, leaving it to the model to identify from the data itself the relevant information.

\cite{Pasquet2019} highlight however an important consideration when using a CNN approach. Whereas flux measurements can be standardized to account for varying noise and PSF, a CNN based only on the raw postage stamps, without additional information, is blind to these observing condition factors. The authors report for instance a noticeable bias as a function of seeing for the CNN, and mention the fact that this information could be provided to the model in future work to let the model compensate for these factors.

In a number of subsequent works \citep{Menou2019, Henghes2021, Zhou2021}, it was proposed to extend this pure convolutional approach to photometric redshift estimation to hybrid models, combining an MLP branch tasked with processed photometric features (e.g. colors) and a CNN branch having access to the full image of the galaxy. It was found in these multiple studies that directly providing the model with highly informative features through the MLP branch improved the overall accuracy. Although all the relevant information is in principle already included in the images themselves, this approach reduces the amount of automatic feature extraction that the convolutional branch needs to perform. In all cases, the best results are achieved when both photometric features and images are provided jointly to the model. 

\paragraph{Improving Scaling with Size of Training Set}

One particular aspect that may limit the applicability of these deep learning methods is the need for large training samples, and so in this case the need for large (expansive) spectroscopic samples to properly train these supervised methods. As a possible mitigation technique for this issue, \cite{Hayat2021} demonstrated that a self-supervised encoding provided by contrastive learning retained significant redshift information.  We direct the interested reader to \autoref{sec:discovery} for more details on contrastive learning. Here, the authors proposed a two-step approach, first training in complete self-supervision (without needing any spectroscopic redshifts) a 1d encoding of galaxies postage stamps. And in a second step, training in a supervised way a shallow MLP to predict redshifts on a small training set of available spectroscopic redshifts. They find two surprising results: 1. This fine-tuned self-supervised approach always outperforms a conventional supervised training, 2. The accuracy of self-supervised estimated redshifts scales very well into the low-data regime. They find for instance that their self-supervised approach using 20\% of labeled data achieves similar accuracy to a fully supervised training using labels for the entire dataset. 

With a different strategy for limiting the amount of data needed, \cite{Dey2021} proposed to replace a conventional convolutional architecture by deep capsule networks. Compared to CNNs, capsule networks are robust to rotations and changes in viewpoints and thus provide a more natural representation for randomly oriented objects like galaxies (see also \autoref{sec:low_level}). The hope is therefore to not require as much training data if the model already provides the built-in invariances of the problem. In their propsosed architecture, the capsule network generates a low dimensional encoding of the input image, which is then used to perform two tasks jointly: reconstructed in the input image with a CNN, and estimating the redshift of the galaxy with an MLP. In addition, their model also classifies, at the level of the capsule outputs, the morphology type of the galaxy (elliptical or spiral). The authors find that this approach leads to a better scaling with size of the training set than \citep{Pasquet2019}, especially at very small training set sizes, but the benefits are not as significant as the ones offered by the self-supervised approach of \cite{Hayat2021}.

Another complementary approach to reduce the dependence on large spectroscopic datasets is to use transfer learning, to build a model on simulated data, and fine-tuning it on survey data. This approach was for instance explored in \citep{Eriksen2020} using a MDN trained on a combination of FSPS simulations and data from the PAU Survey. Using a pretraining on simulations was found to reduce the photo-z scatter by as much as 50\% for faint galaxies.

Finally, the question of robustness and stability of these deep neural networks was investigated in  \cite{Campagne2020} which highlighted that inception models like the one proposed in \cite{Pasquet2019} can be highly sensitive to adversarial attacks. Although these attacks are unlikely to happen on astronomical data (see however~\citealp{Ciprijanovic2021a}), this result underlines again the fact that these black-box methods are not as interpretable as more conventional approaches like template-fitting (interpretability is discussed in~\autoref{sec:challenges}).


\subsubsection{Galaxy Structural Parameters}
In addition to classification, galaxy morphology can be also quantified with some parameters that define an analytical description of the surface brightness distribution of galaxies. The so called Sersic models are defined by three quantities: the effective radius ($r_e$), the Sersic index ($n$) and the axis ratio ($b/a$). By combining these parameters with a normalization factor to account for the different galaxy luminosities, one can describe most of the surface brightness distributions of galaxies. The standard way to measure these parameters is by fitting PSF convolved analytic models to the 2D surface brightness distributions of galaxies. The task can be formulated as a mapping between pixel values and real numbers,  which characterize the galaxy shape. It is therefore well suited for a supervised regression problem, provided that a reliable training set is available. Given that the input data are galaxy images, Convolutional Neural Networks are the most common approach.~\cite{Tuccillo2018} first used a CNN to estimate galaxy structural parameters. In this first work, the authors used a simple training set made of analytic profiles with noise added and demonstrated that CNNs achieve comparable or better performance than standard methods, with the key advantage of being a factor of $\sim50$ faster. As previously said, computational efficiency is one of the main motivations behind these works aiming at emulating existing software.~\cite{Tuccillo2018} also attempted to apply the trained CNNs to observed galaxies with HST. However, because the training set was too simple, the results did not appear to be satisfactory. In particular, the authors did not include foreground and background galaxies in the training set which constitutes an important difference with observations. The authors proposed a transfer learning step using measurements performed with standard approaches. Although the results quickly improve, the final results necessarily propagate the systematics of existing software, which cannot be improved by construction. In that respect, the main gain is speed. \cite{ghosh2020} built on this and showed that with a transfer learning step, CNNs can provide reliable structural parameters for both low and high redshift galaxies. The authors estimate structural parameters for $\sim120.000$ galaxies in the SDSS and CANDELS surveys. 

\begin{figure}
\centering
    \includegraphics[width=\linewidth]{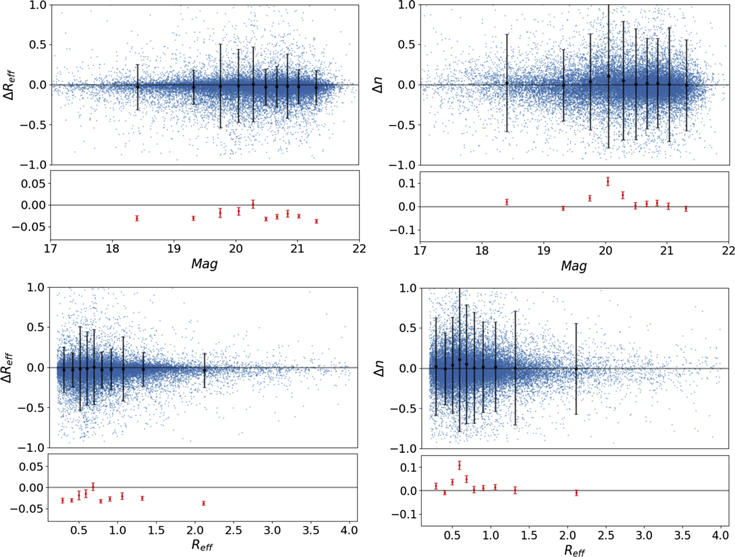}

    \caption{Convolutional Neural Network to measure galaxy structure. The left column shows the difference between true and estimated radius. The right column is the same for Sersic index. Top row shows the results as a function of magnitude while the bottom row is as a function of radius. Figure adapted from~\cite{Li2021}.}
    \label{fig:li_sersic}
\end{figure}

More recently,~\cite{Li2021} attempted a similar approach applied to ground-based observations. The training is still done on simulations but with realistic backgrounds. Additionally, the PSF is included as an input to the CNN, so that the networks can learn the effects of varying PSFs across the field of view. The authors show, that by including these improvements, the CNNs generalise well to observations without need of transfer learning and achieve comparable results to standard approaches, with the advantage of being $\sim 3$ times faster (see~\autoref{fig:li_sersic}). Other works have attempted to decompose the galaxies into bulges and disks. This is an equivalent problem but the number of parameters is increased by a factor of two.~\cite{Grover2021} showed that CNNs can estimate the bulge-to-disk ratio - i.e. luminosity ratio between the bulge and the disk components - of $20.000$ galaxies in less than a minute. The main motivation is, once more, a gain in computational time.~\cite{Tohill2021} explored the use of CNNs to estimate other morphological parameters of galaxies which quantify the distribution of light (i.e. Concentration-Asymmetry-Smoothness - CAS -  system;~\citealp{Conselice2003}). The conclusion is very similar; neural networks accurately reproduce measurements compared with standard algorithms, but faster. Interestingly, they also show that using CNNs makes the measurements more robust in the low signal-to-noise regime, which is one of the main issues of the CAS system. 

Overall, these approaches look very promising to deal with large amounts of photometric data such as the datasets that will be delivered by Euclid and the Rubin Observatory for example. The limitations are similar to other problems. The networks need to be trained on simulations by definition. The extrapolation to real observations is always complicated as one needs to make sure that the training set covers all the observed parameter space. As this is a common challenge, we discuss it in~\autoref{sec:final_thoughts}. This is particularly challenging for space based observations in which the enhanced spatial resolution increases the differences between the simulated datasets used for training and the observations. A possible solution is the inclusion of some sort of uncertainty estimation which could help identifying failures. This has been recently done by~\cite{ghosh2022} who showed that a combination of MDNs and MonteCarlo Dropout can provide well calibrated uncertainties of galaxy structural parameters.  ~\cite{AragonCalvo2020} explored a self-supervised approach to avoid using a fully supervised training based on simulations. However, the approach has not been applied so far to large samples of galaxies. In addition, since the inference time is very short, the bottleneck of this type of approaches is in the training. In current approaches, a specific training set needs to be built for every different application which is not an optimal solution.

\subsubsection{Stellar populations, Star formation histories}

A similar type of application of deep learning is to derive the properties of the stellar populations of large samples of galaxies. This is also a regression type of application, in which a mapping between the galaxy photometry and properties like the stellar mass, the metallicity or stellar age, is sought. As for the previous applications, it exists a standard approach based on the fitting of the Spectral Energy Distributions (SEDs). However, it is typically slow and not adapted to the large volumes of data that will be delivered by future surveys. Deep learning is thus used as an accelerator. Most of the published works so far follow a similar approach. A supervised Neural Network is trained for regression between photometric values and stellar population properties. For example,~\cite{Surana2020} used fully connected Artificial Neural Networks applied to data from the GAMA survey to derive stellar masses, star-formation rates and dust properties of galaxies. The training is performed on stellar population models and the results are compared to standard approaches. The conclusions are also very similar to other applications falling in the same category. Deep Learning performs similarly to standard methods, but a factor of a few faster. Similarly,~\cite{Simet2019} used neural networks to estimate the stellar population properties of high redshift galaxies from the CANDELS survey. The training is in this case performed on semi-analytical models. The conclusion is that galaxy physical properties can be measured with neural networks at a competitive degree of accuracy and precision to template-fitting methods. It is worth noticing that Neural Networks are not the only approach to address this problem, although in this review we primarily focus on deep learning techniques. As this is essentially a mapping between two sets of real numbers, other Machine Learning techniques can be employed -\citealp{Gilda2021, Davidzon2019} used for example Boosting Trees and SOMs respectively for a similar task. 

\begin{figure}
\centering
buck21    \includegraphics[width=\linewidth]{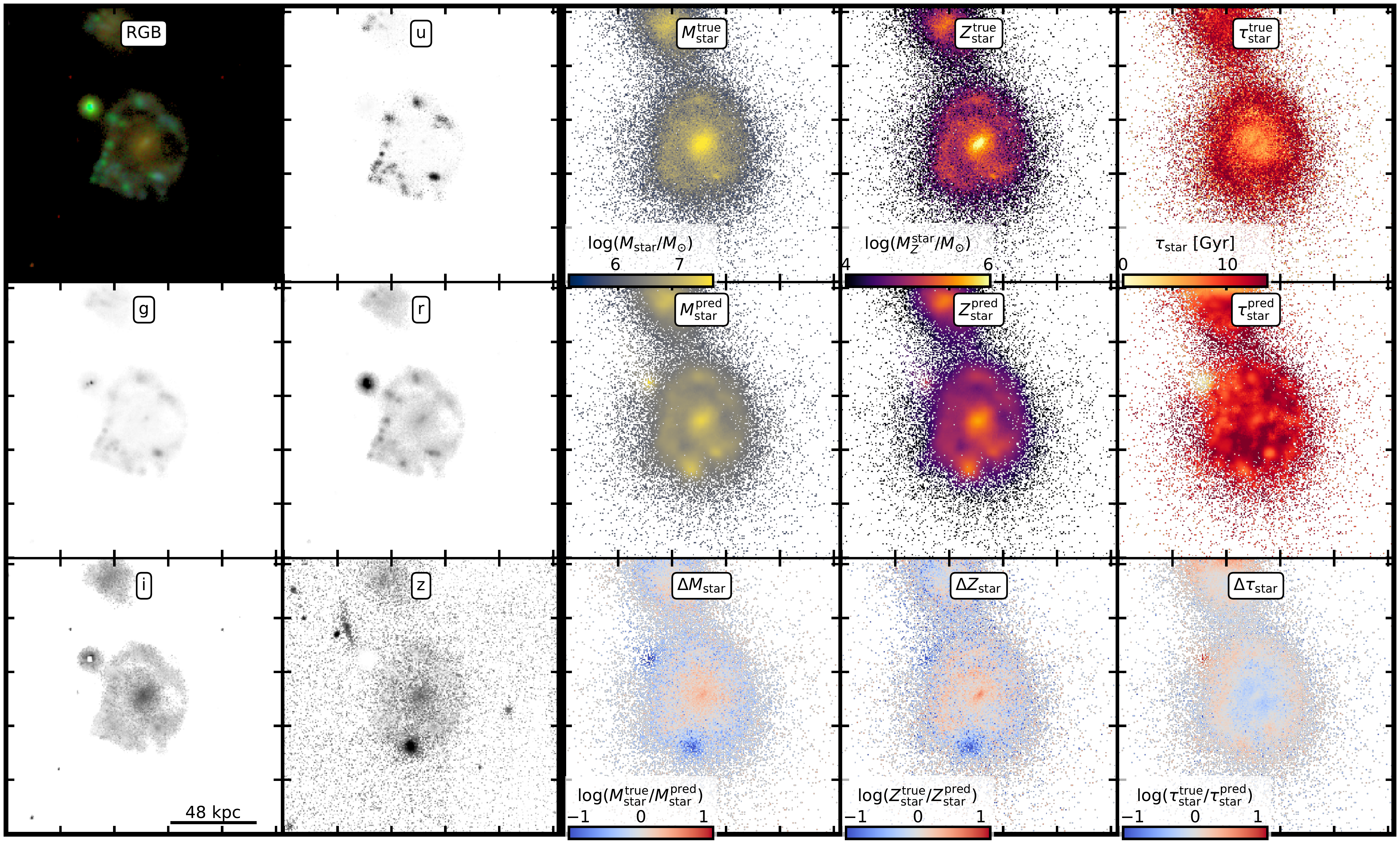}

    \caption{CNN applied to estimate resolved stellar populations of galaxies. The input are images in different wavebands (2 left columns) and the output are different stellar population maps (3 right columns). In the stellar population maps, the top row is the ground truth, the middle one the prediction and the bottom row shows the residual. Figure adapted from~\cite{Buck2021}.}
    \label{fig:2D_maps}
\end{figure}

In a recent work, ~\cite{Buck2021} pushed this idea further by trying to predict resolved stellar population properties instead of integrated quantities (\autoref{fig:2D_maps}). In that case, the mapping is made between broad band photometric images of galaxies and 2D maps of stellar mass, metallicity and other stellar population properties. This is equivalent to a regression at the pixel level. The architecture for this type of work is en encoder-decoder Unet type of network as the ones used for segmentation (see \autoref{sec:low_level} and~\autoref{fig:unet}) but with a mean square error loss to work in regression mode. 

\begin{figure}
\centering
    \includegraphics[width=\linewidth]{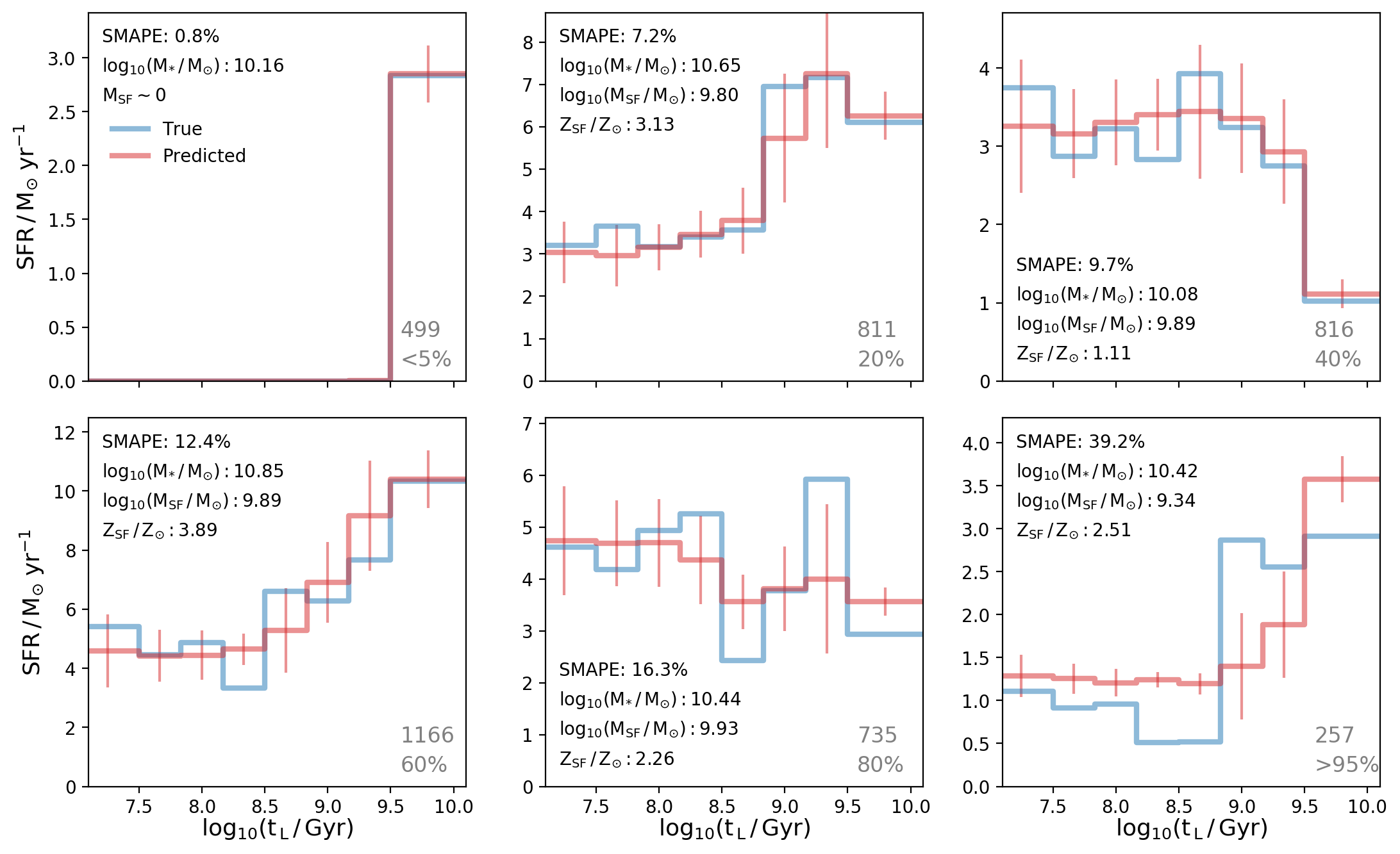}

    \caption{CNN applied to reconstruct the Star Formation Histories of galaxies.  The input are photometric images and the output the Star Formation Rates in different bins of time. The different panels illustrate some examples of predictions (red lines)  together with the ground truth from cosmological simulations (blue lines). Figure adapted from~\cite{Lovell2019}.}
    \label{fig:SFH}
\end{figure}

In addition to stellar population properties at the time of observation, one can use the photometry of galaxies to infer the star formation histories (SFHs), i.e. the star formation rate as a function of cosmic time. There are several established approaches using either parametric or non-parametric methods. However, the problem is known to be significantly degenerate and the star formation activity at early times is in general poorly constrained. Therefore the final estimation heavily relies on established priors. ~\cite{Lovell2019} first attempted to use CNNs in a supervised regression setting to estimate the SFHs of galaxies in the EAGLE cosmological simulation (\autoref{fig:SFH}). One advantage of training on hydrodynamic simulations is that the prior is learned in a data-driven way by using fairly realistic distributions from the simulations. The authors of this first work show a reasonable reconstruction of the SFH and a decent robustness to domain changes.~\cite{Qiu2021} uses CNNs for the opposite task, i.e estimate the galaxy SED from the galaxy SFH taken from simulations. In this case deep learning acts as an emulator of radiative transfer codes.  

As other applications of this kind, the results strongly rely on simulated training sets and on the implicit assumption that the simulations properly cover the observed parameter space. This is particularly critical here since the mocking from numerical simulation is usually done with existing stellar population models which are unavoidably a simplification of reality. Another important limitation is that, up to now, none of the published works on this front properly accounts for uncertainties, even though uncertainty estimation is far from being a solved issue with traditional methods. Both uncertainty estimation and domain shift are common challenges to many applications which are discussed in~\autoref{sec:final_thoughts}.

\subsubsection{Strong lensing modeling}

\begin{figure*}
\centering
    \includegraphics[width=\linewidth]{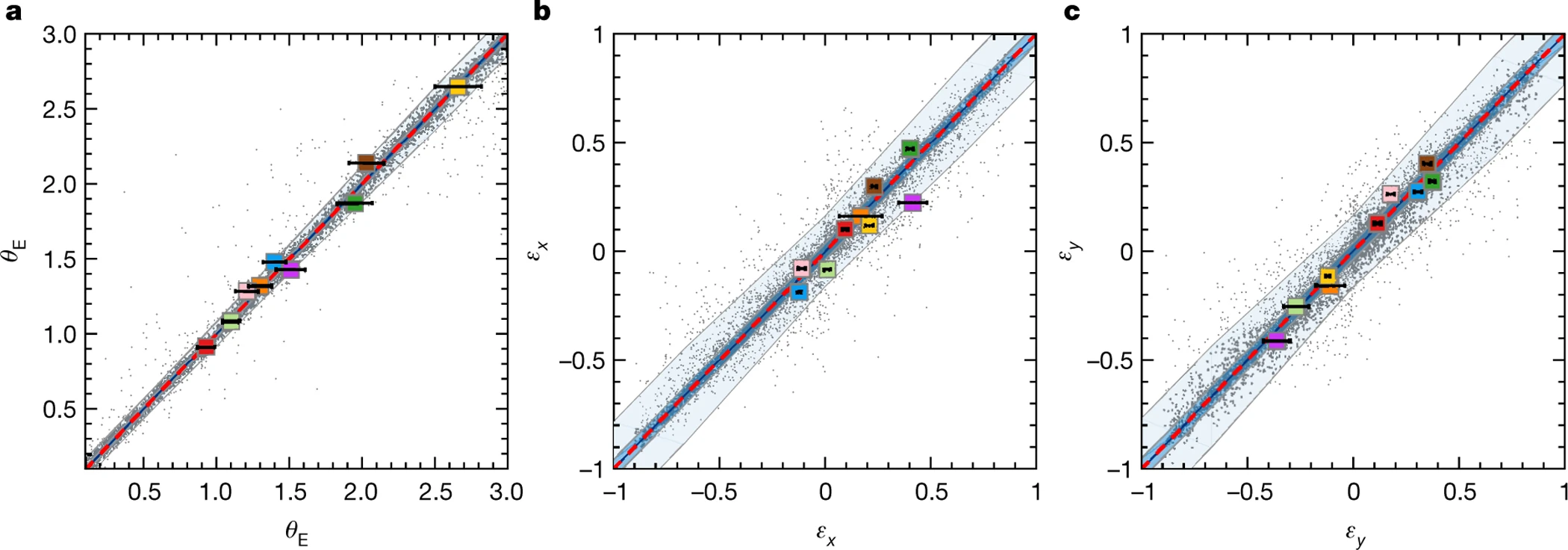}

    \caption{Estimation of strong lensing parameters with CNNs. Each panel shows a different parameter (ground truth versus CNN estimation). Figure from~\cite{Hezaveh2017}}
    \label{fig:lens_model1}
\end{figure*}

Another science case in which deep learning techniques have been extensively tested over the past years is the modeling of strong gravitational lenses - the formation of multiple images of distant sources due to the deflection of their light by the gravity of intervening structures. In~\autoref{sec:class} we have discussed efforts done to find these lenses on large datasets. The goal here is to characterize the lenses. This generally means quantifying image distortions caused by strong gravitational lensing and estimating the corresponding matter distribution of these structures (the 'gravitational lens'). Similarly to the previous applications in this category, there exists a method to perform this analysis, based on maximum likelihood modelling of observations. The process is however time consuming requiring complex dedicated software. Deep learning appears therefore as an appealing approach for accelerating the inference. The first work in exploring this is by~\cite{Hezaveh2017}. The authors test CNNs to estimate the lensing parameters from the images - Einstein radius, complex ellipticity, and the coordinates of the center
of the lens. They show that CNNs can recover the parameters with similar accuracy than standard methods but ten million times faster (\autoref{fig:lens_model1}). An obvious caveat of the deep learning approach for inference is the lack of reliable uncertainties.~\cite{PerreaultLevasseur2017} is one of the first works exploring Bayesian Neural Networks to estimate uncertainties in the modeling of strong lenses, and in astrophysics in general. They use the technique of Monte Carlo dropout to approximate Bayesian Neural Networks~\citep{Gal2015,Charnock2020} and show that, in that particular case, it results in accurate and precise uncertainty estimates, significantly faster than Monte Carlo Markov Chains.  

\begin{figure}
\centering
    \includegraphics[width=\linewidth]{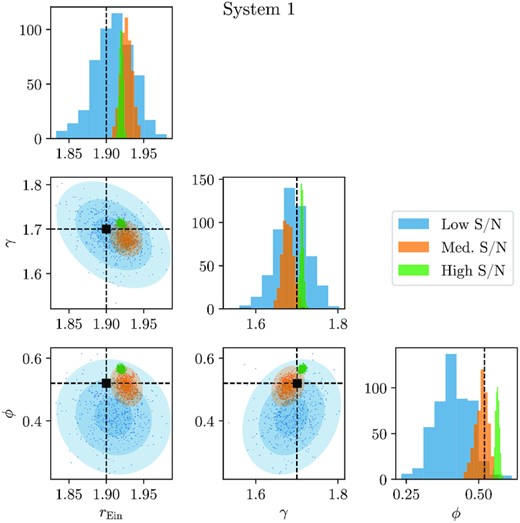}

    \caption{Lens modelling with a fully differentiable approach. The panels show the posterior distribution of different parameters of the model for different SNRs as labeled, together with the ground truth (black circle).  Figure adapted from~\cite{Chianese2020}}
    \label{fig:lens_model2}
\end{figure}

These two pioneer works have set the route for a fair amount of publications exploring the use of deep learning for lens modeling along similar lines. The most typical approach is the use of CNNs on images with an approximate Bayesian component to estimate uncertainties. For example, ~\cite{Madireddy2019} proposed a complete deep learning based pipeline including detection and classification of lenses followed by a modeling phase.~\cite{Bom2019} apply Residual Neural Networks to simulated images of the Dark Energy Survey to predict Einstein Radius, lens velocity dispersion and lens redshift within $\sim10-15\%$. See also~\cite{Schuldt2021} for similar conclusions on simulated Hubble Space Telescope and Hyper Suprime-Cam Survey images.~\cite{Pearson2019a} applied CNNs to simulated LSST and Euclid images. They find that including color information results in a $\sim20\%$  increase in accuracy compared to single band estimates. In a follow up paper,~\cite{Pearson2021} perform a systematic comparison between CNN based estimation and conventional parametric modelling on increasingly realistic datasets going from smooth parametric profiles to real observations from the Hubble Ultra Deep Field. The main conclusion is that CNNs outperform traditional methods not only in terms of speed but also in accuracy by $\sim20\%$. However, the work also concludes that combining both approaches reduces further the errors.  In addition, the use of CNN priors reduces the computational time of parametric modelling by a factor of $\sim2$.  ~\cite{Chianese2020} goes a step further by proposing a fully differentiable framework for lensing modeling. The pipelines combines a data-driven model based on a VAE for the source and a physical model for the lens which allows a forward modeling of the system to be compared with the observations. The main novelty, is that thanks to the differentiable programming framework, it becomes possible to compute the derivatives of the likelihood function with respect to both the parameters of the source and the lens, allowing for fast inference (\autoref{fig:lens_model2}). Along similar lines, ~\cite{Morningstar2019} combines a physical model with a Recurrent Neural Network to iteratively reconstruct the lens model which is then fed to a CNN to estimate the lens parameters.~\cite{Morningstar2018} applies the same methodology to analyze interferometric data.  The modeling of lenses can be combined with a direct inference of cosmological parameters such as the Hubble constant~\citep{Park2021}. We will address these applications in more detail in \autoref{sec:models}.~\cite{Maresca2021} proposed to use CNNs, not for reconstruction, but to identify unphysical models from parametric fitting.     

\subsubsection{Other properties}

Deep learning has also been applied to measure other galaxy properties, in addition to the major categories described in the previous subsections.~\cite{Stark2018} used a Generative Model to measure the photometry of AGN hosts. The neural network is used in this case to separate the light of the quasar from the emission of the host galaxy. In that respect it is similar to a deblending problem discussed in \autoref{sec:low_level}. In line with works under the same category, the authors demonstrate that their approach is more than 40 times faster than standard model fitting approaches.~\cite{YaoYuLin2021} also addressed the issue of AGN quantification by inferring the mass directly without going through photometry. The input in this case are quasar light time series. The work shows that neural networks reach similar accuracy than traditional methods that use SDSS spectra which are more time consuming to obtain. Also within the time domain community,~\cite{Stahl2020} developed a deep learning framework to measure the phase and the light curve shape of Type Ia supernova from an optical spectrum. ~\cite{CabayolGarcia2020} and~\cite{Cabayol2021} explored CNNs to systematically measure photometry on the narrow band PAU survey. They find in this case that the deep learning approach improves the photometric accuracy by up to $\sim40\%$.  

\subsection{Galaxy physical properties from simulations}

\label{sec:unobs}

In the previous section we have discussed applications where deep learning is used as an accelerator to replace existing methods. The main motivation of these works is therefore efficiency for dealing with large amounts of data. Because neural networks are universal approximators they can also be used to unveil new correlations between observables and physical quantities. In this case, deep learning is not replacing existing methods, but rather used as an exploration tool to unveil new patterns in the data which can be informative about underlying physical processes and/or about physical properties of galaxies from simulations. 

\subsubsection{Physical processes of galaxy formation}

Deep Learning offers a new way of establishing correlations between image features and physical processes driving galaxy formation in general. The general procedure is that simulations are used to identify a physical process without ambiguity. For example, galaxy mergers are straightforward to identify galaxies in a simulation but challenging to find in observations. Mock observations can therefore be produced to train a CNN at identifying the physical process. The advantage is that the network is let free to identify the optimal features that characterize a physical process given the observables. This way of proceeding is partly new. It has been driven by both the emergence of deep learning techniques and large sets of numerical simulations.  

\paragraph{Galaxy mergers}
A central example of this type of application which has caught the attention of many groups in the past years is the characterization of galaxy mergers. Mergers of galaxies are arguably a key driver of stellar mass assembly in galaxies across cosmic time. Identifying and characterizing the properties of large samples of mergers to assess their impact on diverse assembly processes has remained a key open issue in the field of galaxy formation for many years. The precise measurement of the merger rate - number of mergers per unit time and unit volume - is also an indirect probe of the cosmological model. Since galaxy mergers tend to disrupt the surface brightness distribution of galaxies because of the gravitational interaction, it is an old idea to use measurements of perturbations in the luminosity profiles of galaxies as indicators of a merger activity. A popular approach in the early 2000's was to measure some moments of light that are sensitive to asymmetries in the light distribution (e.g~\citealp{Conselice2003}). The problem is that the link between these perturbations and the actual merger activity is loose. The observability timescale of a given feature such a tidal structure depends on the type of merger, the cosmic epoch and other properties, which makes it very difficult to establish a direct link between image features and merger status. 

In that context, numerical simulations offer an attractive way of connecting measurements on images to a phase in the merger since the dynamical process of merger can be entirely tracked down in the simulation. Early efforts had indeed tried to establish some first order calibration using numerical simulations of galaxy pairs (e.g~\citealp{Lotz2008}). The work by~\cite{Lotz2008} associated variations in the moments of light - concentration, Gini, asymmetry also known as CAS parameters -  with the merger phases. However this was done manually and with a limited set of simulations.~\cite{Snyder2019} improved on these early works by exploring Random Forests applied to moments of light on numerical simulations. More recently~\cite{Whitney2021} used galaxies from the TNG simulation to calibrate the observability timescale of the so-called CAS parameters.   

\begin{figure*}
\centering
    \includegraphics[scale=0.5]{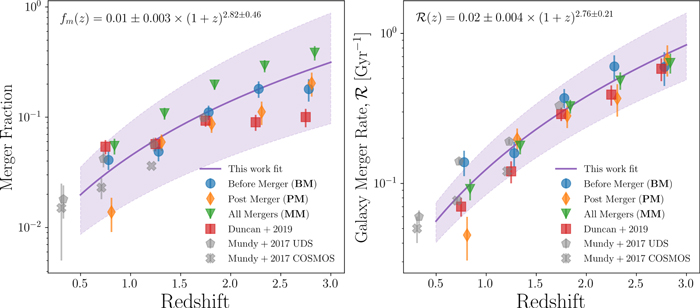}

    \caption{Supervised deep learning trained on simulations used to infer galaxy merger rates. The panels show the inferred merger fractions (left) and merger rates (right) with deep learning as compared to other independent techniques. Figure adapted from~\cite{Ferreira2020}}
    \label{fig:ferreira}
\end{figure*}

Deep learning offers however a new way of looking at this problem of detecting and characterizing galaxy mergers by bypassing summary statistics and manually engineered features. ~\cite{Pearson2019} first applied a CNN to the identification of galaxy mergers using a training set of mergers from the EAGLE simulation mimicking the SDSS observational properties. The key difference with previous works is that no manual features are extracted; the images of the different mergers are provided as input. In this first work, they found that CNNs did not achieve very high performance which is interpreted as a signature that the images used did not present significant perturbations. One possible reason for this poor performance is that mergers were selected in an non homogeneous way over a large range of times. The importance of the training set was carefully analyzed in the work by~\cite{Bottrell2019}. The authors explored how the performance of CNNs changed depending on the realism of the training set used for training. The main conclusion is that it is more important that images used for training reproduce the observational properties of the sample to be analyzed, i.e. PSF, noise, background sources, than using full radiative transfer to improve the conversion from stellar particles to light. 
\cite{Ferreira2020} followed up on this idea and used deep learning for the first time to compute the merger rate up to $z\sim3$. They trained a supervised CNN to identify mergers labeled in the TN300 simulation and selected over a fixed time window of $\sim1Gyr$. They showed that the CNNs can distinguish mergers from non mergers based on multi-wavelength imaging with an accuracy of $\sim90\%$. When applied to data from the CANDELS survey, their results reconcile measurements of the merger rate performed with pair counts and photometry (see \autoref{fig:ferreira}). These results confirm that the discrepancy between the two approaches was caused mainly because of calibration issues. Using deep learning on simulation allows one to properly calibrate the observability timescale based on the simulation metadata and therefore obtain a more reliable measurement of the merger rate. A similar approach was followed by~\cite{Bickley2021}, who trained a CNN to focus only on post mergers identified in the TNG simulation. They confirm that deep learning techniques outperform moment based identifications of post mergers and applied their model to the CFIS survey. However, the authors highlight a common problem with very unbalanced classification problems as this one. Since the number of post merger galaxies is very small compared to the global population of galaxies, even a small fraction of false positives, strongly impacts the purity of the post merger sample (see \autoref{sec:stronglenses}). 

Because neural networks are very flexible, they can be easily used to combine the information of different types of inputs which is more difficult with other techniques.~\cite{Bottrell2021} recently explored the combined information provided by photometry and kinematics to detect mergers. Both maps are fed into deep learning network which combines the information in an unsupervised way. Interestingly, the authors conclude that kinematics does not bring significant additional information.   

Other works have attempted to go beyond the classification and use neural networks to regress on the properties of the mergers.~\cite{Koppula2021} used a residual deep neural network to estimate the time to / from the first passage in a merger sequence. Using information from the Horizon-AGN simulation, the authors produce a large sample of images of major mergers at different stages. They show that, based on a single snapshot, the CNN is able to recover the position of the stamp in the merger sequence with an error of $\sim40$ Myrs. This is interesting because deep neural networks are used to provide temporal constraints on phases of galaxy formation, based on a single snapshot. A similar approach was followed by~\cite{Cai2020}. They trained a combination of Autoencoders and Variational Autoencoders to infer the properties of galaxy mergers. They conclude as well that with a single image, the dynamical status of the mergers can be inferred, bypassing dynamical modelling. Even more recently~\cite{Eisert2022} used an invertible neural network to infer several mass assembly indicators of galaxies (i.e. mass of accreted stars, time since the last major merger) using a variety of observable quantities from the TNG simulation. 

\paragraph{Other physical processes}

\begin{figure}
\centering
    \includegraphics[width=\linewidth]{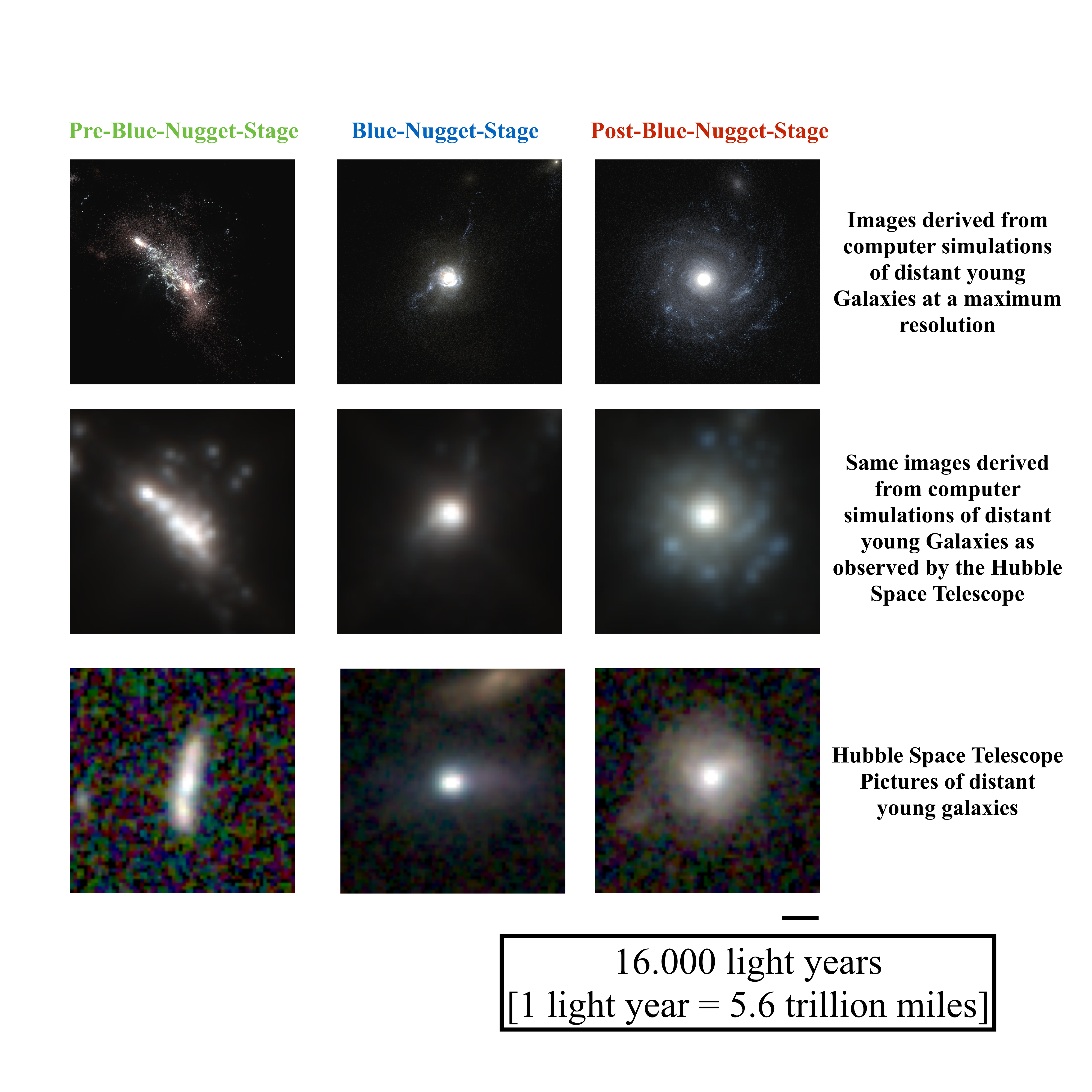}

    \caption{CNN applied to reconstruct to classify galaxies in different evolutionary stages defined by cosmological simulations. Each column shows a different phase of evolution as defined by the simulation. The top row is the high resolution from the simulation. The middle row shows the same galaxy with observational realism added. The bottom row shows real observed galaxies identified by the CNN as being in one of the three different phases.  Figure adapted from~\cite{HuertasCompany2018}}
    \label{fig:BNs}
\end{figure}

~\cite{HuertasCompany2018} first applied this idea to the identification of a physical process called \emph{compaction} or \emph{blue nugget} phase. Several observational works suggested that diffuse star forming galaxies become compact star-forming galaxies called 'blue nuggets' (BN) which subsequently quench ('red nuggets') following a sudden gas inflow towards their center. The VELA  zoom-in simulations~\citep{Ceverino2015} show also rapid gas inflows leading to central starbursts, and several mechanisms were identified that lead to this compaction phenomenon including major gas-rich mergers or disk instabilities often triggered by minor. The authors tested whether deep learning can detect the blue nugget phase of galaxy evolution purely identified in the simulation metadata. Mock HST images from the simulated galaxies were labelled according to their evolutionary stage from the simulation metadata (before, during, or after the BN phase), i.e. the labeling is exclusively done based on physics. The result is that galaxies can be successfully classified into evolutionary stages without identifying specific features, i.e. just using the pixel space (see \autoref{fig:BNs}).

~\cite{Diaz2019} applied 
a similar idea to the classification of formation mechanisms of lenticular galaxies. They identified different evolutionary tracks leading to the formation of S0 galaxies - isolated, tidal interaction in a group halo, and spiral-spiral merger -  and trained a CNN to identify them based on stellar density and two-dimensional kinematic maps. They found that the CNNs are able to distinguish the different formation scenarios and conclude about the  potential of deep learning to classify galaxies according to their evolutionary phases.~\cite{Schawinski2018} used a Generative Adversarial Network to try constraining the physical processes leading to the quenching of star formation in galaxies.~\cite{Ginzburg2021} used a CNN trained on zoom-in cosmological simulations to infer the longevity of star-forming clumps in high redshift galaxies. Instead of relying on the conversion between photometry and age of the stellar populations, they defined two types of clumps - short and long lived - based on the information from the simulation and trained a supervised neural network in a binary classification mode to distinguish between the two types. This is yet another example where neural networks are used as universal approximators to find not obvious links between observables and physical properties. 

\paragraph{Open issues}
Using deep learning trained on simulations to constrain the phases of galaxy formation is becoming increasingly popular in the community and the example of mergers clearly illustrates this. All these approaches suffer however from an obvious limitation, the so called domain gap between observations and simulations. As this is a recurrent problem, more information is provided in~\autoref{sec:final_thoughts}. Since simulations do not perfectly reproduce observations, applying a trained network on observations will induce some biases. Moreover, since observations are generally not labeled at all, especially when trying to infer physical properties, it is impossible to evaluate the effect of the domain gap and the results need to be accepted blindly. As already mentioned in previous sections, including uncertainty quantification in the neural networks is an option to mitigate this effect. However the error induced by changes in the domains is in general difficult to capture by uncertainty quantification methods. Other options consist in performing the domain adaptation during training to ensure that the features learned by the neural networks are not specific to the simulation domain. There are different approaches to do so since it is a problem that exists in many fields of application (see e.g.~\citealp{Wang2018}). In extragalactic astronomy there has been limited exploration of these approaches.~\cite{Ciprijanovic2021} recently explored the impact of domain adaptation during training for the identification of galaxy mergers. They tested different domain adaptation techniques such as Maximum Mean Discrepancy and Domain Adversarial Neural Networks and concluded that including these leads to an increase of target domain classification accuracy of up to $\sim20\%$. This is a promising result for future applications of neural networks trained on simulations. It is likely that future works will start to incorporate these approaches more often. 

\subsubsection{Dark Matter}

In a similar spirit of finding new correlations, deep learning has been increasingly used in the past years to estimate the dark matter masses and properties of galaxies and clusters, based on observable quantities. 

\paragraph{Dynamical masses of clusters of galaxies}

The earliest applications focused on galaxy clusters, which are the largest gravitationally bound objects. This is because there exist alternative methods to measure dynamical masses of clusters. However standard approaches are based on simple scaling relations based on the virial theorem. The simplest approach is to use a power law relation between the dispersion of the line of sight velocities (LOS) and the cluster mass. This was indeed one of the first probes of dark matter by~\cite{Zwicky1933}. Other classical approaches consist in using scaling relation between the X-ray luminosity of the gas and the mass of the cluster or the Sunyaev-Zeldovich effect. However, these scaling laws are all based on strong assumptions about the physical status of the cluster, the  most important being that the system is in virial equilibrium. This entails some inherent biases in the estimated cluster masses.

Deep learning and machine learning in general, calibrated on simulations where dark matter properties are known, offer an interesting approach to look for additional correlations which could help in reducing the scatter in the dynamical mass estimates. Early efforts were done before the emergence of deep learning, especially in the works by~\cite{Ntampaka2015,Ntampaka2016}. The authors explored Support Distribution Machine class of algorithms to predict cluster masses using LOS velocities and radial positions of cluster members. They reported an improvement of a factor of two (see also~\cite{Armitage2019} for similar conclusions). 

\cite{Ntampaka2019} applied CNNs to cluster mass estimation in which is the first work using deep learning for this purpose. The training is performed with mock 2D X-ray images of Chandra observations. There is indeed a well known correlation between X-ray luminosity and cluster masses. They report a $\sim10\%$ smaller scatter than standard luminosity based methods, even without using any spectral information. Interpretability techniques based on attribution methods, reveal that the CNNs tend to ignore the cluster centers because they likely lead to more biased estimates. This is another example of CNNs used to find new correlations in the data and one of the few examples were basic attribution techniques provide meaningful information. Similarly,~\cite{Yan2020} used a combination of feature maps (stellar mass, soft X-ray flux, bolometric X-ray flux, and the Compton y parameter) and reach comparable results. This work illustrates an other advantage of deep learning over traditional approaches, which is the possibility of combining multiple observables in a transparent way. 

Along the same lines, ~\cite{Ho2019} explored CNNs to estimate cluster masses. The training is based on LOS velocities and radial positions of galaxies in the cluster. They explored both 1D and 2D CNNs. They also report a factor of $\sim2$ improvement with respect to power-law based estimates. Interestingly CNNs also improve the results of more classical ML approaches explored in previous works.  

Convolutional Neural Networks have also been explored on the third observable usually employed to infer cluster masses, the Sunyaev-Zeldovich effect.~\cite{Andres2021} used CNNs on mock maps of the Planck satellite from numerical simulations. The advantage of using deep learning is again that no assumptions on the symmetry of the cluster's gas distribution nor on the cluster physical state are made.

\begin{figure}
\centering
    \includegraphics[width=\linewidth]{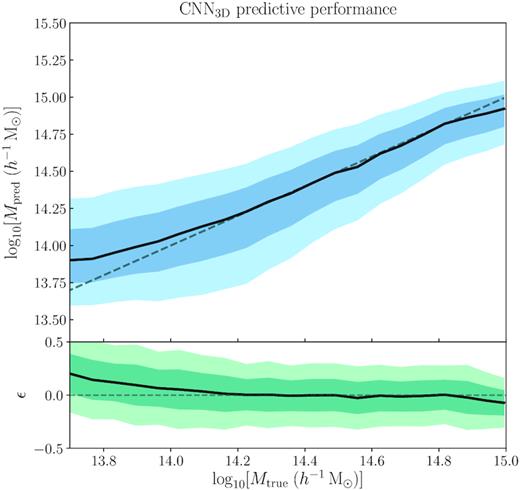}

    \caption{ The figure shows the predicted dark matter mass as a function of the true mass along with uncertainties using a Normalizing Flow.  Figure adapted from~\cite{KodiRamanah2021}}
    \label{fig:arf_clusters}
\end{figure}

These previous works, although promising, remain at the exploratory level and suffer from the same limitations than other similar approaches. Namely, the results are restricted by the prior inferred from the simulations and they usually lack of uncertainty estimation with some exceptions. The works by~\cite{KodiRamanah2020,KodiRamanah2021} make a step forward to address some of these issues. In these works, the authors explore, for the first time, flow based neural networks (see~\autoref{fig:flow_only}) trained of the phase space of cluster galaxies to infer cluster masses (\autoref{fig:arf_clusters}). The key addition of their approach is that the network provides therefore a full probability distribution of the cluster mass instead of a single point estimate, which can therefore be used to account for uncertainties. This is a key step forward towards an application of deep learning based approaches for estimation of cluster masses in large surveys. The authors claim a factor of 4 improvement compared to scaling relations  based estimations. They then apply their model to a sample of observed clusters with well calibrated dynamical masses and show that the neural network provides both unbiased measurements and well calibrated uncertainties. ~\cite{Ho2021} also investigate the use of Bayesian CNNs to include uncertainty measurements. They show that BNNs  recover well calibrated 68\% and 90\% confidence intervals in cluster mass to within 1\% of their measured value. 

\begin{figure}
\centering
    \includegraphics[width=\linewidth]{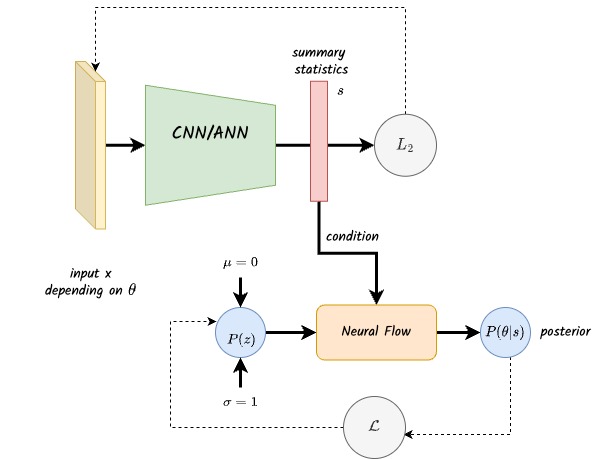}

    \caption{Illustration of a Neural Flow for estimating posterior distributions. This type of approach is starting to become common in simulation based inference approaches for estimating cluster masses or cosmological parameters. A first Neural Network with a standard $L_2$ loss is used to obtain some summary statistics of the data which are then use as a condition for a Neural Flow mapping a simple distribution into an approximation of the posterior.  }
    \label{fig:flow}
\end{figure}

\paragraph{Halos of galaxies}

\begin{figure}
\centering
    \includegraphics[width=\linewidth]{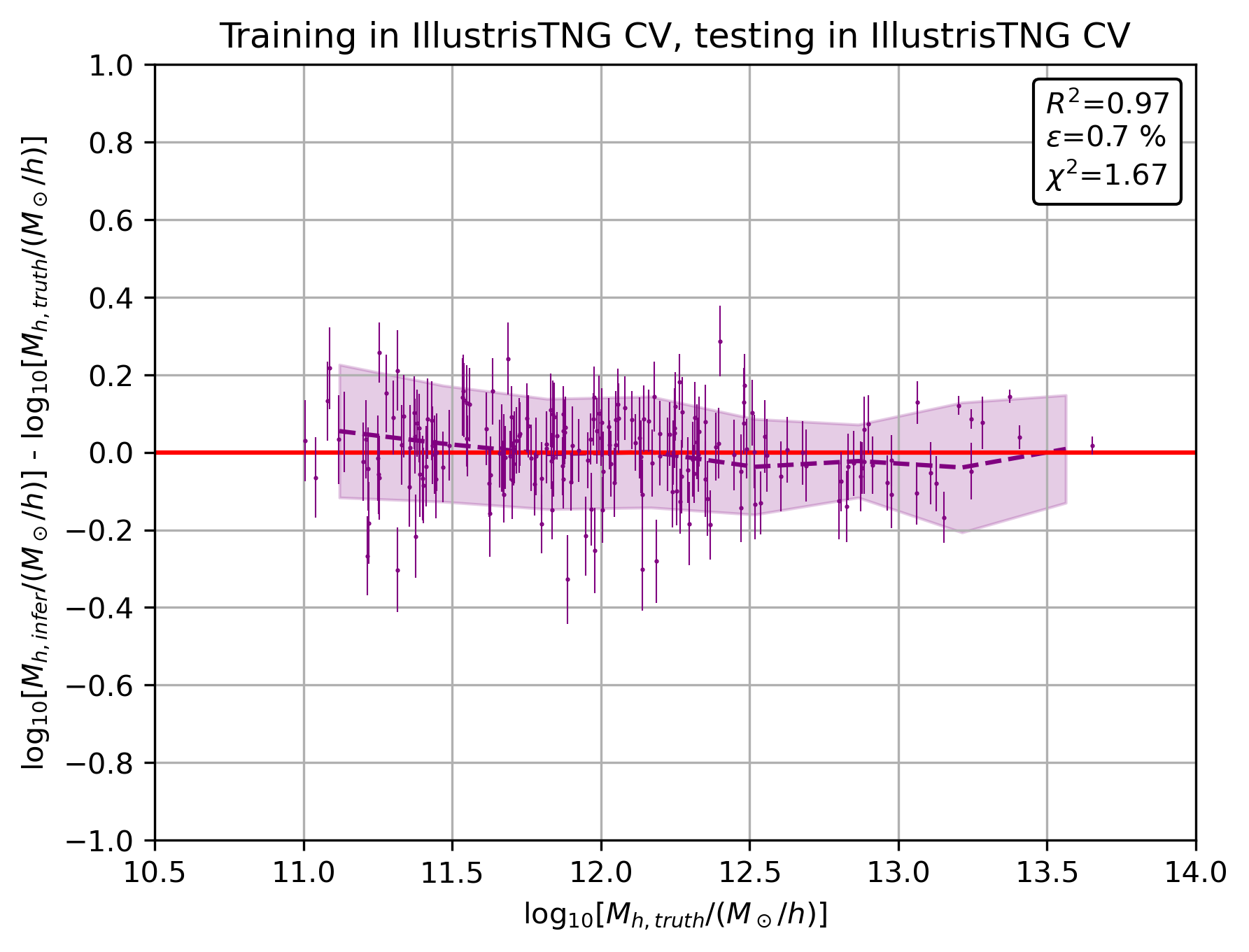}
     \includegraphics[width=\linewidth]{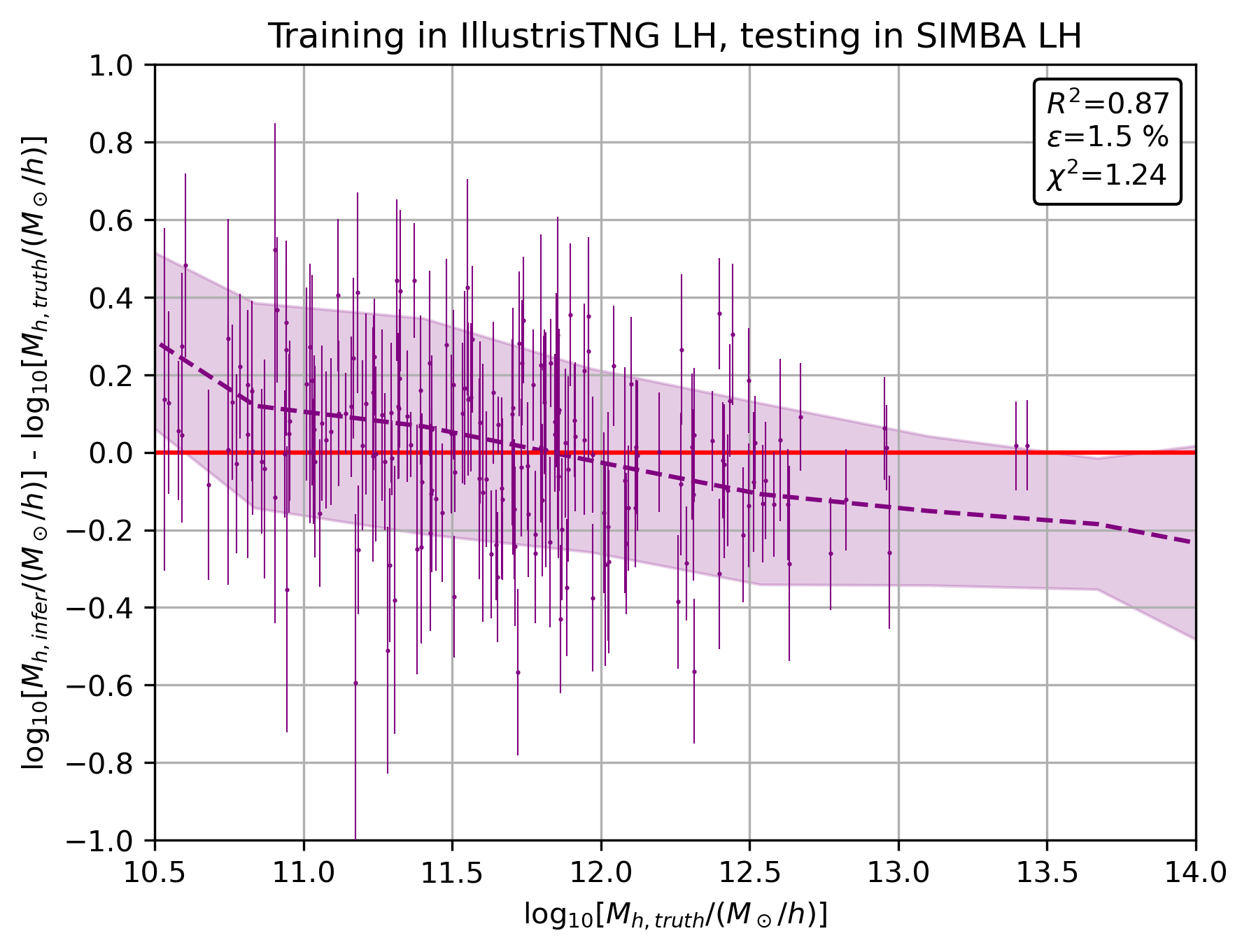}

    \caption{The top panel illustrates the accuracy obtained on simulations, when the training and testing is done on datasets coming from the same underlying cosmological simulation. The bottom right panel is when two different simulations are used for training and testing respectively. Figure adapted from~\cite{VillanuevaDomingo2021}}
    \label{fig:GNN_halos}
\end{figure}

Deep learning can be also extended to estimate dark matter halo masses of less massive galaxies than clusters. In low mass halos, there are also less galaxies and therefore it is more challenging to use LOS velocities. They do not present any X-ray emission either. The most standard way to proceed is by using Abundance Matching techniques. Abundance Matching main assumption is - with some variations - a monotonic relation between the stellar masses of galaxies and dark matter halos.  By using dark matter halo mass functions from N-body simulations, one can then assign halo masses to galaxies. In this context, deep learning can be used 
to look for additional correlations between galaxy properties and dark matter beyond simple abundance matching assumptions.~\cite{Calderon2019} explored several machine learning algorithms, including neural networks, for estimating the dark matter halo masses of galaxies in the SDSS. They used ML to explore how much information about halos is provided by several galaxy properties from synthetic catalogs. They conclude that including more physical properties is translated into a better accuracy than the one reached by abundance matching. A problem with this approach, acknowledged by the authors, is that secondary dependencies of halo masses on galaxy properties might be very model dependent. This can therefore induce systematic biases in the inferred dark matter masses which do not exist in simpler approaches.~\cite{Shao2021} investigated how subhalo masses can be estimated using neural networks trained on a number of physical properties of galaxies (i.e. black hole mass, gas mass, stellar mass etc) from numerical simulations. They used the CAMELS simulation suite which is a series of numerical simulations performed with different codes and cosmologies, specially designed for ML. We will describe the simulations into more detail in \autoref{sec:models}. They found that subhalo masses can be predicted accurately ($\sim0.2$ dex) at any redshift from simulations with different cosmologies, astrophysics models, subgrid physics, volumes, and resolutions. The authors argue that the neural networks might have found a universal relation, which turns out to be a generalized version of the virial theorem involving radius, velocity dispersion  and maximum circular velocity. This is a good example of deep neural networks used to find hidden correlations which can be even translated into analytical expressions. We will discuss this further in \autoref{sec:discovery}. 
In a recent work,~\cite{VillanuevaDomingo2021} explored the use of Graph Neural Networks (GNNs) to estimate halo masses of galaxies. GNNs are a special type of neural networks that are built on graphs, and therefore allow one to account for the relations between neighbouring halos. The authors used the halos from cosmological simulations as nodes of the graphs and encoded the gravitational interaction between them in the edges of the graph. The nodes include the positions of the halos, the relative velocities, the stellar mass and the half-mass radius. They show that the model is able to estimate halo masses with a $\sim0.2$ dex uncertainty (\autoref{fig:GNN_halos}). The model is also built to account for uncertainties and is shown to generalize reasonably well between different simulated datasets. 

In a follow-up work~\citep{VillanuevaDomingo2021a}, the authors apply the trained GNNs to measure the halo masses of the Milky-Way and Andromeda galaxies. They show that the inferred constraints are in good agreement with estimates from other traditional methods.

\subsubsection{Deep learning generated observations}
Some recent works have tried to push even further the ability of neural networks to establish not obvious non-linear mappings between domains to bypass telescope observations. Spectroscopic observations are more expensive in terms of observing time than imaging ones. However, spectroscopy is significantly richer in terms of available information about astrophysical processes. Some works have therefore explored if deep neural networks can infer spectra from images without the need for observing time.~\cite{Wu2020} showed that deep neural networks efficiently learn a mapping between galaxy imaging and spectra. They showed that SDSS spectra can be inferred from 5-band images with very high accuracy. The authors argue that this approach could be used as an exploration tool for future surveys such as LSST.~\cite{Holwerda2021} followed up on this work by applying the same network to estimate the spectrum of an AGN and compare it to existing spectra from the literature. Although the neural network predicted spectrum is in overall good agreement with the observed ones there are some differences in the strength of emission lines. This suggests that the ML approach might be an interesting way of identifying specific types of objects such as AGN and/or as a first order exploration. It is difficult to use the inferred spectra to analyze the physical properties though. This is somehow expected since spectra contain more information than imaging, as previously stated. A similar experiment is performed by~\cite{Hansen2020} who infer kinematic information of galaxies (velocity dispersion and rotation maps) from single band imaging.  

\vspace{.3cm}
\begin{mdframed}[backgroundcolor=orange!30] 
{\bf Summary of Deep learning for inferring physical properties of galaxies} \\
\begin{itemize}
    \item Over the past years, deep learning has been tested as a tool to infer physical properties of galaxies in large surveys. The most common applications are: photometric redshift estimation, galaxy structure and stellar populations and strong lensing modeling. 
    \item The main motivation for using deep learning for these tasks is computational speed. Deep learning is used here as a fast emulator for existing methods. Overall, the most common conclusion of these tests is that deep learning approaches achieve state-of-the art performance, several orders of magnitude faster.
    \item The typical approach used is a supervised regression (C)NN trained on simulated datasets, for which the ground truth is known. 
    \item A major challenge of these applications is robustness. Since the models are predominantly trained on simulations, the representativity of the training set is a major issue. Extrapolation with neural networks is in general problematic. Therefore, making sure that the training samples properly cover the inference dataset is a key challenge.    
    \item Uncertainty quantification is also particularly important for this type of applications. Bayesian Neural Networks and Density estimators are among the most commonly employed solutions.  
    \item As an extension to the derivation of physical properties, deep learning has also been explored as a tool to identify new correlations between observable and other physical properties of galaxies, which are generally not accessible with existing methods. Examples of this type of application include the inference of phases of galaxy evolution such as interactions, or the estimation of the dark matter content of galaxies.
    \item The main motivation is that neural networks are universal approximators. Deep learning is therefore employed to unveil hidden correlations using simulation based inference. By definition, the training is performed on (cosmological) simulations in which all the physical properties and evolutionary phases of galaxies are accessible. 
    \item A key challenge for these type of approaches is robustness against the representativity of training sets and domain shifts. The effect of these issues is particularly dramatic here since cosmological simulations are known to be approximations to the observed universe. Moreover, in general, \textit{pre-deep learning} approaches to compare with do not exist, as opposed to the previous type of application. 
    \item Because deep learning is typically used \emph{blindly} informed by simulations, interpretability becomes a key limitation. Solutions based on saliency maps or symbolic regression - when possible - have been explored. However these approaches still present a limited informative power. There is significant room for improvement in the future.
\end{itemize}
\end{mdframed}

\section{deep learning for discovery}
\label{sec:discovery}

In this section, we focus on efforts done by the community to use deep learning as a discovery tool. Applications typically include dimensionality reduction to visualize complex datasets and identify groups of objects, anomaly detection to automatically find potentially interesting objects in large datasets and some early efforts to automatically learn fundamental laws of physics. 

\subsection{Visualization of large datasets}

In addition of increasing in volume, datasets in astrophysics are becoming increasingly complex and of high dimensionality. Machine Learning can be employed to visualize datasets in a low dimension space to look for trends and correlations in the data, which otherwise are difficult to extract. It can also be used to identify classes of objects that share some properties which can help with the scientific analysis. These applications use typically unsupervised learning approaches, as opposed to what has been previously discussed. In unsupervised learning, data is unlabeled and therefore we seek a representation of the data instead of a mapping between data points and labels. We emphasize again that the present review focuses on deep learning applications and, therefore, we will not describe in detail works using other ML approaches for data visualization. There exist however a large variety of techniques which do not involve neural networks and that have been applied to astronomy. For example, \cite{Baron2021} used graph representations to find structures in imaging and spectroscopic data. The works by~\cite{Hocking2018} and~\cite{Martin2020} also explore clustering coupled with graph representations to group images of galaxies that look similar. Self Organizing Maps have also been used to represent images of galaxies in the radio domain (see e.g.~\citealp{Galvin2020}) and for spectral classification (e.g.~\citealp{Rahmani2018}). Other non neural network based dimensionality reduction techniques such as Principal Component Analysis (PCA), t-SNE~\citep{vanDerMaaten2008} or UMAP~\citep{McInnes2018} are also used in several works to explore data.

\begin{figure}
\centering
  
    \includegraphics[width=\linewidth]{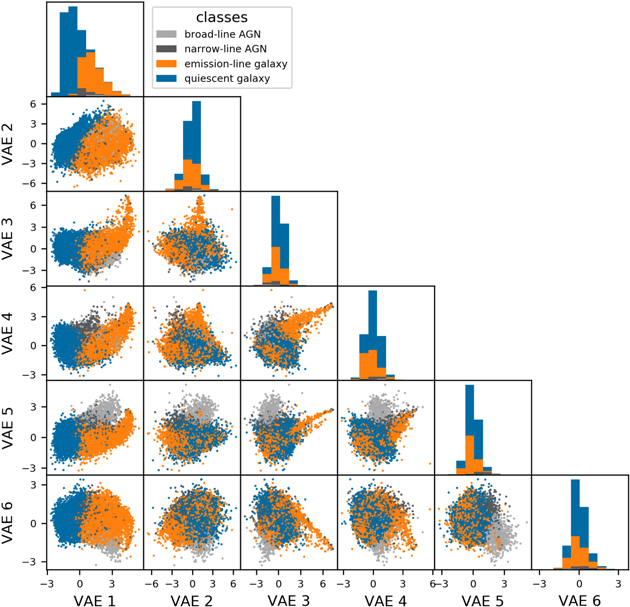}

    \caption{Variational Autoencoder for dimensionality reduction and data visualisation. 
     The figure show how the different types of spectra (labeled with different colors) populate different projections of the latent space. Figure adapted from~\cite{Portillo2020}}
    \label{fig:portillo}
\end{figure}

Deep learning offers several approaches for dimensionality reduction and visualization. The most standard and widely used are Autoencoders, a particular type of encoder-decoder architectures (see \autoref{sec:low_level} and~\autoref{fig:unet}) in which the inputs and outputs are identical. The networks therefore simply learn how to reproduce the input data. However, since these architectures - which can be deterministic or probabilistic - typically present a bottleneck at the junction between the encoder and the decoder, they naturally represent the data in a low dimension space. Exploring the distribution in this bottleneck layer is useful to find structures in the data.
~\cite{Ma2019} used a Convolutional Autoencoder (CAE) to explore a sample of radio active galactic nuclei (AGNs). By feeding the CAE with images of radio AGNs with hosts of different morphologies, they showed that the network naturally clusters similar objects together in the bottleneck. In that particular work, the low dimensional representation is also used for a downstream supervised classification using the learned feature space as input for a supervised network. This is also a common application of the dimensionality reduction approach. When only a reduced subsample of labeled examples is available, the reduced dimensionality space can be used to train a supervised network with smaller training sets. A similar approach was followed by~\cite{Cheng2020}. They used a CAE to represent images of galaxies in a low dimension space with the goal of finding strong gravitational lenses. Sine the samples of labeled lensed systems are typically small (see \autoref{sec:stronglenses} for more details), unsupervised representation offers an alternative way to find lenses without labels. The authors perform a clustering step in the latent space learned by the neural network to automatically find groups of objects which similar properties. They find that the CAE based method successfully isolates $\sim60$ per cent of all lensing images in the training set. In~\cite{Cheng2021} they extend the same approach to the unsupervised exploration of galaxy morphology. Using a modified version of a Variational Autoencoder, they obtain an unsupervised representation of nearby galaxies from the SDSS survey. They then perform a hierarchical clustering in the latent space to conclude that the neural network representations share some properties with the classical Hubble sequence but provide a more meaningful representation, especially for ambiguous intermediate morphological types. See also the work by \cite{spindler2021} for an application of VAEs to representation of galaxy morphology. A similar approach is presented in the work by~\cite{Zhou2021}. The authors apply a combination of CAE based representation with a multi-clustering model to study the morphologies of high redshift galaxies from the CANDELS survey. ~\cite{Portillo2020} also used the same type of approach involving a Variational Autoencoder to represent spectra of nearby galaxies. The authors projected SDSS spectra into a latent space of 6 dimensions and showed that the different types - i.e star-forming, quiescent - are naturally separated without labeling (\autoref{fig:portillo}). Interestingly, the non-linear components of VAEs seem to enable a better separation than a simple PCA decomposition if the latent space remains of dimension lower than $\sim10$. The conclusion is that dimensionality reduction with neural networks is a sensitive way of exploring data of high dimension. Notice however that they did not use convolutional layers. 

In a recent work,~\cite{Teimoorinia2021} follow a similar approach, but with additional layers of complexity, to explore Integral Field Unit (IFU) data from the Manga survey. In their approach, called DESOM, the authors propose to combine a convolutional Autoencoder and a SOM to represent spectra. The spectra are fed to a CAE and projected into a latent space of lower dimension. In the same training loop, the representations are used to train a SOM that further clusters similar objects. That way, all spectra of a galaxy can be passed into the machinery to obtain a \emph{fingerprint} for every object which corresponds to the final projection into the SOM plane. The authors propose going a step forward by passing again the obtained \emph{fingerprint} into the DESOM to obtain a single location for a galaxy based on the 2D distribution of all spectra beloging to the same galaxy.   

Other works have also applied deep learning based dimensionality reduction to assess data quality. For example,~\cite{Mesarcik2020} used an Autoencoder to explore radio data and identify possible technical failures in the observations. This illustrates another interesting use of deep learning dimensionality reduction techniques to quickly explore complex datasets. 

\begin{figure}
\centering
    \includegraphics[width=\linewidth]{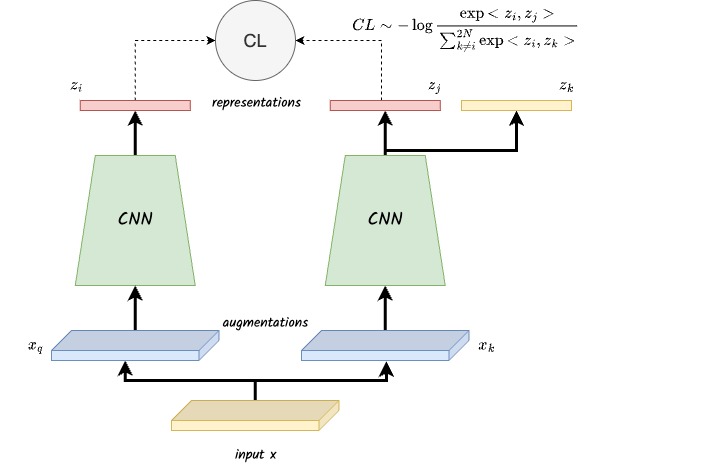}
      
    \caption{Illustration of a self supervised contrastive learning architecture. Multiple random augmentations of the same image (positive pairs) are fed to two different CNNs which map them into a latent representation space. Also during training, pairs of completely different images (negative pairs) are also fed to the two CNNs. The contrastive loss is optimized to increase (decrease) the dot product of representations of positive (negative) pairs. Contrastive learning is starting to be used for dimensionality reduction and as a generalized feature extraction process for multiple downstream tasks such as galaxy classification or photometric redshift estimation.}
    \label{fig:contrastive}
\end{figure}

\begin{figure}
\centering
 
      \includegraphics[width=\linewidth]{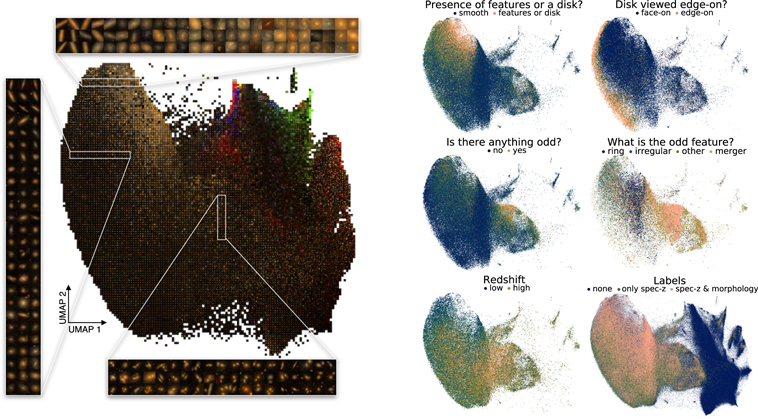}

    \caption{Self-supervised learning applied to multi band SDSS images. The left panels shows a UMAP of the representations obtained with contrastive learning. The panels on the right show the same UMAP color coded with different galaxy properties. Similar images are clustered together. Figure adapted from~\cite{Hayat2021}}
    \label{fig:self}
\end{figure}

Another deep learning approach for dimensionality reduction which is increasing in popularity in the recent years, is what is generally known as self-supervised learning through contrastive learning. As opposed to the Autoencoder approach, where the projection depends on the architecture, in contrastive learning, the computation of representations is more data oriented. The general idea is to apply some perturbations to the input data so that the networks learn to ignore those and cluster together data points coming from the same parent input data. This is obtained by what is called a contrastive loss term. We emphasize that it is not the goal of this review the technical details of the different deep learning approaches, but to review how they are being used in astronomy. We refer the reader to~\cite{Chen2020} and references therein for more details (see \autoref{app:acron} for references on the different deep learning methods mentioned in this review). \autoref{fig:contrastive} shows a very schematic representation of a contrastive learning setting. The output is in essence similar to the one obtained with an Autoencoder - i.e. a representation of data in a reduced latent space - but the underlying idea is significantly different. One of the key advantages of a contrastive approach is that the perturbations applied to the input data can be tuned for a science case and turn the representations independent to a known undesired effect. In astronomy, it can enable to mitigate the effects of instrumental or selection biases for example. 
Contrastive learning has only started to be applied in astrophysics relatively recently. The first work exploring self-supervised learning is by~\cite{Hayat2021}. The authors used an existing network to compute representations for multi-band SDSS images. Among the perturbations applied to the images, they included standard rotations and cropping, but also some adapted to astronomy, such as extinction. They showed that the contrastive learning model successfully clusters galaxies with similar morphological properties and therefore constitutes a promising way for data exploration in astrophysics (\autoref{fig:self}).

\begin{figure}
\centering
    \includegraphics[width=\linewidth]{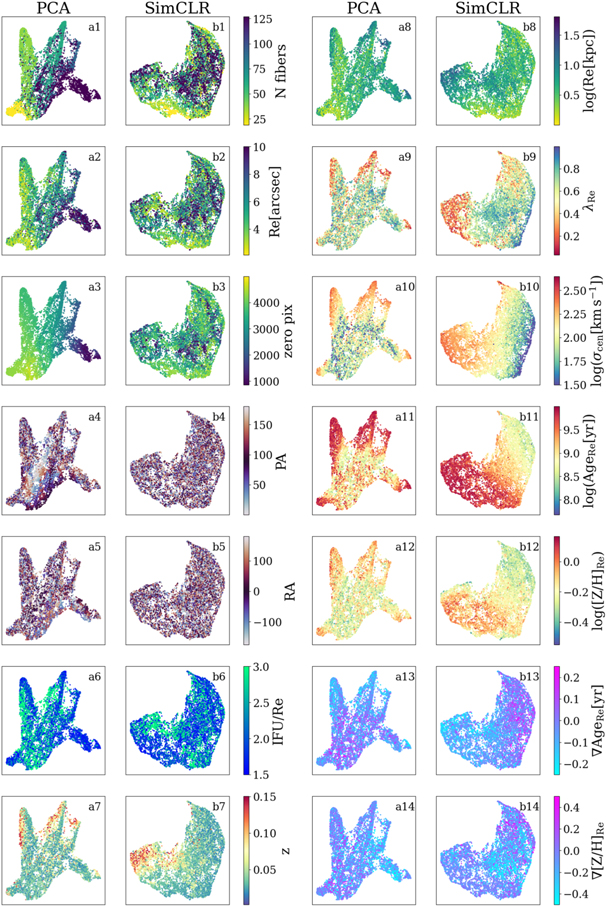}

    \caption{Representations of Manga maps using PCA and contrastive learning, and projected into a UMAP. The two leftmost columns show the plane color coded with non-physical parameter (e.g. number of fibers in the IFU). The rightmost columns show the same maps color coded with physical properties. Self-supervised representations cluster galaxies according to physical parameters while PCA focuses mostly on the number of fibers. Figure adapted from~\cite{Sarmiento2021}}
    \label{fig:sar}
\end{figure}

~\cite{Sarmiento2021} also applied contrastive learning to visualize data of nearby galaxies from the Manga survey. Instead of images, they used post processed maps of stellar populations properties (metallicity, age) as well as stellar kinematic maps. They also show that the self-supervised learning setting is able to condense the information from this high dimensional dataset into a subset of meaningful representations which contain information about the physical properties of galaxies. Interestingly, this is a case in which other simpler dimensionality reduction techniques such as PCA, or even Autoencoders, fail, given the large amount of instrumental effects present in the data. The authors show that more standard techniques tend to organize galaxies based on properties of the instrument (i.e. fiber size) instead of physical ones. Because the contrastive setting allows one to tune the augmentations to a specific problem, it can be trained so that the representations become independent of instrumental biases (\autoref{fig:sar}).  The inferred representations can be used for example to perform a clustering step and identify different classes of objects. ~\cite{Sarmiento2021} show that well-known types of galaxies appear naturally without any human supervision from a data-driven perspective.  

 In addition to visualization, a common application of self-supervised representations is to use them as input for other downstream tasks. For example, the representations can be used for a subsequent supervised classification. The fact that objects have already been clustered together, helps to converge faster, which makes it specially appealing when a small amount of labeled data is available (see \autoref{sec:low_level}).~\cite{Hayat2021} showed indeed that, by using the latent space for morphological classification of galaxies, they can reach a similar accuracy as with a fully supervised CNN but using $>10$ times less labeled data. In the follow up work by~\cite{Stein2021a}, they also explore how the self supervised representations can be used to find strong gravitational lenses reaching similar conclusions. Similarly to the work by~\cite{Cheng2020}, the representations are used with a small sample of lenses to train a simple linear classifier and find new strong lenses candidates.
 
 \subsection{Outlier detection}

 Anomaly or outlier detection is a fundamental aspect of discovery in physical sciences. A fair amount of new results in astrophysics have been triggered by serendipitous discoveries through the exploration of observational datasets. With the arrival of new big-data surveys, finding potentially interesting objects becomes increasingly difficult with purely human based approaches. In the past year, unsupervised deep learning has been explored by several groups as a way to assist astronomers in the search of potentially interesting objects. 

An anomaly or outlier is usually defined as a data point which properties deviate from the average properties of objects in the sample, under some metric. The visualization techniques described in the previous subsection, which tend to cluster together data points with similar properties, can therefore be useful as well to identify deviant objects. 

From a probabilistic point of view, an outlier can be also defined as an object which probability of observation under the probability density distribution of a data set is smaller than a given $\epsilon$. In that context, modern generative models (e.g. VAEs, GANs), can approximate the probability density function $p(X)$ of a dataset $X$ with increasing accuracy. Therefore, they can be employed to look for objects with a small probability of observation (see~\cite{Chalapathy2019} for a generic review of deep learning techniques applied to anomaly detection).

\subsubsection{Transient astronomy}

The field of transient astronomy has been particularly active in this context. As previously summarized, the field is about to experience a data revolution. The forthcoming LSST survey will observe all the Southern Hemisphere sky every $\sim2-3$ nights, producing an unprecedented real time \textit{movie} of the night sky. The community expects to discover a significant amount of new types of variable objects using this dataset~\citep{Li2022}. Therefore there have been over the past years a number of works exploring deep learning and machine learning in general, to identify anomalous light curves in preparation of LSST 
We emphasize again that deep learning is not the unique machine learning approach to identify anomalies.~\cite{Malanchev2021} performed a comparison of several anomaly detection algorithms - isolation forests, on-class SVMs, Gaussian Mixture Models and Local Outlier Factor - to identify outliers in the Zwicky Transient Facility (ZTF). See also the work by~\cite{MartinezGalarza2021} which used decision trees and manifold learning.~\cite{Pruzhinskaya2019} uses Isolation Forests as well on a set of features derived from the light curves using interpolation with Gaussian Processes. In the following, we will however focus  on efforts relying on deep learning.

~\cite{Villar2021} used a Variational Recurrent Autoencoder (VRAE) network described in~\cite{Villar2020} to identify anomalous light curves from the simulated PLAsTiCC dataset. The proposed methodology is based on three main steps involving three different ML algorithms. First the light curves are interpolated using Gaussian Processes (GPs). The resulting interpolations are then fed into a Variational Autoencoder. The temporal aspect is encoded by appending the time step to the elements representing the time series. The low dimension representation of the time series is finally passed through an Isolation Forest algorithm to assign an anomaly score. As opposed to well defined supervised problems, evaluating and comparing anomaly detection algorithms is always difficult since by definition the objective is not well defined. In that particular work, the authors quote a $\sim95\%$ purity in identifying light curves others than the ones generated by well known types of objects. However, by definition the exercise is incomplete, since the sensitivity to unknown unknowns cannot be assessed. This work illustrates an interesting way of combining multiple ML approaches though. In particular, the introduction of a GP for preprocessing turns the model agnostic to the sampling frequency of the time series which is a very interesting feature for astronomical applications. A Variational Recurrent Autoencoder  is also used by~\cite{SanchezSaez2021} to identify changing-look Active Galactic Nuclei. The approach is analogous but it is applied to real observations from the ZTF survey. The VRAE is trained with light curves to obtain a representation in a low dimension space and then Isolation Forest is used to associate anomaly scores. ~\cite{Boone2021} also employs a VAE to learn how to reconstruct light curves and then uses the latent space to assign anomaly scores to potentially deviant curves. The architecture used is slightly different than in the previous two works. Namely, they add a layer introducing some physical information about the light curve so that the setting can also be used for classification. However, the overall idea is analogous in essence.   

\begin{figure}
\centering
    \includegraphics[width=\linewidth]{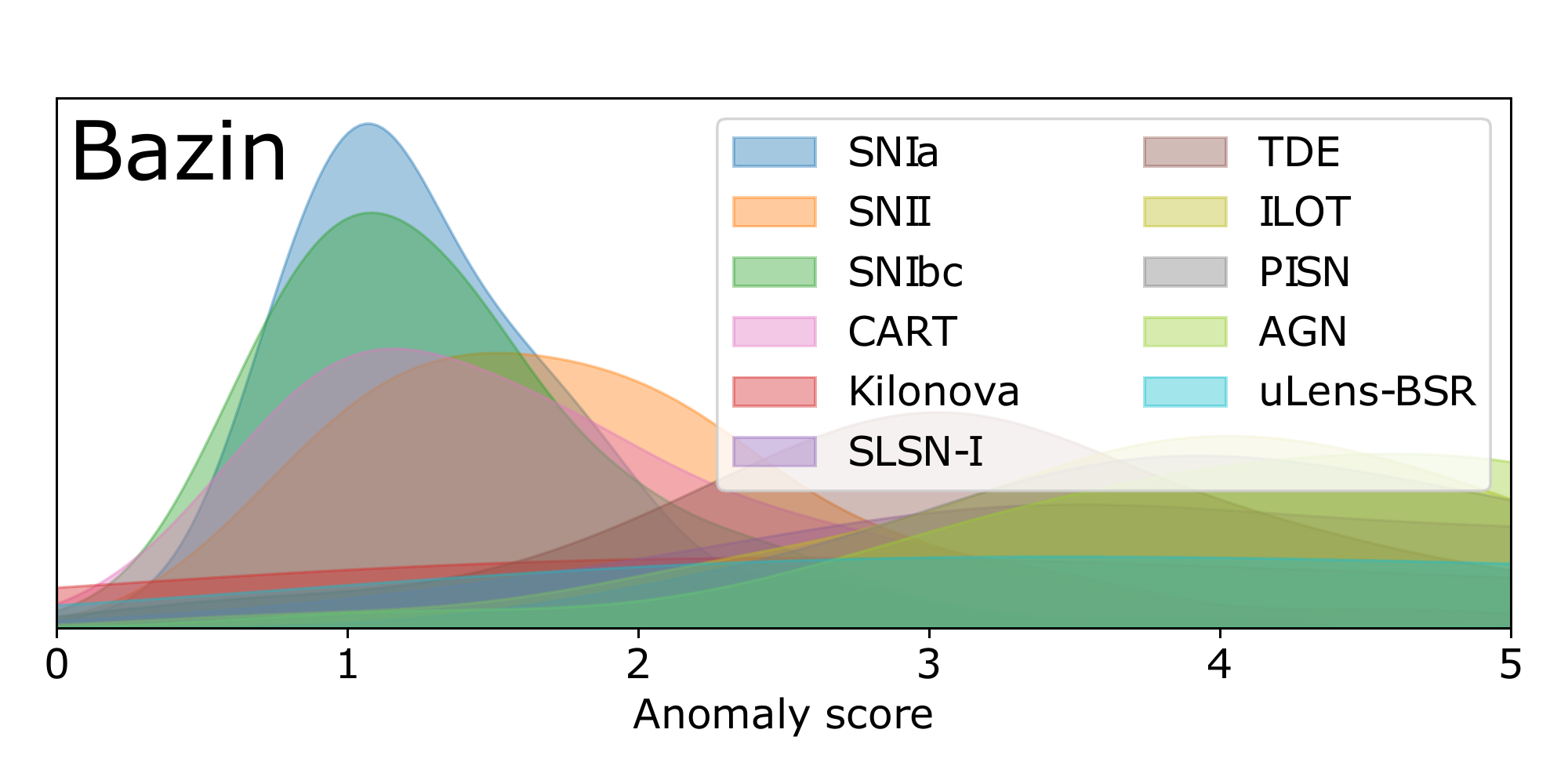}
    \includegraphics[width=\linewidth]{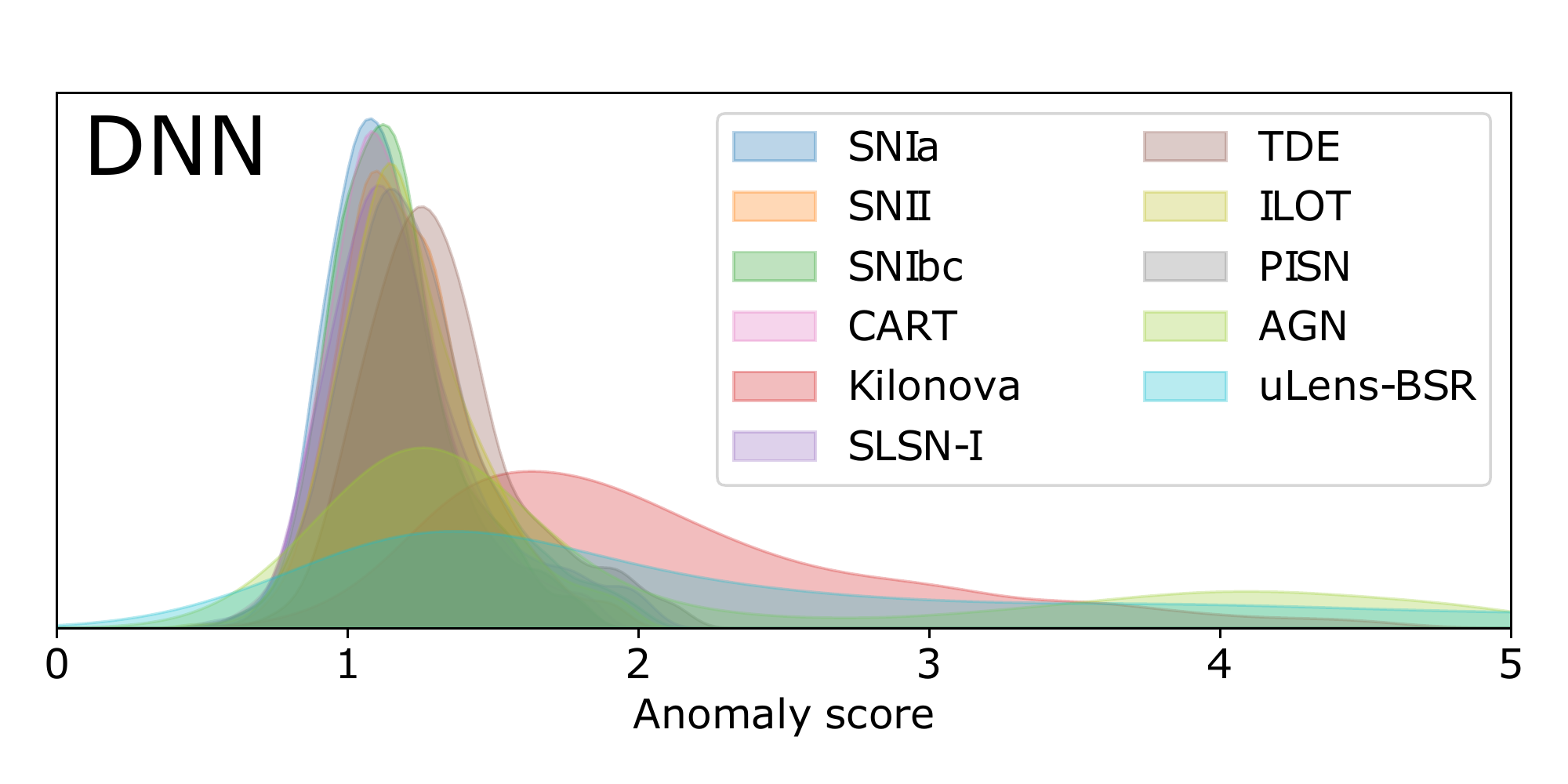}

    \caption{Anomaly scores for different types of light curves obtained with deep AutoRegressive Model (bottom panel) and with a Bayesian Reconstruction algorithm (top panel). Unknown light curves not used for training have larger anomaly scores when using the Bayesian method than with the Neural Network. Figure from~\cite{Muthukrishna2021}}
    \label{fig:autor_transient}
\end{figure}

~\cite{Muthukrishna2021} explored a different approach. They trained instead an Autoregressive Generative Model to generate three known types of light curves (SNIa, SNII,
SNIb). They try then to use the trained models to reconstruct other light curves and use as anomaly score the $\chi^2$ difference between the input light curve and the reconstructed one. The underlying idea is that \textit{common} light curves will be well reconstructed by the Neural Network if they have properly learned $p(X)$ - the probability density function of the data distribution -  while rare events will have larger reconstruction errors. Interestingly, they find that Autoregressive models used that way are not very efficient to identify outliers as compared for example to a Bayesian reconstruction of light curves. The explanation put forward is that neural networks are \textit{too} efficient and are therefore able to generate, with descent accuracy, even light curves which were not part of the original dataset (\autoref{fig:autor_transient}). This behavior of AutoRegressive models has also been reported in the ML community~\citep{Ren2019}. These models are indeed able to easily reconstruct less structured data than the data used for training, leading to small out-of-distribution probabilities. Some solutions have been suggested, which will be discussed in the following subsection. 

Although these works are very recent and the community is still at an exploration phase, they confirm that outlier detection is in general a very complex task. Deep learning offers interesting options, especially because of the potential it has to accurately model the probability distribution of the data. However, there is still no satisfactory solution available and all require some level of human interaction to identify the most interesting objects. This is eventually a structural issue, since the definition of \textit{interesting} depends on the scientific case. Following these conclusions, the group of M. Lochner and collaborators have created dedicated tools to apply different outlier detection methods and enable a human inspection of potentially interesting candidates (see~\citealp{Lochner2021}). One interesting feature this approach puts forward is the fact that another layer of ML is added to adapt the definition of \textit{interesting objects} to each user.  

\subsubsection{Imaging and spectroscopic outliers}

Anomalies can also be found in \textit{static} data, i.e. spectra or images of galaxies. This is again specially relevant in the context of future large imaging and spectroscopic surveys in which a manual inspection of all data points is prohibitively time consuming. Therefore the community has also explored several approaches to identify outliers in large imaging surveys. The underlying idea is analogous to what has been described for time series, i.e. identifying objects which present some deviations from the general properties of the sample. Here again, there exist a large number of ML algorithms that can be employed; many of them not based on deep learning. For example, the recent work by~\cite{Shamir2021} uses a set of manually engineered features to find outlier candidates in HST images. Dimensionality reduction techniques such as Self Organizing Maps have also been explored as a way to identify anomalous spectra~\cite{Fustes2013}. ~\cite{Baron2017} used unsupervised Random Forests to isolate the rarest spectra in the SDSS by using individual fluxes as input. As done for the previous sections, we will focus here on deep learning applications.      

\begin{figure}
\centering
    \includegraphics[width=\linewidth]{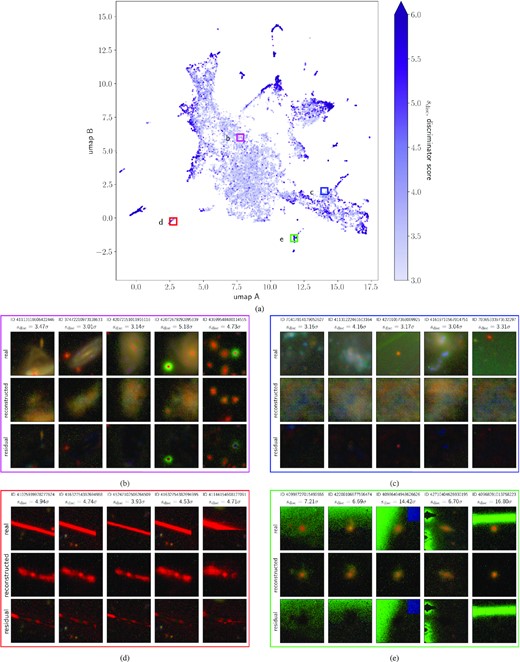}

    \caption{Example of anomalous objects identified with a combination of WGAN and a CAE applied to the HSC survey. The top panel shows the latent space from the image residuals of the WGAN reconstruction obtained with the CAE. The images below show examples of different regions of the parameter space. Figure from~\cite{StoreyFisher2021}}
    \label{fig:gan_anom}
\end{figure}

\cite{StoreyFisher2021} trained a Generative Adversarial Network using postage stamps of observed galaxies from the Hyper Suprime Cam (HSC) survey. They selected galaxies above an apparent magnitude limit and trained a Wassertsein Generative Adversarial Network (WGAN) to generate realistic images of galaxies. The underlying idea is that the model will learn how to accurately reproduce common galaxies but will fail when confronted to objects which appear with a small frequency in the training set. Once trained, WGANs do not provide an explicit latent space to sample. In order to associate anomaly scores to all galaxies, the authors perform an iterative search to identify the closest object that the WGAN can generate. They compute then an anomaly score based on a combination of the quadratic difference between the real and reconstructed image and an additional L2 difference of the features of the last layer of the critic network of the WGAN. They show that the framework is able to identify potentially interesting objects. However, a significant fraction of them are only image artifacts. The authors propose to add another dimensionality reduction layer with a CAE trained on the residual images (difference between the WGAN reconstruction and the original image). They show that, after this additional step, the different types of anomalies cluster together and a visual inspection is proposed to identify the most interesting anomalies (\autoref{fig:gan_anom}). Interestingly, the work also compares the anomalies obtained with a less complex approach based on a CAE to reduce the dimension of the data. The WGAN is able to find more subtle anomalies because of the improved quality of the reconstruction. However, ~\cite{Tanaka2021} showed on the same dataset, that a Convolutional Autoencoder is also able to identify interesting anomalies. They quantify the performance of the anomaly detection algorithm on a set of known extreme emission lines galaxies and quasars. They report that $\sim60\%$ of the objects belonging to these under represented classes are identified.

\begin{figure}
\centering
    \includegraphics[width=\linewidth]{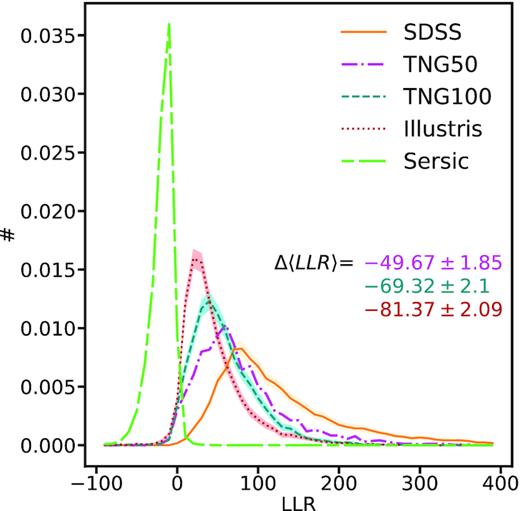}

    \caption{Distribution of likelihood ratios obtained with two pixelCNN networks of observations (SDSS) and different models as labeled. The closer the histograms from simulations are to the one from SDSS, the more realistic the simulation is. The unsupervised model is able to capture the improvement of simulations with time.  Figure from~\cite{Zanisi2021}}
    \label{fig:zanisi}
\end{figure}

~\cite{MargalefBentabol2020} use the same WGAN approach outlined in~\cite{StoreyFisher2021} but in a slightly different context. In this work, the anomaly detection setting is used to assess the realism of galaxies produced by cosmological simulations. In that context the WGAN is trained with mock images from simulations. The trained model is then confronted with real observations from the HST. The authors compare then the anomaly scores from both datasets and conclude that the neural network struggles to reconstruct some of the observed galaxies, meaning that they do not exist among the simulated galaxies.~\cite{Zanisi2021} also explores whether anomaly detection approaches with deep learning can be used to compare galaxy images from cosmological simulations to observations. They use however a different approach based on the Autoregressive Model pixelCNN. Similar to GANs, pixelCNN is a generative model which can be use to learn the probability density function of data and generate new samples. However it provides an explicit expression of the likelihood function built in an Autoregressive fashion; i.e. the values of a given pixel are determined based on the values of the previous ones. They apply the method to the comparison of SDSS and TNG galaxies. Similarly to what was reported by~\cite{Muthukrishna2021} for time series, they find that the model can easily learn to generate simple objects and therefore very smooth galaxy profiles or even pure noisy images achieve high likelihoods of observation under the regressive model. To correct for this effect, the authors use instead the likelihood ratio presented in~\cite{Ren2019} which forces the metric to become sensitive to the fine grained structure of galaxies. They can then show how the likelihood ratio metric is able to measure the improvement in the realism of cosmological simulations from the first Illustris model to the updated TNG one (\autoref{fig:zanisi}). 

An additional way to identify outliers is through the representation space learned by contrastive learning, as one would do with a latent space from an Autoeconder.\cite{Stein2021} explored this approach on images from the DESI survey and demonstrated the efficiency of self-supervised representations to identify outliers and perform similarity searches. See also the work by~\cite{Walmsley2021} which uses the tools developed by~\cite{Lochner2021} for similarity search and anomaly detection on the representation spaces.

\subsection{Discovery of Physical Laws}
\begin{figure}
\centering
    \includegraphics[width=\linewidth]{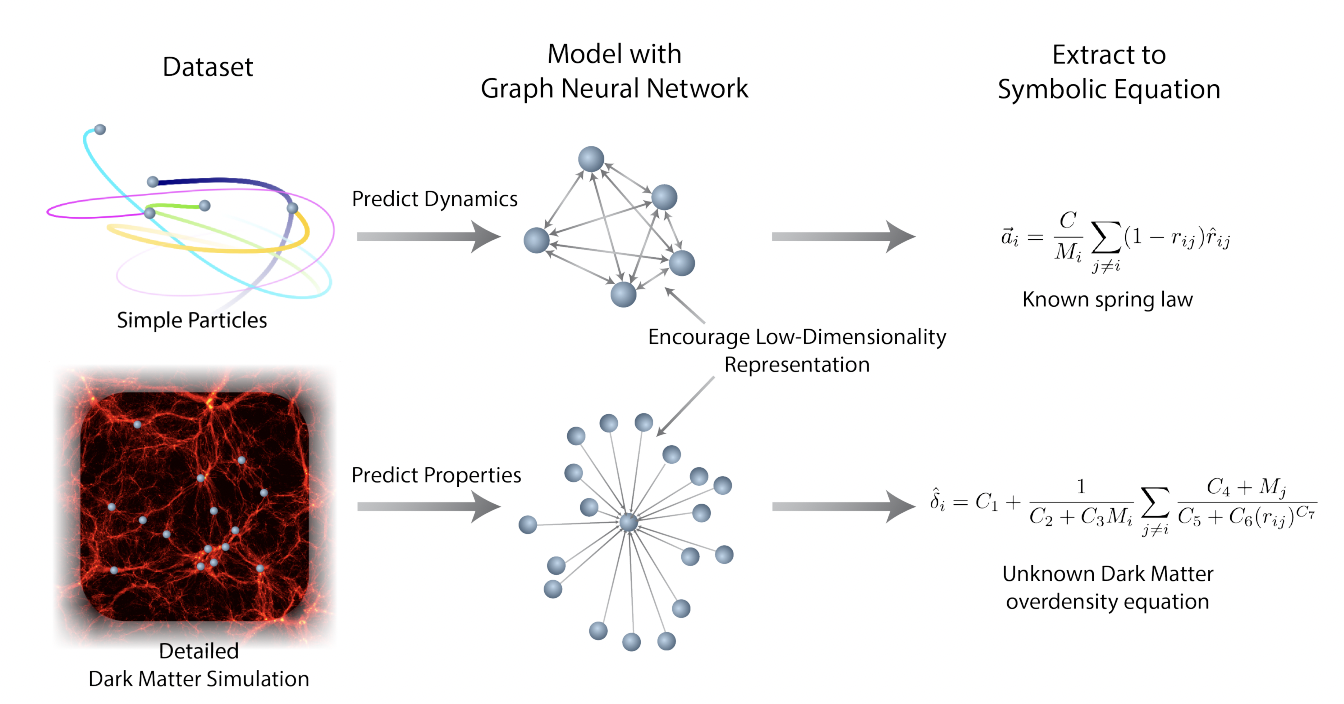}
    \caption{Cartoon illustrating the method to extract physical equation from a deep learning model applied to different datasets.  Figure from~\cite{Cranmer2020}}
    \label{fig:cranmer_laws}
\end{figure}

Arguably one of the final goals of science is to find universal physical laws which can explain a broad set of observations. As said several times in the previous sections, deep neural networks offer an excellent predictive power but their interpretability is low as compared to model-driven approaches. Symbolic regression is the general term employed for the ensemble of techniques that aim at uncovering an analytical equation from data. They can be seen as a generalization of polynomial regression to the space of all possible mathematical formulas that best predict the output variable taking as input the input variables. See~\cite{Schmidt2009} for more details. One way to enable discovery with deep learning is therefore to apply symbolic regression techniques to the trained deep neural network model. This is usually very challenging given that neural network models are usually parametrized by a large number of parameters. There is one work in astrophysics attempting this by~\cite{Cranmer2020}. The authors train a GNN in a supervised manner to predict the properties of some dataset encouraging a sparse representation by the neural network. They then apply symbolic regression to the trained model (\autoref{fig:cranmer_laws}). The authors show that they are able to recover for example some known Newtonian laws by predicting the movement of particles. More interestingly, they discover a new analytic formula which can predict the concentration of dark matter from the mass distribution of nearby cosmic structures. The formula is learned by applying symbolic regression to a GNN which learned the properties of a Dark Matter only simulation. In a follow up work,~\cite{Lemos2022} apply a similar approach to study orbital mechanics. 

\vspace{.3cm}
\begin{mdframed}[backgroundcolor=orange!30] 
{\bf Summary of Deep learning for discovery}\\
\begin{itemize}
    \item Deep learning has been explored as a discovery tool. The main motivation is that future big data surveys are too large and too complex for efficient human-based exploration. Deep learning is therefore mainly used for visualization and anomaly - outlier - detection. The transient astronomy community has been a particularly active field on this front in preparation for LSST.
    \item As opposed to previous applications, these applications rely on unsupervised deep learning. There is a variety of different approaches which have been tested: Autoencoders, Generative Models - GANs, VAEs, Autoregressive Flows. Self-supervised approaches using contrastive learning have also started to be used for this task.
    \item For visualization, the usual approach is to use deep learning to obtain a low dimensional representation of the data which can be explored more easily.
    \item Anomaly detection implies learning a probabilistic description of the data and identifying objects with low likelihoods, i.e. which can hardly reproduced by the trained models.
    \item A common result is that deep learning techniques correctly identify some anomalies, however the quantification of performance is challenging because by definition the problem is ill-posed. Some works find that complex deep learning networks might not be the most efficient way of detecting outliers because they are flexible enough to properly reproduce \emph{simpler} data than the data used for training. 
    \item In addition, filtering out interesting anomalies from artifacts remains an unsolved issue. Current solutions consists in providing anomalous candidates for further inspection.  
     \item The issue of discovering physical laws from  deep learning models has been just recently explored by applying symbolic regression methods on the trained models. It is difficult to generalize at this stage given the large amount of parameters of current deep learning models and the limited interpretability.

\end{itemize}
\end{mdframed}

\section{Deep learning for cosmology}
\label{sec:acc}

In addition to galaxy formation, a key goal of modern deep surveys is to constrain cosmology. Deep learning is playing an increasingly large and promising role in at least two fronts: accelerating simulations - which are needed for efficient cosmological inference - and direct inference of cosmological parameters. This section is focused on these applications. We first review approaches aimed at producing simulations and then we move to the inference of cosmological parameters.  . 

\subsection{Accelerating simulations}
Numerical simulations play an important role in our understanding of galaxy surveys, from shedding light into the physics of galaxy formation and evolution to the modeling of the large scale structure of the Universe and its connection to galaxies. Fast and efficient simulations are needed to interpret the data. However, a major bottleneck is computational type. Ideally one would like to have high resolution and large volume simulations including both N-body and hydrodynamics. However, this is usually prohibitively time consuming and for that reason fast emulators are desirable. In the past years, deep learning has appeared as a promising solution, In this first subsection, we review how it has started to impact the field of cosmological simulations.

\subsubsection{Learning N-body Simulations}

We begin by reviewing different strategies to either partially or entirely learn the physics of N-body simulations and to some extent complement a physical model to provide a significant acceleration compared to a full simulation. We will split these methods into two categories, depending on whether they act on and model matter density fields, or Lagrangian particle displacements.

\paragraph{Lagrangian displacement models}

N-body simulations typically model the matter distribution using tracer particles and evolving in time the position and velocity of these particles under the effect of gravitational forces. In this Lagrangian approach, the full output of a simulation can be seen as the displacement that each particle has undergone from its initial position on a regular lattice, along with its final velocity. The methods described here all aim at modeling this displacement field and therefore are not acting on 3D density fields, but on this displacement sampled at the initial particle positions on a regular grid.

{\bf Modeling residual displacements against fast simulation:} In \cite{Dai2018}, the authors propose a physically motivated post-processing technique, dubbed Potential Gradient Descent (PGD), able to recover the small-scales of fast Particle-Mesh (PM) simulations, and mimic the output of high-resolution N-body simulations, or even mimic the baryonic feedback from hydrodynamical simulations. The advantage of these fast PM simulations (such as FastPM \citep{Feng2016} or COLA \citep{Tassev2013}) is that they can be run inexpensively on very large comoving volumes, but their lack of force resolution and their coarse time stepping limit their resolution, typically leading to a lack of power on small scales and inaccurate halo profiles. The method proposed in \cite{Dai2018} is to learn an additional displacement of the particles, moving them deeper into their local gravitational potential, which has the effect of sharpening the halo profiles. To compute this displacement, instead of using Convolutional Neural Networks, the authors use a physically motivated parameterization in Fourier space, defined by an overall amplitude and a band-pass filter applied to the gravitational potential for a total of only 3 parameters, which respects the translational and rotational invariance of the problem. Training of the parameters of this model is done by either minimizing the Mean Square Error (MSE) on the power spectrum or on the density field between a reference simulation and a fast simulation ran from the same initial conditions. With this simple, yet powerful, scheme the authors can emulate to within 5\% accuracy the Illustris-3 simulation from only a 10 step FastPM simulation.

\begin{figure}
 	\includegraphics[width=\columnwidth]{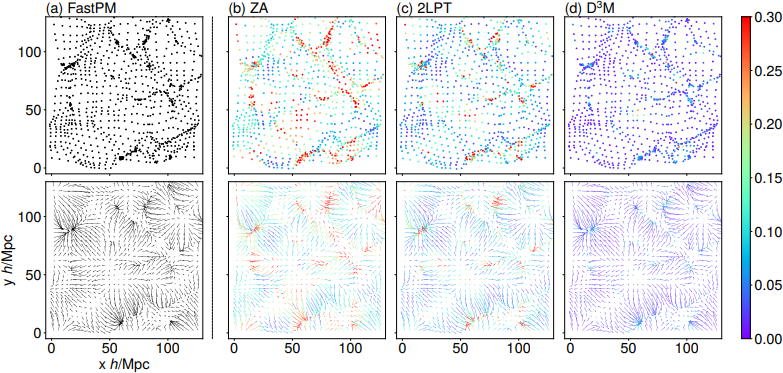}
	\caption{Illustration of learned displacement field in an N-body simulation from \cite{He2018}. The first column is the reference simulation (FastPM), the second column shows a simple linear displacement (ZA), the third column shows a second order Lagrangian Perturbation Theory displacement (2LPT) and the last column shows the output of the 3D U-Net (D$^{3}$M). The top row shows final particle positions, the bottom row shows the final displacement field. The color scale shows the error in position or displacement vectors between each approach and the reference FastPM simulation.}
	\label{fig:He2018}
\end{figure}

In one of the first works applying deep learning to N-body simulations, \cite{He2018} proposed a model based on a 3D convolutional U-Net (see \autoref{sec:segmentation}) that learned to predict full non-linear particle displacements given as an input analytic Zel'dovich Approximation (ZA) displacements (which corresponds to a single step of a FastPM algorithm). The 3D CNN takes as inputs a 3-channel 3D field providing this ZA displacement field sampled at the initial particle positions, and outputs the final displacement vector, still as a 3-channels 3D field. The model is then trained by Mean Squared Error loss on the reference displacement field provided by a FastPM simulation. An important realization from that work was that a 3D CNN was able to accurately model this displacement field. \autoref{fig:He2018} illustrates the approximation error of this method (right column) compared to other fast approximations for the displacement field. It is found to be significantly more accurate than analytic solutions. 

With a similar model, \cite{Giusarma2019} showed that it was possible to learn the residual displacements between  a $\Lambda$CDM N-body simulation and a simulation with massive neutrinos. They used a modified version of D$^{3}$M to learn this residual displacement at $z=0$ between the two sets of simulations, and found excellent results down to $k < 0.7 h$Mpc$^{-1}$.

\begin{figure}
 	\includegraphics[width=\columnwidth]{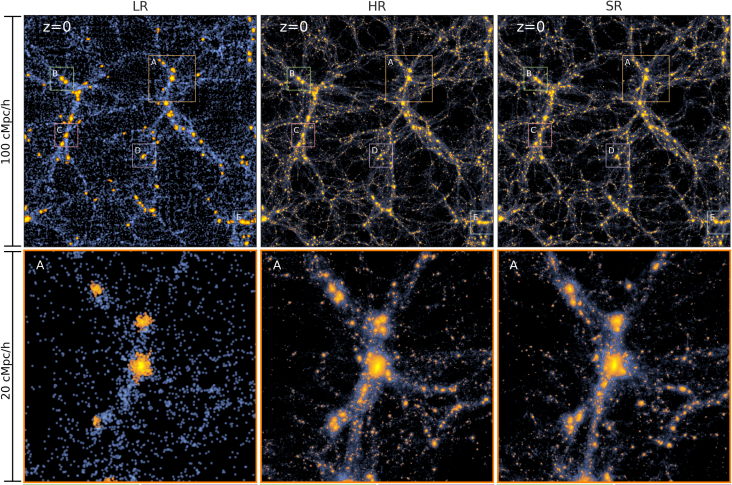}
	\caption{Illustration of N-body simulation super-resolution from \cite{Li2020a} showing from left to right, the Low Resolution (LR) input, High Resolution (HR) target, and Super Resolution output of the model. The bottom row is a zoom-in on the region marked A.}
	\label{fig:gan_sr_li}
\end{figure}

{\bf Upsampling the displacement field from a low resolution simulation:} Recognizing that upsampling the displacement field is equivalent to increasing the number of particles in a simulation, \cite{Li2020a} proposed a Super-Resolution technique based on a conditional GAN directly inspired from a StyleGAN2 \citep{Karras2019} architecture. The generator takes as an input a low resolution displacement field, and outputs an upsampled high-resolution displacement field. This approach achieves impressive results up to an upsampling factor of $\times8$ translating into a direct computational speedup of a factor $\times 1000$ in a setting where the goal would be to produce a 100$h^{-1}$Mpc simulation with $512^3$ particles. A visual illustration of this model is shown on \autoref{fig:gan_sr_li} where the rightmost panel is the output of the model.
As a further extension of this approach \cite{Ni2021} trained a similar Lagrangian conditional GAN to model not only displacements but also velocities, yielding a full phase-space information that a real N-body simulation would produce. They further tested this model to demonstrate accurate matter power spectrum recovery up to within 5\% up to k $\geq$ 10 $h^{-1}$ Mpc for an upsampling factor of $\times8$ of a 100 $h^{-1}$ Mpc box, and validated the recovery of halo and sub-halo abundance.

\paragraph{Density field models}

The first method for simulation super-resolution, proposed by \cite{Ramanah2020}, relied on a 3D Convolutional WGAN with a generator taking as inputs the density field from a low resolution simulation run and a high resolution set of initial conditions, and tasked with outputting a high resolution final density field. The 3D convolutional discriminator compared the high resolution density fields from the full simulation to the generator output. With this approach, the authors were able to upsample by a factor of 2 the resolution of the final density field, while reproducing faithfully a number of field properties including the power spectrum, the density contrast probability density function, and the bispectrum. This upsampling ratio represented a computational speedup of about $\times 11$ for a 1 $h^{-1}$Mpc and $512^3$ particles simulation. More recently, \cite{Schaurecker2021} proposed a similar model acting directly at the level of the density field, but only using the low-resolution final density field as an input (without needing the high-resolution initial conditions of  \cite{Ramanah2020}).

\subsubsection{N-body emulation by Deep Generative Modeling}

All of the models from the previous section had the particularity of trying to model the residuals compared to a physical model instead of completely supplanting the physical simulation. In this section, we now cover the works that have taken the approach of trying to learn from scratch the entire simulation using a Deep Generative Model. 

A number of works started to apply to the emulation of cosmological fields the new Deep Generative Models that were gaining in popularity at the time, and especially GANs. In the first instance of such an application, \cite{Mustafa2019} applied a conventional DCGAN to the modeling of weak-lensing convergence maps and demonstrated that the model was able to accurately reproduce a number of statistics of these fields, including their power spectra and Minkowski functionals. Shortly after, \cite{Rodriguez2018} presented a similar application of using a DCGAN to model slices of N-body simulations with results demonstrating that these models were able to capture most of the relevant statistics of the cosmic web.

\begin{figure}
	\includegraphics[width=\columnwidth]{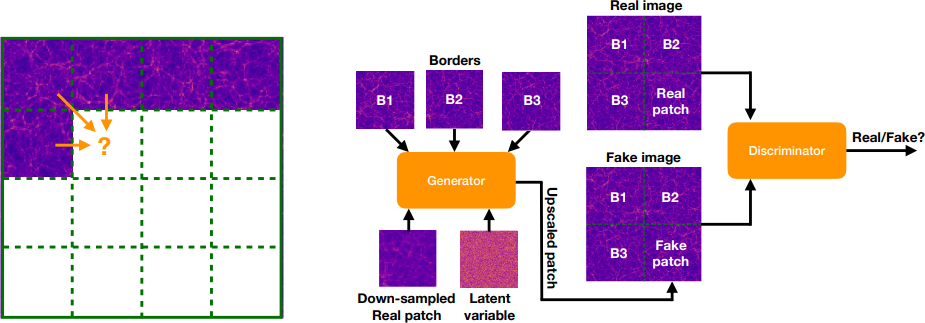}
	\caption{Sequential generation and upsampling strategy of the scalable GAN for N-body modeling presented in \cite{Perraudin2019} scaling up to $256^3$. The left illustration shows how the sequential generation of a large volume would proceed. The right plot illustrates the proposed architecture where the generator is conditioned on both neighboring patches, and on the lower resolution patch, which at sampling time would be generated by a separate GAN trained on coarser volumes. Distribute under the Creative Commons CC BY licence (\url{http://creativecommons.org/licenses/by/4.0/}).}
	\label{fig:perraudin2019}
\end{figure}

These early works on generative modeling for cosmological fields were quickly confronted to the difficulty of building high quality models for very large or even 3D fields. \cite{Perraudin2019} explored more in depth the limitations of simple DCGANs for generating large 3D N-body volumes and highlighted two key strategies enabling high quality results in this setting: 1. generating the field by patches, 2. using a multi-resolution approach based on a Laplacian pyramid. To generate N-body meshes of size $256^3$, their proposed model uses 3 independent GANs that are trained on 3 increasingly high data resolution ($32^3$, $64^3$, $256^3$). The first model trained on the coarsest resolution is a conventional DCGAN while the other models are conditioned on lower resolution inputs. In addition, the GANs are made conditional on the neighboring patches in the simulation, which allows at inference time to generate a large volume patch-by-patch in a sequential fashion, thus avoiding the need of storing the entire volume in GPU memory. \autoref{fig:perraudin2019} illustrates the proposed strategy, which is shown in the paper to be able to recover both the power spectrum and peak counts to satisfying levels whereas a non-multiscale approach fails significantly. 

Additional works have investigated other possible improvements to GANs for N-body simulations. In particular \cite{Feder2020} proposed modeling the latent space prior of a DCGAN with a heavy-tailed distribution instead of a Gaussian. In that work, using a Student's t-distribution is shown to improve the model's ability to capture the sampling variance of the peaks in the dark matter distribution, and overall improves power spectrum recovery on all scales.

These generative models of the matter distribution start to become useful when they are made conditional on some external parameters, for instance cosmological parameters or redshifts. The GAN model proposed in \cite{Feder2020} was for instance made conditional on redshift by simple concatenation of the conditional variable to the latent vector of the GAN, allowing the authors to generate volumes at intermediate redshifts, which would be useful to create lightcones. \citep{Perraudin2020} proposed a conditional GANs to produce 2D weak lensing mass-maps conditioned on $(\sigma_8, \Omega_m)$ through a remapping of the latent vector of the GAN by a function that rescales the norm of that vector based on the conditional variable. More recently, \citep{WingHeiYiu2021} extended that work to the sphere, using a DeepSphere \citep{Perraudin2019a} graph convolutional architecture, to emulate the KiDS-1000 survey footprint as a function of $(\sigma_8, \Omega_m)$.

While conditional GANs could be useful as emulators, they however cannot be directly used for cosmological inference due to the fact that GANs do not possess tractable likelihoods. One significantly different approach to generative modeling of the dark matter distribution proposed in \cite{Dai2022} relies on a Normalizing Flow approach instead of a GAN to model explicitly the conditional distribution $p(x | \theta)$ where $x$ is the dark matter distribution, and $\theta$ are cosmological parameters. Once trained, such a model can directly be used as the likelihood function of the high-dimensional data in a Markov-Chain Monte-Carlo. In this paper, the authors introduce a Translation and Rotation Equivariant Normalizing Flow (TRENF) model.  It builds an n-d normalizing flow based on learning filters and performs convolutions in  Fourier space which impose by construction translation and rotation equivariance. The authors demonstrate that this approach accurately captures the high-dimensional likelihood of dark matter density fields and that it can be used not only for generating these fields, but also for inferring cosmological parameters.

\subsubsection{Finding Dark Matter Halos}

In the pipeline needed to go from N-body dark matter simulations to observable galaxy distributions, a typically essential step is the identification of dark matter halos, which can then be populated with galaxies under a variety of techniques (e.g. HOD or SHAM). In this section, we review the various approaches which have been proposed to go from the dark matter density field to dark matter halos.

\begin{figure}
    \centering
    \includegraphics[width=\columnwidth]{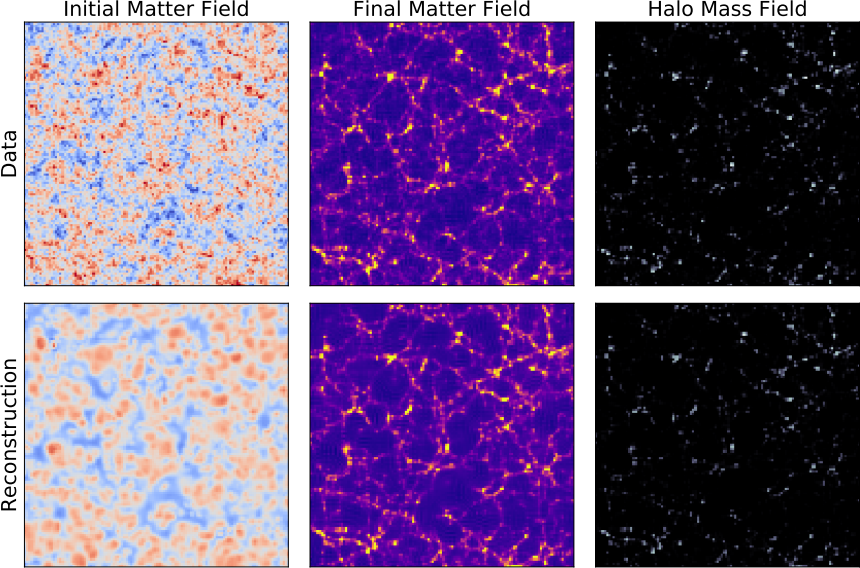}
    \caption{Illustration of an application of differentiable neural mapping between the dark matter density and dark matter halo fields from \cite{Modi2018}. The top row shows the initial conditions dark matter field, final dark matter field (at z=0), and the dark matter halo field obtained by a FoF halo finder. The bottom row shows the result of a reconstruction by gradient descent of these initial conditions, using a neural network to provide the mapping between the final density field and the halo field.}
    \label{fig:Modi2018}
\end{figure}

One of the first approaches to learn this connection was proposed in \cite{Modi2018}, and assumed a shallow neural network mapping between the local 3D dark matter density and "halo mask" and "halo mass" fields. The binary halo mask field was essentially used to model whether a given voxel actually contained a halo, while the halo mass field was predicting in each voxel a likely total halo mass. Given this model, a halo field could be recovered by multiplying these two outputs of the neural network. The actual neural network was based on a simple MLP taking as inputs a 3x3x3 voxel region of the dark matter density field itself, the field smoothed on a given scale, and the difference between the field smoothed on two difference scales. The authors find that the predicted halo-mass field exhibits over a 95\% correlation with the true field up to $k = 0.7 h$Mpc$^{-1}$. Perhaps most interestingly, this model provided effectively a differentiable mapping between dark matter and halos, a differentiable halo finder, and the authors demonstrated that this mapping could be used in a reconstruction scheme to infer initial conditions from a halo field by gradient descent through the neural network and a differentiable N-body simulation as illustrated on \autoref{fig:Modi2018}.

\cite{Charnock2019} proposed a different approach to the same problem, building a probabilistic model based on a Mixture Density Network for the halo mass distribution in each voxel conditioned on the underlying dark matter density. The overall goal of the authors in that paper was to make the model minimalistically parametric  while respecting the important physical properties of the problem, namely that the halo bias should be globally rotational and translational invariant. To reach this goal, instead of relying on standard CNNs, the authors proposed a model they refer to as the Neural Physical Engine (NPE), which applies a set of convolutional kernels parametrized to exhibit by design the desired symmetries. This leads to a reduced number of parameters compared to a similar 3D CNN. They apply this NPE on the dark matter density field, and use its output to condition a simple Gaussian MDN tasked with representing the local halo mass distribution. They demonstrate that a minimal model with only 17 parameters is able to accurately capture the halo mass distribution and its dependence on local environment, and further present an application where the model is used as part of a Bayesian reconstruction of initial conditions in a simulation from a given halo distribution.

Going in a deeper direction, a number of papers have looked into identifying dark matter halos, not from  Eulerian space, but directly as Lagrangian patches at the level of the input density field of the simulation using a deep CNN model. \cite{Berger2018} introduced the first deep CNN model to identify dark matter halos with this approach, relying on V-net model illustrated in \cite{Berger2018}. They used as a training set simulated initial density fields of size 128$^3$ with a 1 $Mpc$ voxel resolution, along with a binary $128^3$ segmented map indicating the position of Lagrangian patches identified as proto-halos by the Peak Patch semi-analytic code \cite{Stein2019}. The network is then trained with a binary classification loss to predict the presence or not of a halo in a given voxel. To build an actual halo catalog given a segmented volume outputted by the trained model, the authors then use a hierarchical Lagrangian halo-finding procedure that ultimately returns a list of halos. This entire procedure is found to lead to a halo-mass function and power spectrum within 10\% of the ground truth simulation. 
As a variation of this approach, \cite{Bernardini2019} proposed to replace a binary segmentation by a regression problem, where the target value corresponds to a normalized distance to the center of the halo, leading to similar performance but with a smaller model. 

In related work, \cite{luciesmith2020, etezadrazavi2021} propose to use a 3D CNN to predict from a initial conditions centered on the location of halos, the mass of the collapsed halos at $z=0$. They use these models to investigate the relative importance of various properties of the initial conditions for halo formation. \cite{luciesmith2020} reports for instance that removing anisotropies in the initial conditions does not significantly affect the masses predicted by the model, hinting that initial shears may not be a significant factor in the halo formation process. \cite{etezadrazavi2021} reports that velocity information becomes more important to accurately predict masses at lower values of $A_s$.

\subsubsection{Painting baryons on N-body simulations}

As a way to bridge the gap between full hydro-dynamical simulations and cheaper Dark Matter Only (DMO) simulations, a number of works have investigated the possibility of "painting" baryons on top of DMO simulations.

A first category of papers proposed to model this mapping between dark matter density and baryonic fields in a probabilistic fashion to account for the inherent uncertainty. In the first work to attempt such modeling, \cite{troester2019} investigated the use of both conditional VAEs and conditional GANs to learn from 2D dark matter density slices a probabilistic map to 2D thermal Sunyaev-Zeldovich (tSZ) maps, which capture the electron pressure. They found excellent agreement with ground truth simulations, at different redshifts, indicating that this approach was very promising, but did report a tradeoff of GANs leading to more accurate results but being harder to train and less stable than VAEs. 

More recently, \cite{bernardini2021} explored the use of a conditional WGANs to learn a similar mapping to predict gas and H$_I$ density on 2D maps. To achieve the generation of high-resolution and large maps, the authors adopt a multi-resolution strategy in which a U-Net generator outputs maps at 3 different resolutions that the critic compares to similarly downsampled versions of the target fields. This strategy allows them to successfully train the model on images of size $512^2$, but after training this purely convolutional model can be applied on much larger fields. The authors report accurate H$_I$ power spectra to within 10\% accuracy up to the ~10 kpc scales while being able to map simulation boxes of 100 $h^{-1}$Mpc on the side.

Extending these conditional generative approach to 3D fields, \cite{horowitz2021} proposed a conditional VAE with a customized U-Net architecture. Contrary to a standard conditional VAE, this model features skip-connections between the conditional branch (the dark matter encoder) and the generative branch (the hydrodynamics decoder) which create a U-Net structure. It remains however probabilistic thanks to the variational bottleneck block which contains stochastic latent variables capable of capturing the aleatoric nature of the mapping. The resulting model inherits from the stability and robustness of VAEs but also benefits from this U-Net structure to enable high resolution mapping. Again, the model is kept strictly convolutional to make it insensitive to the size of the input field, so that it can be applied on larger volume than it is trained on. This model can also be made conditional on redshift and the authors demonstrate the possibility of generating lightcones with this approach. 

A second class of papers propose similar but deterministic mappings, using 3D U-Nets trained to regress particular baryonic fields. \cite{thiele2020} proposed a 3D U-Net to learn a similar mapping, although no longer probabilistic, between the 3D dark matter field and electron density, momentum, and pressure. This work reports two significant challenges in learning this mapping in 3D, one being the sparsity of interesting voxels (as in 3D most of the voxels are in empty regions), and the other being the high dynamic range of the fields to model. They address these challenges by biasing the loss function towards high density regions, and applying range-compression schemes. Overall they report better agreement with the reference simulations compared to semi-analytical models. In similar work, \cite{Wadekar2020} trained a U-Net to output H$_I$ density maps and again reported better quality results than a standard HOD approach, while \cite{Harrington2021} used a U-Net to predict hydrodynamical fields (density, temperature, velocity) subsequently used to model Ly$\alpha$ fluxes with a physical prescription. \cite{Zhang2019} proposed a two-step approach to map the dark matter field to a 3D galaxy distribution, where a deep 3D CNN would predict a mask of the likely non-empty voxel, and a second CNN would regress the number of galaxies in these regions. They find that their CNN model is able to predict a galaxy distribution recovering the expected power spectrum to within the $10 \%$ level up to k= $10 h$Mpc$^{-1}$ scale.

Finally, a singular approach was proposed in \cite{Dai2020} which extended the PGD method of \cite{Dai2018} - mentioned in the previous section - to parametrize a mapping between a dark matter only simulation and various fields accessible in hydrodynamical simulations using a combination of particle displacements and voxel-wise non linearities. This method, dubbed Lagrangian Deep Learning, reused the same Fourier-based parametrisation to displace particles from a dark matter only simulation (e.g. FastPM), thus using a very small number of parameters  (order 10) and providing translation and rotation equivariance, before painting them on a 3D mesh and applying a non-linear transform. The few parameters can be fitted by gradient descent on a single pair of dark matter only and hydrodynamical simulations. This scheme was successfully demonstrated to reproduce a range of maps from IllustrisTNG, including stellar mass distribution and electron momentum and pressure.


%
%
%

\subsection{Deep learning for cosmological inference}
\label{sec:models}

 In this section we will review in particular how emulators and Likelihood-Free Inference techniques are enabling the inference not only of cosmological parameters but also of cosmological fields.

\subsubsection{Field-level Cosmological Constraints}
%
%
%

\paragraph{Weak Gravitational Lensing}

It was realized early on that given access to simulations to act as a training set, neural networks can be used to extract cosmological information from high-dimensional data such as maps of weak gravitational lensing maps. The first example, presented in \cite{Schmelzle2017}, demonstrated that a CNN-based classification model could discriminate between different discrete cosmological models, especially along the $\sigma_8 -\Omega_m$ degeneracy that conventional 2pt correlation functions are unable to resolve. Similar results in \cite{Peel2019, Merten2019} showed that a CNN classifier was able to distinguish between $\Lambda CDM$, modified gravity and massive neutrinos models from weak lensing maps, with better discriminating power than more conventional higher-order statistics such as peak statistics. These results sparked a lot of interest into the potential use of CNNs to extract cosmological information from weak lensing surveys, which resulted in a number of subsequent publications with the ultimate goal of yielding proper Bayesian posteriors on cosmological parameters. 

Going beyond a classification task between discrete models, a second class of papers \citep{Gupta2018, Fluri2018, Ribli2019a} built CNN regression models where the network is tasked with directly predicting $(\sigma_8, \Omega_m)$, either using a Maximum Absolute Error (MAE) loss \citep{Gupta2018, Ribli2019a}, or Gaussian-parameterized negative log likelihood loss \citep{Fluri2018}. It is important to note, as reported in all these papers, that the output of the network trained for regression will not be an unbiased estimator for the cosmological parameters, but should be interpreted as a low-dimensional summary statistic, which can then be used for inference in a second step, independent from the network training. To retrieve cosmological parameters, these papers assume a Gaussian likelihood on the output of the network and characterize the mean and covariance of that likelihood on a set of simulation, similarly to what is conventionally done for peak count statistics or other Higher-Order Statistics without an analytic likelihood. With this approach, all these papers reported the ability to extract more information than a 2pt function analysis, even on realistically noisy data, with \cite{Ribli2019a} reporting a factor of $\sim2$ smaller contours on $(\sigma_8, \Omega_m)$ in a Euclid or LSST setting.

Building on these promising results, the next phase of papers deployed these approaches to actual survey data. \cite{Fluri2019} followed a similar strategy to \cite{Fluri2018} and trained a ResNet model, on a suite of tomographic lensing simulations mimicking the KiDS-450 survey and spanning a range of $(\sigma_8, \Omega_m, A_{IA})$ values, where $A_{IA}$ is the amplitude of the intrinsic galaxy alignment signal. This study found constraints broadly consistent with the fiducial 2pt function analysis of KiDS-450 \citep{Hildebrandt2017} but when compared with an internal power-spectrum analysis yielded a 30\% tighter posterior. In the most recent extension of that work, \cite{Fluri2022} performed a $w$CDM analysis of the KiDS-1000 weak lensing maps including a large number of refinements. In particular, they used a spherical CNN architecture, DeepSphere \citep{Perraudin2019a}, in order to process spherical fields and extended their simulation suites to include a baryonic prescription, and left the dark energy equation of state parameter $w_0$ free to vary along with 5 other cosmological parameters. They find again broad agreement with KiDS-1000 $w$CDM, and internally consistent results with a power spectrum analysis, but with only a meagre $11\%$ improvement on S8 constraints. The most likely reason for the limited constraining power of the CNN analysis comes from the low-resolution of maps used in the analysis (HEALPix nside=1024), but available non-Gaussian information content when baryonic systematics are taken into account is also an open question discussed below.

\cite{Jeffrey2021} performed an analysis of the DES Science Verification (SV) data using a slightly different approach to previous works. Instead of training a CNN for regression, the authors introduced an information loss which explicitly trains the network to compress the input lensing maps into a low dimensional (asymptotically) sufficient statistic, that can further be used for inference with a Likelihood-Free Inference approach. More specifically, they used a Variational Mutual Information lower bound to train the model, which relies on using a Normalizing Flow (NF) to approximate the posterior distribution on cosmological parameter from the low dimensional output of the convolutional compressor network, and training both models jointly as to minimize the negative log likelihood of the NF. Once trained under this information loss, the model can be applied to data, and a robust estimate of the posterior was achieved by Neural Likelihood Estimation using the pyDELFI package \citep{Alsing2019}. Compared to previous papers, this approach has asymptotic optimality guarantees, and does not rely on any Gaussian assumptions for the likelihood of the summary statistic. In this paper, the authors found consistent but tighter constraints with this approach compared to a power spectrum analysis, but the constraints remained very large due to the small size of the SV dataset.

One question remains unclear, however, regarding the amount of additional information deep learning can extract over the power spectrum or simpler higher-order statistics, when systematics like baryonic effects are taken into account. \cite{Lu2022} investigated this question using a simple baryonic correction model \citep[BCM][]{Arico2020a} for dark matter only simulations and trained a deep CNN to infer both cosmology and baryonic parameters from simulated lensing maps under a realistic HSC-like setting. The authors find that using a CNN instead of a power spectrum ($100 < \ell < 12000$) improves the constraining power on $(\Omega_m, \sigma_8)$ (in terms of 1-sigma area) by a factor of a few if the astrophysical parameters are kept fixed. However, the improvement degrades significantly when marginalizing over astrophysical parameters. Indicating that there is some amount of non-Gaussian information left even after marginalizing on baryons, but how sensitive are the resulting constraints to the specific baryonic model assumptions is uncertain. 

\bigskip

With the success of these methods, several papers have attempted to introspect the CNN trained on weak lensing maps, to try to identify or recognize what features of the data are being used to extract the cosmological information. \cite{Ribli2019} proposed using at the first convolutional layer of the model a large 7x7 kernel, and recognized after training that this first layer was learning a kernel close to a Laplace operator and concluded that this operator allowed the network to be sensitive to the steepness of the peaks in convergence maps. Building on that insight, the authors handcrafted a new summary statistic based on histograms of Sobel-filtered lensing-maps, which are sensitive to peak steepness. This simple statistic was found to outperform a deep CNN on noiseless data, while deteriorating in the presence of noise, but still outperforming conventional peak counts.

Using a different methodology, \cite{Matilla2020} performed a similar study using saliency methods to identify the features of the lensing maps relevant to the cosmological inference task. They found that in all cases, the most relevant pixels in the input maps were the ones with extreme values. In noiseless maps, regions with negative convergence accounted for the majority of the attribution, while on realistically noisy maps, the high value convergence regions (positive peaks) account for the majority of the attribution. 


\paragraph{Large Scale Structure} 

Although not as actively researched, cosmological information can also be extracted from the galaxy distribution with deep learning. This was first illustrated by \cite{ravanbakhsh2017} which used a 3D CNN to regress cosmological parameters $(\sigma_8, \Omega_m)$ from the 3D dark matter density in a suite of N-body simulations. Although not directly applicable to actual surveys, this work demonstrated that convolutional approaches where able to retrieve cosmological information from the 3D large-scale structure. Addressing the same problem, but with the computational aspects of training 3D CNNs on large-scale High Performance Computing (HPC) systems in mind, \cite{Mathuriya2018} presented a similar result on cubes of size 128$^3$, and showcasing distributed training on 2048 and up to 8192 CPU nodes on the NERSC Cori machine. 

Going further in that direction, \cite{Ntampaka2019} presented a 3D CNN model acting this time on the 3D distribution of galaxies, with spectroscopic surveys in mind. This work relied on a suite of 40 dark matter simulations, populated with galaxies with a range of various HOD models (15 different models) as a way to marginalize over uncertainties on the galaxy-halo connection. Galaxies in a given comoving volume were painted on 3D slabs of size 550 x 550 x 220 $h^{-1}$Mpc to estimate a galaxy density field. A 3D CNN was then tasked with outputting $(\sigma_8, \Omega_m)$ and the model was trained by Maximum Absolute Error. The authors also proposed a variants on that model, with an MLP branch taking directly as an input the power spectrum of the volume, and combined or not with the CNN branch to provide the cosmological parameter estimates. The main takeaways of that paper were that the CNN was able to extract more information than the power spectrum alone, and that the model trained in this fashion generalized well to unseen HOD models.

\paragraph{Robustness to Baryonic Effects}
\label{sec:baryons}

Because deep learning-based cosmological inference schemes remain opaque, one important question is how to robustify such an analysis to modeling uncertainties and systematics. Just like in modern 2 point function analyses, one of the most prominent questions is how to account for uncertainties in Baryonic physics.

Answering this question is one of the motivations for the Cosmology and Astrophysics with Machine-learning Simulations (CAMELS) suite \citep{VillaescusaNavarro2021b}, a set of about 4,000 simulations of $25 (h^{-1}$Mpc$)^3$ volumes which break down into $2,000$ dark matter only simulations, $1,000$ hydrodynamical simulations following the IllustrisTNG model, and $1,000$ hydrodynamical simulations using the SIMBA model~\citep{Dave2019}. Each of these 2 different hydrodynamical sets of simulations varies not only cosmological parameters $(\Omega_m, \sigma_8)$ but also astrophysical parameters, namely $(A_{SN1}, A_{SN2})$ which regulate supernova feedback and $(A_{AGN1}, A_{AGN2})$ which parameterize AGN feedback. This suite of simulations therefore not only allows the study of the dependency between cosmological parameters and a given set of astrophysical systematics, but also can be used to check the robustness to an assumption of a particular baryonic feedback model (IllustrisTNG or SIMBA). On this dataset, the authors demonstrated that not only does the total matter distribution contain significant cosmological information accessible by deep neural networks, but also baryonic fields such as the $H_{I}$ density \citep{VillaescusaNavarro2021a}, and that even individual galaxies bare some imprints of cosmological parameters \citep{VillaescusaNavarro2022}. The question however is: can this information be retrieved under the uncertainties of the baryonic model?

In \cite{VillaescusaNavarro2020}, the authors illustrated on an analytically tractable toy model that a Neural Network can be trained to optimally marginalize over baryonic effects as long as the training data left the associated parameters free to vary according to a given prior. While promising, this result didn't necessarily imply that this implicit marginalization would be robust to a change in baryonic model. To investigate precisely this question, \cite{VillaescusaNavarro2020} trained a CNN on 2D projected density fields from the CAMELS dataset to regress $(\Omega_m, \sigma_8)$. They showed that models trained on IllustrisTNG lead to almost unbiased results on SIMBA and vice-versa, implying that in the process of learning a summary statistic that marginalizes over baryonic effects, the neural networks are discarding the part of the signal affected by baryons, and are therefore no longer extremely sensitive to the details of the modeling of those effects. This result remains of course limited, but is very encouraging for the analysis of data.

\subsubsection{Dark Matter substructures from Strong Gravitational Lensing}

In previous sections we have described how deep learning has significantly impacted the detection and characterization of strong gravitational lenses. Strong lensing systems can be used as well to constrain the substructure of dark matter on extended arcs which contains a wealth of information about the properties and distribution of dark matter on small scales and, consequently, about the underlying nature of the dark matter particle. The information can therefore be used to distinguish between various dark matter models - warm or cold dark matter for example.

However, probing this effect is challenging since the likelihood function for realistic simulations of population-level parameters is intractable. \cite{Alexander2020} followed a simple approach which consists in converting the inference problem into a classification of CNNs in classification mode to distinguish various types of dark matter models. They show they can reach AUC scores above $90\%$, for images with no substructure, spherical sub-halos, and vortices on idealized simulations.~\cite{Varma2020} also used a CNN for multi-class classification in seven different categories corresponding to different lower mass cut-offs of the subhalo mass function. They report being able to correctly identify the lower mass cut-off within an order of magnitude to better than $\sim90\%$ accuracy.     

\begin{figure}
\centering
    \includegraphics[width=\linewidth]{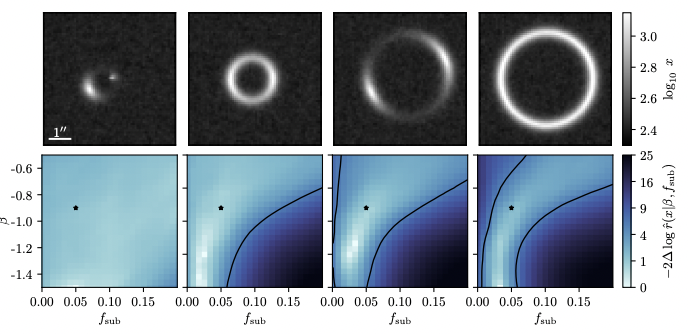}

    \caption{Illustration of lensed systems and the corresponding likelihood ratio maps estimated with simulation based inference and deep learning. The black crosses show the true values. Figure from~\cite{Brehmer2019}}
    \label{fig:dm_sub1}
\end{figure}

Other works have attempted to go a step beyond by estimating the parameters describing the dark matter substructure in a regression mode using simulation based inference with deep learning. The first work exploring this is by~\cite{Brehmer2019}. The authors characterize substructure with a set of parameters and show, in a proof-of-concept application to simulated data, that neural networks can be trained to accurately estimate likelihood ratios associated to the dark matter substructure parameters (\autoref{fig:dm_sub1}). They conclude that $\sim100$ strong lenses might be enough for characterizing the abundance of substructure down to $\sim10\%$.~\cite{Coogan2020} also used a likelihood-free approach to infer posterior distribution of the mass and positions of subhalos in simulated systems.~\cite{Vernardos2020} applied a similar approach combining a simulator based on Gaussian Random Fields for the potential but combined with images of real galaxies for the lensed source and show they can also constrain the substructure parameters.

\subsubsection{Reconstructing cosmological fields}

Applications in cosmology also go beyond cosmological parameter estimation, and an increasingly large number of works explore applications of deep learning for inferring latent cosmological fields from observations.

\paragraph{Weak Lensing Mass-Mapping}

One active research question in weak gravitational lensing is the reconstruction of the matter distribution that gives rise to the measured lensing effect, a task known as mass-mapping. This problem is made particularly difficult by the noisy nature of the observations (intrinsic galaxy ellipticities being much larger than the weak gravitational shear) and the need to invert a linear operator mapping shear to projected mass (also known as convergence) which becomes ill-posed in the presence of survey masks. This is therefore an instance of an ill-posed inverse problem, which does not have any unique solution, in the sense that different mass-maps can lead to a shear signal equally compatible with the data. For these problems, the hope of Deep Learning approaches is that they can learn, implicitly or explicitly, a prior on the signal to recover from training data, and use that prior to solve the inverse problem in an optimal fashion.

The first class of methods, explored in \cite{Shirasaki2019, Shirasaki2021}, used a conditional adversarial network adapted from the pix2pix model \citep{Isola2016} for image-to-image translation. In this approach, a first network with a U-Net structure is tasked with taking a noisy convergence obtained by a rough direct inversion of the shear field as an input, and outputting an estimate of the true convergence map. To train this denoiser, a second network is introduced to act as a discriminator, taking as an input both noisy and denoised convergence maps, coming either from the denoiser or from the training set, and outputting a probability between 0 and 1 of the denoised image being real. The model is then trained with a combination of a standard adversarial loss and an l1 loss between the recovered denoised mass-map and the truth from simulations. It is to be noted here that this adversarial model is not a generative model, the denoiser does not take random variables as an input and is therefore deterministic. Instead the adversarial loss can be understood as a learned similarity metric to compare recovered to true map. Training such a model typically requires a set of ray-traced lensing simulations, that are corrupted to include the same noise properties and masks as present in the data. \cite{Shirasaki2021} generated mock HSC observations including photometric redshift uncertainties, shape measurement uncertainties, realistic galaxy ellipticity noise and distribution on the sky, and actual HSC survey masks. On simulations the authors find that about 60\% of the peaks identified on the denoised maps have significant clusters counterparts, against about 85\% of positive matches on true maps, highlighting that the recovered map still correlate well with real structures, and the authors further show that the 1-point statistics of the recovered map shows stronger cosmological dependence than the noisy maps, hinting at interesting applications in constraining cosmological parameters. 
While this approach provides empirically good results, one drawback of this pix2pix training is that the recovered map does not have a clear Bayesian interpretation. As we will see below, subsequently developed techniques abandon this effective adversarial training but gain a proper Bayesian interpretation of the output of the models.

In \cite{Jeffrey2020}, the authors introduce a method, called \texttt{DeepMass}, using a similar Unet architecture but trained under a simple Mean Squared Error loss between true convergence map and output from the Unet. As highlighted by the authors, a regression model trained under an MSE implicitly learns to predict the mean of the posterior distribution of the target given the input. In the present case, the authors generate a suite of ray-tracing lensing simulations matching the DES Science Verification setting, the noiseless convergence maps from these simulations provide an implicit prior, while the simulated shear observations (including realistic noise and masks) provide an implicit likelihood. By training the model to reconstruct the true convergence given simulated shear data under an MSE loss, the model will therefore learn to output the mean posterior convergence map, under the implicit prior and implicit likelihood that are provided by the training set. In this DES SV setting, the authors demonstrate that this approach leads to an 11\% improvement in MSE evaluated on simulations compared to a standard Wiener filter approach.

\begin{figure}
	\includegraphics[width=\columnwidth]{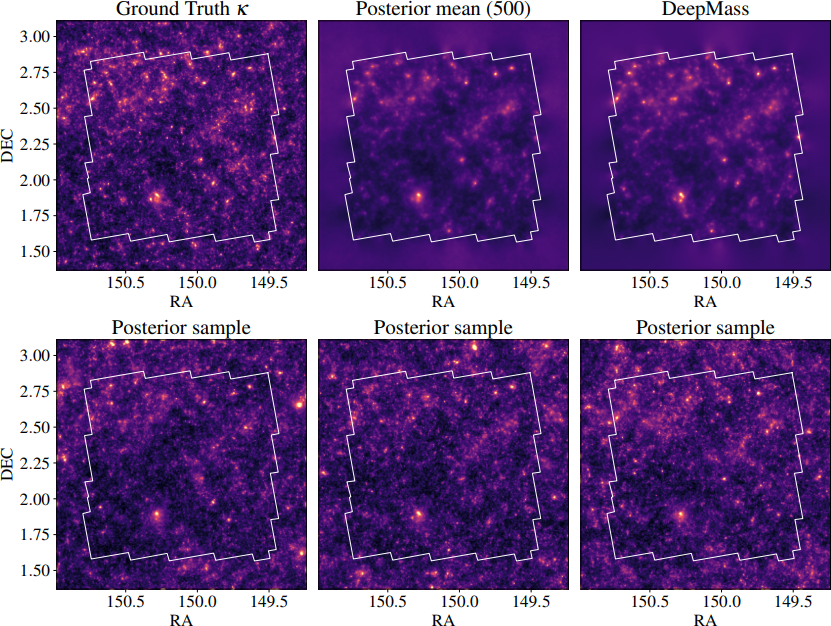}
	\caption{Illustration of weak lensing mass-map reconstructions in a simulated COSMOS survey setting with the posterior sampling method of \cite{Remy2022} and the \texttt{DeepMass} direct posterior mean estimation method  of \cite{Jeffrey2020}. As can be seen, individual posterior samples (bottom row) are visually similar to a real convergence map (e.g. ground truth at the top left) but exhibit variability on structures not strongly constrained by data (e.g. outside of the survey region marked by the white contours). The top row illustrates that the \texttt{DeepMass} estimate indeed recovers the Bayesian posterior mean.}
	\label{fig:remy2022}
\end{figure}

While \cite{Jeffrey2020} had the benefit of providing a Bayesian understanding of the network output, it did not provide any sort of uncertainty on the recovered map, which makes the interpretation and scientific exploitation of these results difficult. To overcome these limitations, \cite{Remy2022} introduced an approach allowing to sample from the full posterior distribution of the mass-mapping problem. They proposed to use a similar U-Net architecture, not to directly estimate the convergence map, but to learn from simulations a generative prior on convergence maps using a Denoising Score Matching technique \citep{Song2019}. With this approach, it can be shown that a neural network trained as a Gaussian denoiser under a simple MSE loss will actually learn the score of the data distribution, i.e. the gradient of the log likelihood of the data. Once trained on simulations, this U-Net gives explicit access to the prior. The authors show that it is possible to combine this learned prior with an explicit data likelihood in an Hamiltonian Monte-Carlo sampling procedure to sample from the full posterior distribution of the problem. \autoref{fig:remy2022} illustrates on the bottom row posterior samples achieved with this method on a simulation of the HST/ACS COSMOS field, compared to the ground truth (top left). Most interestingly, it is shown that the mean of the posterior samples indeed converge to the same solution as the \texttt{DeepMass} estimate.


\paragraph{Initial Conditions Reconstructions}

One particularly interesting problem for the analysis of galaxy surveys is the reconstruction of the initial density field from the observed Large Scale Structure. This can for instance be used to refine Baryonic Acoustic Oscillations (BAO) measurements \citep[e.g.][]{Schmittfull2017}, or part of a Bayesian forward modeling inference scheme \citep{Seljak2017}. 

In a first example of applying deep learning to this problem, \cite{Mao2020} proposed a 3D CNN trained on N-body  simulations to recover the initial density field at $z=10$ given the final density field at $z=0$ under a density weighted mean squared error loss. They find that their convolution model is capable of beating a standard linear reconstruction on scales smaller than $k \leq 0.2h$Mpc$^{-1}$, but under performs on larger scales. Interestingly   they find that their learned inversion can extrapolate to some extent to other cosmological parameters; a model trained on WMAP7 cosmology is capable of reconstructing initial conditions on WMAP5 simulations that lead to a slightly biased BAO signal, but still significantly different from the WMAP7 signal of the training data. 

\begin{figure}
\includegraphics[width=\columnwidth]{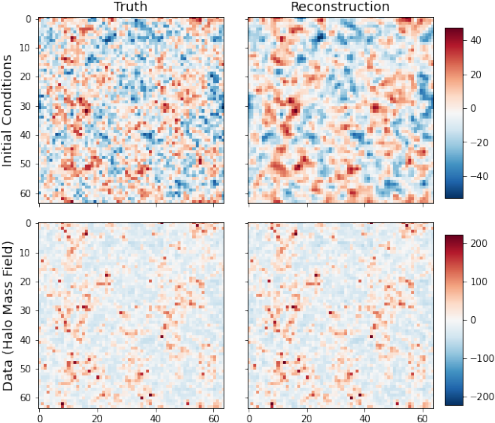}
\caption{CosmicRIM initial conditions reconstruction technique of \cite{Modi2021a} based on a 3D LSTM recurrent neural network and includes explicitly at each iteration the gradients of the data likelihood.  The plot shows a given ground truth initial condition field (top left) and associated final field (bottom left), along with the reconstructed initial conditions (top right) and reconstructed final field (bottom right).}
\label{fig:modi2021}
\end{figure}

Going beyond a direct inversion method, \cite{Modi2021a} proposed an iterative reconstruction scheme based on Recurrent Inference Machines \citep[RIM, ][]{putzky2017}. This approach can be thought of as a learned iterative reconstruction algorithm. At each iteration a recurrent neural network proposes an update of the current reconstruction based on the knowledge of previous iterations and on the gradient of an explicit data likelihood term. In the absence of this neural network, the algorithm would result in a standard gradient descent scheme leading to a Maximum Likelihood Estimation of the initial conditions. By training the Neural Network to minimize at each iteration the Mean Squared Error between the current solution and the true initial conditions, the network will learn both an implicit prior, and a fast inference scheme to minimize the number of updates needed. The result will therefore be a fast convergence towards the mean posterior solution. Most interestingly, in order to compute this explicit likelihood, a differentiable forward model is needed i.e. in this case an N-body simulation. The authors make use of the \texttt{FlowPM} TensorFlow-based fast N-body code \citep{Modi2021} for this likelihood which becomes just a layer within a neural network (\autoref{fig:modi2021}). The authors propose to initialize the reconstruction at the standard linear reconstruction, and show that in only 10 iterations this method yields a better solution than an iterative reconstruction based on 400 iterations of an LBFGS minimizer. 

\vspace{.3cm}
\begin{mdframed}[backgroundcolor=orange!30] 
{\bf Summary of Deep learning for Cosmology}\\
\begin{enumerate}
    \item Emulation
    \begin{itemize}
    \item The main motivation for using deep learning in simulations is to bypass some of the expensive computational steps needed to generate large volume and high-resolution simulations needed to model modern surveys. Applications have targeted in particular: emulating N-body simulations, enhancing the resolution of existing simulations (super-resolution), and learning mappings between the 3D dark matter distribution and dark matter halos or a range of hydrodynamical fields.

    \item Despite many impressive results, these methods have not yet been used for scientific applications. The main difficulty is that deep models can only be used within their training regimes (defined by specific training resolution, specific sets of cosmological parameters, or specific hydrodynamical run), and thus it is unclear whether they will bring concrete computational gains when the cost of the training sets are taken into account. 
 
    \item An alternative class of models, which so far as attracted limited attention, aims instead for a minimal set of parameters, building upon known symmetries and physical insight, greatly reducing the amount of training simulations needed and even opening the possibility of inferring these parameters from the data itself.
    
\end{itemize}
\item Cosmological Inference
\begin{itemize}
    \item Deep learning is opening a new way of comparing observations to theory: 1. it allows for the automatic extraction of cosmological information from high-dimensional data without requiring analytic summary statistics; 2. Neural Density Estimation makes it possible to perform Bayesian inference by leveraging numerical simulations.
    
    \item Although in theory a deep learning approach is statistically sound, it assumes that the simulators provide an accurate physical model of the observations. Any unaccounted for systematics may result in biases, which due to the black-box nature of deep neural networks are difficult to test/detect.
    
    \item A number of papers are starting to apply these methodology on data. In weak lensing, the gains compared to a more standard power spectrum analysis have remained limited on current generation surveys when systematics are included and marginalized over in the analysis.
        
    \item A few applications have proposed to perform high-dimensional inference of cosmological fields (e.g. dark matter maps, reconstructing initial conditions). These works however do not yet attempt joint inference of fields and cosmological parameters.
    
\end{itemize}
\end{enumerate}

\end{mdframed}

\section{Final thoughts: assessing the present and future of deep learning for galaxy surveys}
\label{sec:final_thoughts}

This final section is devoted to extract some indicators about the impact that deep learning has had in the analysis of galaxy surveys. We also attempt to highlight some of the key challenges these methods are facing, which in some cases might prevent or delay the general deployment of deep learning for scientific analysis. Some of these challenges have already been highlighted in the previous sections, but this section tries to extract the most commonly encountered

\subsection{On the penetration of deep learning techniques in astronomy}

We start by questioning what are the deep learning techniques most commonly used in astronomy and how efficiently the rapid progresses made in the ML community reach our community. In the previous sections, we have described different applications of deep learning making use of a variety of techniques. We summarize in \autoref{tbl:deep_techniques}, the broad type of neural network architectures used in the four categories of scientific applications we defined in this review. We have divided the neural network models in seven big groups. CNNs encapsulates all variants of convolutional  neural networks, from Vanilla to more complex Residual Networks. The second group contains, generally speaking, image to image networks such as Encoder-Decoders, Autoencoders but also segmentation specific networks such as Mask R-CNNs or YOLO. The third family of models are Generative Models which include Variational Autoencoders, Generative Adversarial Networks and also Autoregressive models. We then include Bayesian Neural Networks which allow for uncertainty quantification (Mixture Density Networks and Flow models are also included in this category), Recursive Neural Networks and Transformers mostly suited for sequences. The last group is made of Graph Neural Networks. The table first shows that applications in astronomy cover a wide range of deep learning techniques. Although standard CNNs are the most commonly used  - probably because it is the most established approach and because imaging is the most common type of data - other more recent models are regularly applied to astronomical data. On the one side, this reflects the fact that astronomical data is rather diverse - including images, but also spectra, time sequences, simulations and observations. On the other hand, it suggests that the penetration of new ML techniques is efficient. State-of-the art methods are rapidly applied to astronomy. This is likely a consequence of the fact that, even advanced ML methods are becoming increasingly easy to use for non-experts. It is almost straightforward to test a new technique on an astrophysical problem with current high level implementations. The downside is that,  generally speaking, the methods are often applied \emph{blindly} off the shelf, with little domain specific adaptation. Consequently, a feature that is still lacking in a fair amount of the applications of deep learning to astronomy is the inclusion of previous physical knowledge into the data driven models. This can be done by adapting the loss functions or by modifying the neural network architectures to incorporate known symmetries (see work by~\citealp{Villar2021a,Bowles2021}). It obviously requires deeper knowledge of the machine learning aspects which is something that will likely take more time. 

Another interesting feature that emerges from \autoref{tbl:deep_techniques} is that training on simulations is the most common approach in astronomy. Almost all supervised approaches rely at some stage on simulated data. It reflects that the samples of labeled data remain small and/or that the measurements in observations are noisy. Relying on simulations to train the models adds however an important element of uncertainty to all applications. Machine learning approaches are indeed very sensitive to domain shift issues. According to \autoref{tbl:deep_techniques}, almost all recent applications are affected by those at some extent. We will discuss this further in \autoref{sec:challenges}. 

\begin{table*}

\begin{tabularx}{\textwidth}[t]{|lll|X|X|X|X|X|X|X|} \hline
\backslashbox{Application}{Model} &  & & CNNs  & Enc. & Gene. & BNN & RNN & Trans.& GNN \\\hline \hline

1. Computer Vision & Classification & Morphology &  \multicolumn{1}{c}{\cellcolor{blue!25}\checkmark} &  \multicolumn{1}{c}{\cellcolor{red!25}\checkmark} & & & & &\\
 &  & Strong Lenses  &  \multicolumn{1}{c}{\cellcolor{blue!25}\checkmark*} &  \multicolumn{1}{c}{\cellcolor{red!25}\checkmark*} & & & & &\\
  &  & Transients  &   &   & & & \multicolumn{1}{c}{\cellcolor{blue!25}\checkmark*$^)$}&\multicolumn{1}{c}{\cellcolor{blue!25}\checkmark*$^)$} & \\
 & Segmentation &  &   &  \multicolumn{1}{c}{\cellcolor{blue!25}\checkmark*} &\multicolumn{1}{c}{\cellcolor{red!25}\checkmark*} & & & &\\ \hline
2. Galaxy Properties &  & Photoz &  \multicolumn{1}{c}{\cellcolor{blue!25}\checkmark} &   & &\multicolumn{1}{c}{\cellcolor{blue!25}\checkmark} &  & &\\ 
 &  & Structure &  \multicolumn{1}{c}{\cellcolor{blue!25}\checkmark*$^)$} &   & & & & & \\ 
 &  & Stellar Populations &  \multicolumn{1}{c}{\cellcolor{blue!25}\checkmark*} &   & & & & &\\
 &  & Lensing  &  \multicolumn{1}{c}{\cellcolor{blue!25}\checkmark*} &   & &\multicolumn{1}{c}{\cellcolor{blue!25}\checkmark*} &  & &\\ 
  &  & Physical Processes  &  \multicolumn{1}{c}{\cellcolor{blue!25}\checkmark*} &   & &  &  & &\\ 
    &  & Dark Matter  &  \multicolumn{1}{c}{\cellcolor{blue!25}\checkmark*} &   & &\multicolumn{1}{c}{\cellcolor{blue!25}\checkmark*} &  & &\multicolumn{1}{c}{\cellcolor{blue!25}\checkmark*}\\ \hline
3. Discovery &  & Visualization &   \multicolumn{1}{c}{\cellcolor{red!25}\checkmark}&\multicolumn{1}{c}{\cellcolor{red!25}\checkmark} &\multicolumn{1}{c}{\cellcolor{red!25}\checkmark} &  & & &\\ 
 &  & Outliers & \multicolumn{1}{c}{\cellcolor{red!25}\checkmark}  &\multicolumn{1}{c}{\cellcolor{red!25}\checkmark} &\multicolumn{1}{c}{\cellcolor{red!25}\checkmark} &  & \multicolumn{1}{c}{\cellcolor{red!25}\checkmark}& &\\ 
  &  & Laws &   & & &  & & &\multicolumn{1}{c}{\cellcolor{red!25}\checkmark*}\\ \hline

4. Cosmology &  & Emulation &  \multicolumn{1}{c}{\cellcolor{blue!25}\checkmark*} & \multicolumn{1}{c}{\cellcolor{blue!25}\checkmark*}  & \multicolumn{1}{c}{\cellcolor{blue!25}\checkmark*}&  \multicolumn{1}{c}{\cellcolor{blue!25}\checkmark*}& & &\\ 
 &  & Cosmological inference &  \multicolumn{1}{c}{\cellcolor{blue!25}\checkmark*} &   & \multicolumn{1}{c}{\cellcolor{blue!25}\checkmark*}&  \multicolumn{1}{c}{\cellcolor{blue!25}\checkmark*}& & &\\ \hline

\end{tabularx}
\vspace{0.1cm}
\caption{Overview of the different deep learning techniques used in the fields of galaxy formation and cosmology, divided by type of application (see text for details). CNNs: Standard classification and regression Convolutional Neural Networks including modern architectures such as ResNets. Enc: Encoder-Decoder networks and variants. Gene: Generative Models. BNNs: Bayesian Neural Networks; we also include Mixture Density Networks. RNNs: Recursive Neural Networks. Trans: Transformers. GNNs; Graph Neural Networks. A blue (red) background indicates supervised (unsupervised) learning. The star symbol highlights applications which require simulations to train the neural networks. The bracket after the star symbol indicates that the use of simulations is not always mandatory. }
\label{tbl:deep_techniques}
\end{table*}

\subsection{Measuring the impact of deep learning}
\label{sec:impact}

\begin{figure*}
\centering
\vspace{-10 pt}

\subcaptionbox{}{\includegraphics[width=0.45\textwidth]{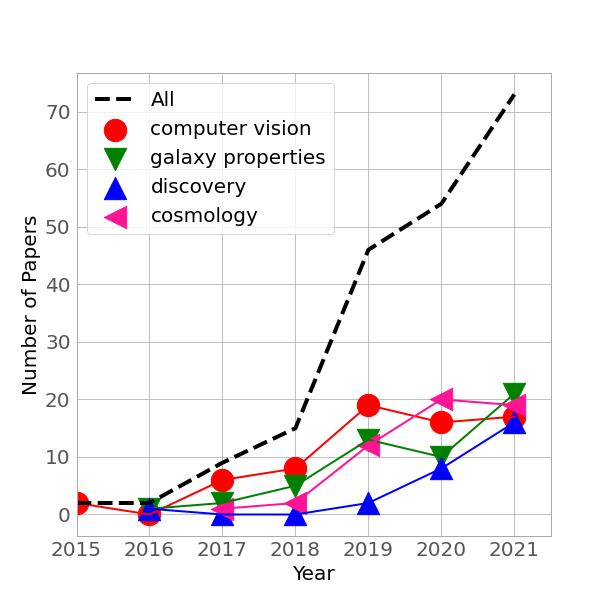}}%
\qquad
\subcaptionbox{}{\includegraphics[width=0.45\textwidth]{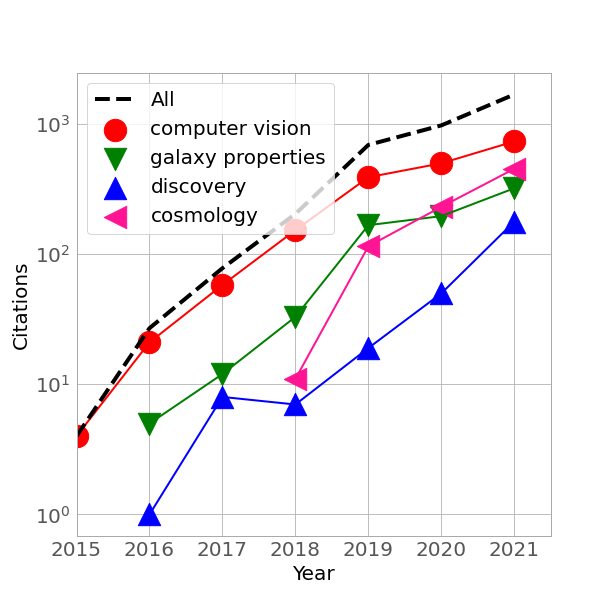}}%
\qquad
\subcaptionbox{}{\includegraphics[width=0.45\textwidth]{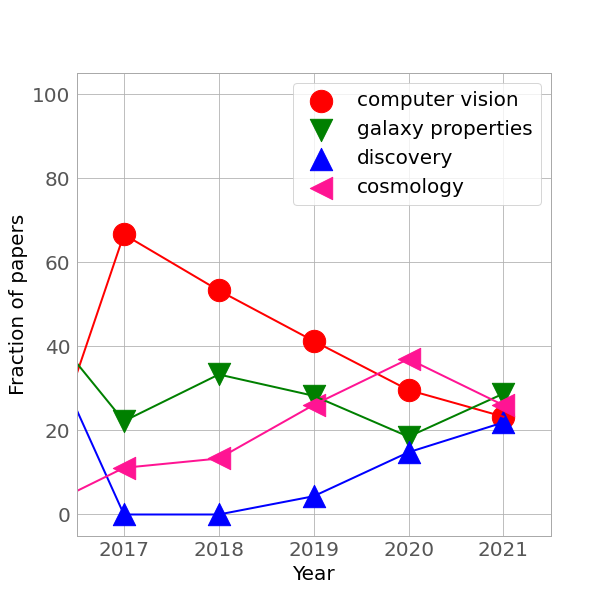}}%
\qquad
\subcaptionbox{}{\includegraphics[width=0.45\textwidth]{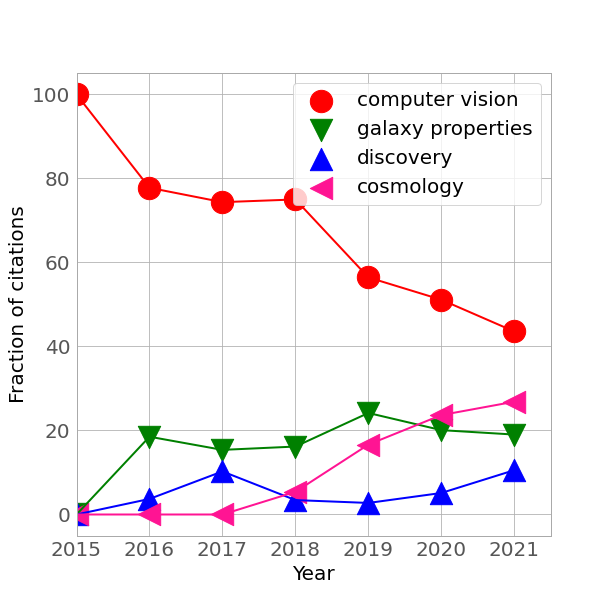}}%
\qquad

\vspace{-5 pt}
\caption{Impact of works using deep learning for galaxy surveys. Each symbol shows a different class of application as labeled (see text for details). The top left and right panels show the number of papers and number of citations as a function of time respectively. The bottom left and right panels show the fraction of  of papers and citations in each class of application.}%
\label{fig:impact1}%
\end{figure*}

We now move to measuring the impact of works using deep learning in the astronomical literature. We have seen in the introduction that the number of papers making use of neural networks has increased exponentially over the past half decade. In this subsection, we try to measure the impact of these works with some standard metrics. \autoref{fig:impact1} shows the evolution of the number of papers, number of citations and average citations per paper in the period 2015-2021. The publications are divided in the four different categories defined in this review, i.e. computer vision, galaxy properties, discovery and cosmology. We have only included in the figure the works explored for this work. As a consequence, it is very likely that the figure is not complete and that the numbers presented are closer to a lower limit. However, it should provide a good overview of the general trends and represents a more controlled experiment than a purely automatic search. We also emphasize that the division in categories is a choice by the authors of this review. Therefore, there is some obvious overlap between the different types of applications. 

Nevertheless, the figure reveals some interesting behaviors. We first confirm the global increasing trend of the number of papers using deep learning for galaxy surveys. Since 2015 there is a clear  exponential increase. In 2021, there are at least 70 papers using deep learning in the context of galaxy surveys, while there were less than 5 in 2015. This is factor of $\sim15$ increase and implies more than a paper per week on average. If we look at the division per type of application, we see that computer vision type of applications (i.e. classification, segmentation) concentrate the largest fraction of publications. All the other remaining classes share similar fractions. However, there is clearly a decreasing trend of the relative importance of classification and segmentation applications. While the fraction of these papers was around $70-80\%$ in 2015-2017, it is only of $\sim20\%$ for papers published in 2021. The trends seem to suggest a diversification of the applications of deep learning to astronomy moving from computer vision tasks - mainly classification and segmentation - to a large variety of different applications.  The number of yearly works for data processing seems to flatten indeed after 2019, while other applications like data exploration (\emph{discovery}) rapidly rise. 

A similar behavior is observed in terms of citations. In 2016, roughly $80\%$ of the citations are for papers using deep learning for computer vision tasks. The fraction is steadily decreasing, but still remains close to $\sim50\%$ in 2021 even though the number of papers is only 20\%. It probably reflects a delay between the publication time and the time papers start to be cited.

\begin{figure}
\centering
    \includegraphics[width=\linewidth]{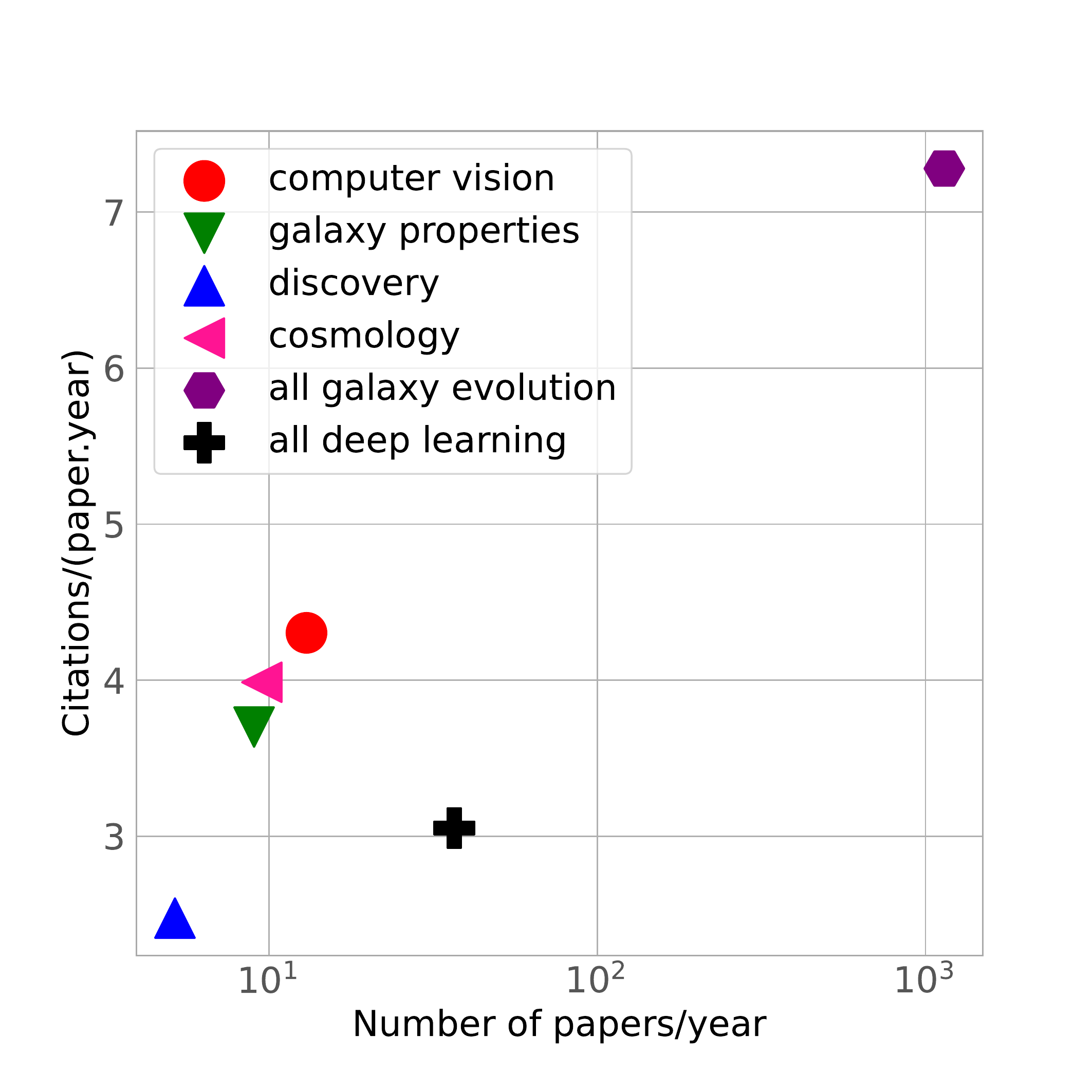}

    \caption{Number of citations normalized by the number of papers and years (from the first publication in that category) as a function of the number of papers per year.  Each symbol shows a different category as defined in this work (see text for details). The "all galaxy evolution" group includes an automatic search of all publications in the field of galaxy formation (with or without deep learning).}
    \label{fig:cit_papers}
\end{figure}

We attempt to remove this effect in \autoref{fig:cit_papers}. We plot the number of papers per year as a function of the number of citations per paper and per year, averaged over the time elapsed since the first citation in a given group. That way, for papers focusing on computer vision, the time window is set to 6 years, while for the others, we consider 4 years. We also show in the figure, for reference, the location of publications flagged with the keywords \emph{galaxy evolution}, according to the NASA ADS search engine. Using this normalization, we observe several interesting trends. Works using deep learning for galaxy surveys, represent roughly $>5$\% of all works focusing on galaxies. This is not a large fraction but it is still remarkable given the relatively recent emergence of deep learning. They receive on average $\sim1.5$ times less citations per publication. It suggests that the impact of deep learning papers remains moderate as compared to the average. We might speculate about possible reasons. One possibility is that most of the works making use of deep learning are thought in preparation of future big data surveys which have not arrived yet (e.g. Euclid, LSST). The works are therefore more a demonstration of feasibility. It could be argued that existing surveys such as DES or SDSS for example are already good targets for data driven science. This is certainly true, and as we have seen in this review, there are many works targeting these surveys. However, the sizes and dimensionality of current surveys still allow one to use relatively optimized codes and leave some room for some manual checks. This is also supported by the trends observed when the deep learning papers are divided by topic. The highest citation rate is measured for works focusing on simulation (i.e. emulation of cosmological simulations), which by definition do not require new observational data.  Publications focused on unsupervised discovery, which strongly rely on new data being available, present the lowest impact, although they show a strong increase in numbers (\autoref{fig:impact1}). It could also be that the technology is still too young, and it is just a matter of time that the impact increases. The majority of the works are still at the proof-of-concept stage and have not reached the deployment stage. This could also mean eventually that there are some challenges / limitations which have not been deeply explored yet and that prevent these new methods to be fully adopted by the community. We explore these challenges more carefully in the following section.

\subsection{Challenges}
\label{sec:challenges}

We list in \autoref{tbl:challenges} what we think are some of the major challenges that deep learning works face and which need to be addressed in the coming years by the community based on the works reported in this review. Some of these challenges are not specific to the astronomical community and can benefit from solutions arising from the field of Machine Learning. However, in some cases, the requirements are more strict in astronomy. The table also provides some possible solutions along with some list of - non exhaustive - references which have explored these solutions. 

\subsubsection{Small (and biased) labeled datasets}
\label{sec:ch1}
A major challenge in applications of deep learning for astronomy is the lack of large enough labeled data sets to train supervised deep learning models. This issue has been highlighted multiple times during the description of works.  We have seen in \autoref{sec:low_level} that deep learning was first applied to classification of galaxy morphologies. This was mainly triggered by the fact that large samples of labeled images existed from previous citizen science efforts. A similar behavior is observed in the radio domain. However, the vast majority of the data are unlabeled preventing the deployment of supervised deep learning methods. Although this issue is common to many disciplines, it is particularly critical in astronomy since the properties of the images change depending on the instrument used. Therefore, a sample of labeled images from one observational dataset cannot directly be exported to another survey, hampering the use of deep learning. We notice however, that large dataset are not always needed. Some works (e.g.~\citealp{Ackermann2018, Walmsley2019}) have demonstrated that reliable results can be obtained with small training samples. This is probably a consequence of the relatively limited complexity of astronomical images compared to natural images. 

Nevertheless, reducing the amount of labels is usually desirable for most supervised applications. There exist several types of solutions listed in \autoref{tbl:challenges}. An obvious one is to manually label more data. It is however very time consuming and cannot be done for every new survey. Active learning approaches allow to label only the more informative examples for the deep learning model, hence reducing the needed time. Active learning has been particularly explored in the framework of galaxy morphology (e.g.~\citealp{walmsley2020}). Another fairly straightforward solution is Transfer Learning. This is usually done by training a model in a similar dataset to the target dataset for which labeled data exist. The neural network weights are then refined with a small sample of labeled examples from the target sample. Transfer learning has also been explored for galaxy morphology (e.g.~\citealp{dominguezsanchez2019}). Other works use simulations to overcome the lack of labels. As seen in the previous section, using simulations is a fairly common approach in astronomy. This is probably because the complexity of astronomical objects - especially images - allows one to obtain quite realistic simulations with fairly simple approaches. This is the case, for instance, for the classification of strong lenses, which exclusively rely on simulated training sets. Although this is a very efficient approach, it also implies some potential issues related to the change of domain. This is a general challenge which we have grouped into Challenge 4 in \autoref{tbl:challenges}. Finally, the ML community has recently started to explore the so-called self-supervised approaches to reduce the amount of labeled examples. The underlying idea is to first compute some meaningful representations of the data in an unsupervised way and then use the obtained representations to train a supervised network. Because the self-supervised step filters out non informative features, the amount of needed examples for training is reduced.~\cite{Hayat2021} and~\cite{Sarmiento2021} have recently explored self-supervised learning for galaxy morphology, photometric redshifts and galaxy kinematics respectively. Contrastive self-supervised learning can also be potentially employed to reduce the gap between simulations and observations as well as Autoencoder representations. 

In addition to the size of the labeled datasets, a major challenge is how representative are those. A common problem in astronomy is that the sample for which labels are available does not always overlap with the inference datasets. This typically could happen because it is easier to obtain labels for a sub population of objects. The extrapolation of the trained neural network results to a dataset which was not exactly used for training is usually a problem. This issue has become very obvious for photometric redshift estimation (see~\autoref{sec:galprop}) in which ML approaches tend to fail for examples poorly represented in the training. The only solution adopted by the community has been to obtain more training data. 

\subsubsection{Uncertainty}
Uncertainty quantification is a major challenge for applications of deep learning to physical sciences. This is a major difference with respect to standard computer vision applications on natural imaging, which usually do not require well calibrated uncertainties. Therefore, standard deep learning methods directly exported to astronomy do not quantify uncertainties and this is generally not acceptable for scientific applications. Some tasks such as classification can sometimes be accepted without precise uncertainty and rely on statistical measurements (i.e. ROC, Precision-Recall curves). However, if deep learning is intended to be used for accurate measurements of galaxy properties or to constrain models, they need to incorporate error measurements. We see a changing trend in the community. The early efforts did not include uncertainty estimation; however more and more works are introducing at least some quantification of errors. The community has explored several solutions, but the problem is far from being solved. A promising approach are Bayesian Neural Networks (BNNs) which aim at measuring posterior distributions from a Bayesian perspective. Several approaches have been proposed over the past years and the implementation has also become more straightforward, which favors the use by non-experts.  For example BNNs are the common approach for the modeling of strong lenses for example (e.g.~\citealp{PerreaultLevasseur2017}). However, one need to keep in mind that BNNs compute an approximation of the true posterior distributions which sometimes might not be accurate enough (see review by~\cite{Charnock2020}). Density Estimator networks such as Regressive Flows or Mixture Density Networks are another approach to sample from complex posterior distributions. Although several works in astronomy pay careful attention to the quantification of uncertainties, they represent still a minority of the deep learning literature in astronomy. We believe this is a major challenge for the future deployment of deep learning in the analysis of deep surveys.  

\subsubsection{Interpretability}
Related to the problem of uncertainty quantification is interpretability. By moving to a data driven approach to data analysis, we unavoidably loose some control on what type of information is extracted and used by the neural network models. This effect is sometimes referred as the \textit{black box effect}. Deep learning models are in general opaque black boxes which perform complex non linear mappings,  difficult to unveil. Although this might be a problem for most applications, it is particularly worrying for scientific ones, and therefore constitutes a major challenge for the acceptance of deep learning by the astronomical community.  For example, not properly understanding the information used can generate some biases, as demonstrated by~\cite{Dhar2022} which show that deep learning based classifications are sensitive to the location of the galaxy in the sky. The field of interpretability of deep neural networks is even less developed than the one of uncertainty estimation and in general the techniques employed provide a limited amount of information. A common approach is to identify the regions of the input data that provide most of the information for the network decisions. These methods can be generally useful to identify biases - e.g. the neural network model focuses on background noise - but are still far from providing any physical interpretation of what is being measured. Some works have looked at ways of enhancing the explainability~\citep{HuertasCompany2018,Bhambra2022} but the amount of extracted information typically consists on the identification of pixels in the input images which contribute most to the decision. Although this is certainly valuable information to identify biases in particular, it does not provide a true explainability in terms of physical meaning. Interpreting the results is easier when the inputs are parameters instead of raw pixel values. In such cases, there exists the possibility of performing symbolic regressions to try to dig into the relations learned by the neural networks (e.g.~\citealp{Cranmer2020,VillaescusaNavarro2022}). An interesting research line to ease interpretability is the inclusion of prior physical constraints in the neural network model. Architectures that preserve known symmetries of the physical problem are for example an interesting way to keep a control on what the networks are extracting (e.g.~\citealp{Scaife2021,Villar2021a, Bowles2021}).     

\subsubsection{Domain shift}

 Another key challenge faced by deep learning applications to astronomy is related to the change of domains between the training and the inference steps. As highlighted in \autoref{tbl:deep_techniques}, a majority of the applications of deep learning to astronomy rely on simulations for training the models. This is typically justified because the availability of labeled samples is limited (see \autoref{sec:ch1}),  because we aim at accessing information that is only available on simulations, e.g. galaxy mergers, dark matter or information about the cosmological model or because the likelihood is intractable but we have an idea on how to simulate the data (see review by~\citealp{Cranmer2019}). For example, the work by~\cite{Bottrell2019} examines very carefully the impact of using more or less realistic simulations for training. The community has explored several solution to mitigate the impact of training on simulations and apply to data. A simple approach is to use transfer learning. This is only possible when there exist some measurements in observations which can be used to fine tune the weights from the neural network model trained on simulations (e.g.~\citealp{Tuccillo2018}). This is not always possible though, especially when we try to infer the parameters of a model and has also the problem of propagating the biases of any existing method previously applied to observations. Domain adaptation techniques are another alternative approach which attempt the make the features learned by the model agnostic to the differences between domains. As opposed to transfer learning, this is done during training so that no domain specific features are learned.~\cite{Ciprijanovic2021a} have recently quantify the gain of such techniques for the identification of galaxy mergers. It remains however an open issue for the future.  
 
 \subsubsection{Benchmarking and Deployment}
 
 A final challenge which has not been discussed much in the literature so far is related to how the different approaches can be robustly compared. As we have thoroughly described in this review, the past years have witnessed an emergence of a large number of deep learning methods applied to a diversity of scientific topics. In many cases, the results are shown for a specific dataset, with a specific configuration, which makes it hard to compare with existing approaches. The ML community has been using since many years, what is called standardized datasets. These are common datasets which are publicly shared and on which any new approach is usually tested. This benchmarking approach has been an important channel for progress in the community. The astronomical one is not used to this type of approach and therefore, with some noticeable exceptions (e.g. PLAsTiCC, Galaxy Zoo for classification), we lack of a coherent way of comparing methods. We argue that this is an important aspect on which to work as a community to boost progress. Having standardized datasets on which test models can not only help comparing methods but also identify pitfalls and biases and therefore contribute to make the neural network models more robust. This is an important step towards a full deployment of these approaches into scientific pipelines.

\begin{center}
\fontfamily{cmss}\selectfont
\begin{table*}
\begin{tabularx}{\textwidth}[t]{XX}
\arrayrulecolor{blue}\hline
\rowcolor{lightBlue} \textbf{\textcolor{blue}{Challenge 1 Small (and biased) labeled datasets}} & \\
\hline
Solution 1.A Transfer Learning & 
\begin{minipage}[t]{\linewidth}%
 \cite{dominguezsanchez2019}~\cite{Samudre2022} ~\cite{Lukic2019}
\end{minipage}\\

\arrayrulecolor{mygray}\hline

Solution 1.B Simulated dataset  &
\begin{minipage}[t]{\linewidth}%
\cite{jacobs2017}~\cite{VegaFerrero2021} 
\end{minipage}\\

\hline

Solution 1.C Self-supervised learning &
\begin{minipage}[t]{\linewidth}%
\cite{Hayat2021}
\end{minipage}\\

\hline

Solution 1.D Active Learning and similar &
\begin{minipage}[t]{\linewidth}%
\cite{walmsley2020}
\end{minipage}\\

\arrayrulecolor{blue}\hline
\rowcolor{lightBlue} \textbf{\textcolor{blue}{Challenge 2 Uncertainty}} & \\
\hline

Solution 2.A Bayesian approximations &
\begin{minipage}[t]{\linewidth}%
\cite{walmsley2020}~\cite{PerreaultLevasseur2017}
\end{minipage}\\

\hline

Solution 2.B Density Estimators & \cite{KodiRamanah2020}
\begin{minipage}[t]{\linewidth}%

\end{minipage}\\

\hline
\rowcolor{lightBlue} \multicolumn{2}{l}{%
\textbf{\textcolor{blue}{Challenge 3 Interpretability}}} \\
\hline
Solution 3.A Saliency maps and similar &
\begin{minipage}[t]{\linewidth}%
\cite{HuertasCompany2018, Bowles2021, Bhambra2022}
\end{minipage}\\

\hline

Solution 3.B Symbolic regression &
\begin{minipage}[t]{\linewidth}%
\cite{Cranmer2020}
\end{minipage}\\

\hline

Solution 3.C Physics informed &
\begin{minipage}[t]{\linewidth}%
\cite{Scaife2021, Villar2021a,Charnock2019}
\end{minipage}\\

\hline
\rowcolor{lightBlue} \multicolumn{2}{l}{%
\textbf{\textcolor{blue}{Challenge 4 Domain shift}}} \\
\hline

Solution 4.A Transfer Learning &
\begin{minipage}[t]{\linewidth}%
 \cite{Tuccillo2018,dominguezsanchez2019,ghosh2020}
\end{minipage}\\

\hline 

Solution 4.A Domain Adaptation &
\begin{minipage}[t]{\linewidth}%
\cite{Ciprijanovic2021}
\end{minipage}\\

\hline
\rowcolor{lightBlue} \multicolumn{2}{l}{%
\textbf{\textcolor{blue}{Challenge 5 Benchmarking}}} \\
\hline

Solution 5.A Standardized datasets &
\begin{minipage}[t]{\linewidth}%
 PLAsTiCC, SKA data challenge, Galaxy Zoo
\end{minipage}\\



\hline
\end{tabularx}
\vspace{0.1cm}
\caption{Major challenges that deep learning works applied to astronomy might suffer and that will need to be addressed in the coming years. We also provide elements of solutions already being explored along with the corresponding references.}
\label{tbl:challenges}
\end{table*}
\end{center}

\section{Summary}

This work reviews the use of modern deep learning techniques for the analysis of deep galaxy surveys. Although machine learning has been used in astronomy for several decades, the recent deep learning revolution as induced an unprecedented number of new works exploring the use of these novel techniques in astronomy. The purpose of this review is to assess how deep learning has been used in astronomy and what are the key achievements and challenges. We do not describe however the technical aspects of deep learning techniques.

We have divided deep learning applications in four broad categories defined by the type of application: 1- computer vision, 2 - galaxy properties, 3 - discovery and 4 - cosmology. The first sections of these review (\autoref{sec:low_level} to \autoref{sec:acc}) describe the most relevant works in each category. A summary of the main points for each type of application is included at the end of the corresponding section. The first category (\autoref{sec:low_level}) includes general computer vision applications such as classifications and object detection. The second (\autoref{sec:galprop}) is related to measure galaxy properties, deep learning acts as a fast emulator and as universal aproximator. The third category (\autoref{sec:discovery}) illustrates all efforts related to visualization and identification on new types of objects. The fourth group (\autoref{sec:acc}) contains publications which use deep learning for cosmology. Namely we include two main applications:  more efficient simulation and cosmological inference.

The last section (\autoref{sec:final_thoughts}) focuses on extracting some lessons about the use of deep learning techniques in astronomy, on the impact they have had so far and on what are - in our humble opinion - the key challenges that will need to be addressed in the near future. We list below key take away messages from this analysis:

\begin{itemize}
    \item The first work using deep learning in astronomy is from 2015. Since then, the number of works using deep learning for galaxy surveys has increased exponentially. There is factor of $\sim15 $ increase between 2015 and 2021. 
    \item The most common deep learning method used are sequential Convolutional Neural Networks with different degrees of complexities. However, there is a good variety of techniques which have been tested for astronomy including recent developments such as Transformers or self-supervised approaches. This reflects a \emph{democratization} of these techniques which are becoming increasingly easy to use. However, the  methods are often applied with a limited amount of physically driven modifications. The combination of previous physical knowledge with data driven models is still an open issue, even if it is a rapidly changing field.
    \item The majority of the works ($>50\%$) focus on what we call computer vision applications - which essentially include classification and segmentation. This is also the field in which deep learning has brought the most important breakthroughs. These are applications which are more prone to a direct import from the ML community. However, we measure a diversification of the applications which span a variety of topics such as the acceleration of cosmological simulations, the inference of galaxy properties or constraints on cosmology cosmology. 
    \item The works using deep learning  represent $\sim 5\%$ of all works on galaxy formation which is remarkable. The receive however $\sim 3$ citations per paper and per year on average. This is roughly $\sim1.5$ times less citations than publications on galaxies for example, although computer vision applications also perform better in this front.  It suggests a moderate impact of deep learning so far which might be explained because most of the works are still at the exploratory stage. 
    \item We have identified a set of 5 major challenges which frequently appear in deep learning applications and that we believe need to be addresses in the nearby future. 1- Small labeled datasets; 2 - Uncertainty estimation; 3- Interpretability; 4 - Domain shift and 5 - Benchmarking.  
    
\end{itemize}

\appendix

\section{Acronyms}
\label{app:acron}

We summarize in this appendix the acronyms used for designating types of deep learning methods, galaxy surveys as well as simulated datasets. For every method we also indicate a reference where more details can be obtained.

Machine Learning:\\
\begin{itemize}
    \item \textbf{ANN:} Artificial Neural Network
    \item \textbf{ARF:} Auto Regressive Flow - \cite{Papamakarios2017}
    \item \textbf{BNN:} Bayesian Neural Network - \cite{Charnock2020, Goan2020}
    \item \textbf{CAE:} Convolutional Autoencoder
    \item \textbf{CNN:} Convolutional Neural Network
     \item \textbf{DT:} Decision Tree
    \item \textbf{(W)GAN:} (Wasserstein) Generative Adversarial Network - \cite{Goodfellow2014, Arjovsky2017}
    \item \textbf{GNN:} Graph Neural Network
    \item \textbf{Mask R-CNN:} Mask Region Convolutional Neural Network - \cite{He2017}
    \item \textbf{MLP:} Multi-Layer Perceptron 
    \item \textbf{MDN:} Mixture Density Network - \cite{Bishop1994}
    \item \textbf{RF:} Random Forest
    \item \textbf{RNN:} Recursive Neural Network
    \item \textbf{SOM:} Self-Organizing Map
    \item \textbf{SVM:} Support Vector Machines
    \item \textbf{VAE:} Variational Autoencoder - \cite{Pu2016}
    \item \textbf{YOLO:} You Only Look Once - \cite{Redmon2015} 
    
        \end{itemize}
        
        \vspace{1cm}
        
        Deep galaxy surveys where deep neural networks have been applied:\\
        
        \begin{itemize}
         \item \textbf{\href{http://arcoiris.ucolick.org/candels/}{CANDELS}: }Cosmic Assembly Near-Infrared Deep Extragalactic Legacy Survey; \cite{Koekemoer2011}
          \item \textbf{DECALS:} The Dark Energy Camera Legacy Survey; \cite{Dey2019}
           \item \textbf{\href{https://www.darkenergysurvey.org/}{DES}:} The Dark Energy Survey; \cite{DESC2016}
           
          \item \textbf{Euclid: } \cite{Laureijs2011}
          \item \textbf{\href{https://hsc.mtk.nao.ac.jp/ssp/}{HSC}: } Hyper Suprime Cam \cite{Aihara2018}
           \item \textbf{{MANGA}: } Mapping Nearby Galaxies at APO; \cite{Bundy2015} 
            \item\textbf{\href{https://panstarrs.stsci.edu/}{Pan-STARRS}:} Panoramic Survey Telescope and Rapid Response System; \cite{Chambers2016}
            \item \textbf{\href{https://pausurvey.org/}{PAU}: } Physics of the Accelerating Universe; 
            \item \textbf{SDSS:} Sloan Digital Sky Survey

            \item \textbf{\href{https://www.lsst.org/}{LSST}: } Legacy Survey of Space and Time; \cite{Ivezic2019}
            
            \item \textbf{\href{https://www.splus.iag.usp.br/}{S-PLUS}: Southern Photometric Local Universe Survey}; \cite{MendesdeOliveira2019}
            \item \textbf{\href{http://www.gama-survey.org/}{GAMA}: } Galaxy and Mass Assembly; \cite{Driver2011} 
             \item \textbf{\href{https://www.ztf.caltech.edu/}{ZTF}: } Zwicky Transient Facility; \cite{Bellm2014} 
            
        \end{itemize}

     \vspace{1cm}
        
         Simulated datasets used to train deep neural networks:\\
    
    \begin{itemize}
        \item \textbf{CAMELS: }\cite{VillaescusaNavarro2021b}
         \item \textbf{PLAsTiCC:} Photometric LSST Astronomical Time-Series Classification Challenge
         \item \textbf{IllustrisTNG: }\cite{Pillepich2018}
        \item \textbf{EAGLE: }\cite{Schaye2015}
        \item \textbf{SIMBA:}\cite{Dave2019}
        \item \textbf{VELA:}\cite{Ceverino2015}
        \item \textbf{Horizon-AGN: }\cite{Dubois2014}

    \end{itemize}

\bibliographystyle{pasa-mnras}
\bibliography{bib_jabref}
\end{document}